\documentclass[a4paper,11pt]{article}
\usepackage{amsmath}
\usepackage{amsfonts}
\usepackage{amssymb}
\usepackage[utf8x]{inputenc}
\usepackage{ae}
\usepackage[T1]{fontenc}
\usepackage[english]{babel}
\usepackage{graphicx}
\usepackage{tabularx}
\usepackage{tabulary}
\usepackage{setspace}
\usepackage{tablefootnote}
\usepackage[flushmargin]{footmisc}
\usepackage[comma]{natbib}
\usepackage{array}
\usepackage{float}
\usepackage{caption}
\usepackage[a4paper,left=2.5cm,right=2.5cm,top=3cm,bottom=3cm]{geometry}
\usepackage[colorinlistoftodos]{todonotes}
\usepackage{pdfpages}
\usepackage{longtable}
\usepackage{adjustbox}
\usepackage{booktabs}
\usepackage{lscape}
\usepackage[colorlinks=true, allcolors=blue]{hyperref}

\begin{document} \doublespacing \pagestyle{plain}

\def\ci{\perp\!\!\!\perp}
\begin{center}

{\LARGE Evaluating (weighted) dynamic treatment effects\\by double machine learning}

{\large \vspace{0.8cm}}

{\large Hugo Bodory*, Martin Huber**, and Luk\'{a}\v{s} Laff\'{e}rs+ }\medskip

{\small {*University of St.\ Gallen, Dept.\ of Economics\\ **University of Fribourg, Dept.\ of Economics\\ +Matej Bel University, Dept. of Mathematics} \bigskip }
\end{center}

\smallskip

\noindent \textbf{Abstract:} {\small We consider evaluating the causal effects of dynamic treatments, i.e.\ of multiple treatment sequences in various periods, based on double machine learning to control for observed, time-varying covariates in a data-driven way under a selection-on-observables assumption. To this end, we make use of so-called Neyman-orthogonal score functions, which imply the robustness of treatment effect estimation to moderate (local) misspecifications of the dynamic outcome and treatment models. This robustness property permits approximating outcome and treatment models by double machine learning even under high dimensional covariates and is combined with data splitting to prevent overfitting. In addition to effect estimation for the total population, we consider weighted estimation that permits assessing dynamic treatment effects in specific subgroups, e.g.\ among those treated in the first treatment period. We demonstrate that the estimators are asymptotically normal and $\sqrt{n}$-consistent under specific regularity conditions and investigate their finite sample properties in a simulation study. Finally, we apply the methods to the Job Corps study in order to assess different sequences of training programs under a large set of covariates.
}

{\small \smallskip }



{\small \smallskip {\scriptsize We have benefited from comments by Jelena Bradic, Saraswata Chaudhuri, Yingying Dong, Arturas Juodis, Frank Kleibergen, Jonathan Roth, Vasilis Syrgkanis, Davide Viviano, and seminar participants at McGill University, the University of Amsterdam, the University of Duisburg-Essen, the University of Bolzano/Bozen, and the University of California Irvine (all online).
Addresses for correspondence: Hugo Bodory, University of St.\ Gallen, Varnb\"{u}elstrasse 14, 9000 St.\ Gallen, Switzerland, hugo.bodory@unisg.ch; Martin Huber, University of Fribourg, Bd.\ de P\'{e}rolles 90, 1700 Fribourg, Switzerland, martin.huber@unifr.ch; Luk\'{a}\v{s} Laff\'{e}rs, Matej Bel University, Tajovskeho 40, 97411 Bansk\'{a} Bystrica, Slovakia, lukas.laffers@gmail.com. Laff\'{e}rs acknowledges support provided by the Slovak Research and Development Agency under contract no. APVV-17-0329 and VEGA-1/0692/20.
}\thispagestyle{empty}\pagebreak  }

{\small \renewcommand{\thefootnote}{\arabic{footnote}} %
\setcounter{footnote}{0}  \pagebreak \setcounter{footnote}{0} \pagebreak %
\setcounter{page}{1} }

\section{Introduction}\label{intro}

In many empirical problems, policy makers and researchers are interested in the causal effects of sequences of interventions or treatments, i.e.\ dynamic treatment effects. Examples include the impact of sequences of training programs (for instance, a job application training followed by a language courses) on the employment probabilities of job seekers or the effect of sequential medical interventions (for instance,a surgery combined with rehabilitation training) on health. As treatment assignment is typically non-random, causal inference about distinct sequences of treatments requires controlling for confounders jointly affecting the various treatments and the outcome of interest. An assumption commonly imposed in the literature is sequential conditional independence, which implies that the treatment in each period is unconfounded conditional on past treatment assignments, past outcomes, and the history of observed covariates up to the respective treatment assignment. Due to increasing data availability, the number of observed covariates that may potentially serve as control variables to justify the sequential conditional independence assumption has been growing in many empirical contexts, which poses the question of how to optimally control for such a wealth of information in the estimation process.

This paper combines the semiparametrically efficient estimation of dynamic treatment effects under sequential conditional independence with  double machine learning (DML) framework outlined in  \cite{Chetal2018} to control for observed covariates in a data-driven way. More specifically, treatment effect estimation is based on the efficient score function, belonging to the class of doubly robust estimation as discussed in \cite{Robins+94} and \cite{RoRo95}, and relies on plug-in estimates of the dynamic treatment propensity scores (the conditional treatment probabilities given histories of covariates and past treatments) and conditional mean outcomes (given histories of treatments, covariates, and past outcomes). We obtain these plug-in estimates by machine learning, which permits algorithmically controlling for covariates with the highest predictive power for the treatments and outcomes.

To safeguard against overfitting bias due to correlations between the estimation steps, the plug-in models and the treatment effects are estimated in different parts of the data, whose role is subsequently swapped to prevent not using parts of the data for effect estimation (and thereby increasing the variance). We show that our estimator satisfies the so-called \cite{Neyman1959} orthogonality discussed in \cite{Chetal2018} and is thus asymptotically normal and $\sqrt{n}$-consistent under specific regularity conditions despite the data-driven estimation of the plug-ins. One restriction is that the convergence of the plug-in estimates to the true models as a function of the covariates is not too slow, which is satisfied if each of the estimators converges at a rate not slower than $n^{-1/4}$. When using lasso as machine learner, this implies a form of approximate sparsity, meaning that the number of important covariates for obtaining a decent approximation of the plug-ins is small relative to the sample size. However, the set of these important confounders need not be known a priori, which is particularly useful in high dimensional data with a vast number of covariates that could potentially serve as control variables.

As a further contribution, we discuss the DML-based estimation of weighted dynamic treatment effects where the weight is defined as a function of the baseline covariates. This permits, for instance, assessing treatment sequences among those treated or not treated in the first period and therefore provides a rather general framework for the definition of interesting subpopulations. Also for this estimator based on a weighted version of the efficient sore function, we show  \cite{Neyman1959} orthogonality and $\sqrt{n}$-consistency under specific restrictions on the convergence rates of the plug-in estimators, which now also include the estimated weighting function.

Furthermore, we investigate the methods' finite sample behavior in a simulation study, and find the point estimators to perform rather decently in the simulation designs considered. As an empirical contribution, we assess the effects of various treatment sequences in the U.S.\ Job Corps study on an educational intervention for disadvantaged youth. We find that attending vocational training in the two initial years of the program likely increases the employment probability four years after the start of Job Corps when compared to no instruction. In contrast, the relative performance of sequences of vocational vs.\ academic classroom training is less clear.


The literature on dynamic treatment effects goes back to \cite{Ro86}, who proposes a dynamic causal framework along with an estimation approach known as g-computation for recursively modeling outcomes at some point in time as functions of the (histories of) observed covariates and treatments under the sequential conditional independence assumption. 
G-computation was originally implemented by parametric maximum likelihood estimation of nested structural models for the outcomes in all periods, requiring the (in general tedious) estimation of the conditional densities of all time-varying covariates. \cite{Robins1998} suggested an alternative, less complex modeling approach based on so-called marginal structural models representing outcomes in specific treatment states as functions of time-constant covariates only. In order to also control for time-varying confounding, such marginal models need (in the spirit of \cite{HoTh52}) to be combined with weighting by the inverse of the dynamic treatment propensity scores, see for instance \cite{RoGrHu1999} and \cite{RoHeBr00}. The propensity scores in each period are typically estimated by sequential logit regressions, but see \cite{ImaiRatkovic2015} for an alternative, empirically likelihood-based approach that aims at finding propensity score specifications that maximize covariate balance. \cite{Lech09} considers inverse probability weighting (IPW) by the dynamic treatment propensity scores alone (i.e. without the use of marginal outcome models), while \cite{LechnerMiquel2010} apply propensity score matching and \cite{BlackwellStrezhnev2020} direct matching on the covariates.

Doubly robust estimators of dynamic treatment effects comprise methods that are consistent if either the sequential treatment propensity scores or nested outcome models are correctly specified. This includes estimation based on the sample analog of the efficient influence function (underlying the semiparametric efficiency bounds) provided in \cite{Robins1999}, which is a function of both the nested treatment and outcome models.\footnote{\cite{YU20061061} discuss an alternative doubly robust approach based on combining propensity scores with the estimation of marginal structural models.} In contrast, \cite{BaRo05} propose a doubly robust estimator that is based on estimating potential outcomes by nested models of conditional mean outcomes (given the covariate histories as well as past and current treatment assignments) in all periods, a form of g-computation that does not require tedious likelihood estimations of conditional densities as initially proposed in \cite{Ro86}. Here, doubly robustness comes from the fact that a weight based on the nested treatment propensity scores is included as additional covariate in conditional mean estimation.

\cite{TargetedMinimumLossBasedEstimation} demonstrate that this approach fits the framework of Targeted Maximum Likelihood Estimation (TMLE) of \cite{vanderLaanRubin2006}, which obtains doubly robustness through updating initial conditional outcome estimates by regressing them on a function of the nested propensity scores in each period, and offer a refined estimator. Specifically, they suggest estimating nuisance parameters by the super learner of \cite{vanderLaanetal2007}, an ensemble method for machine learning. In contrast, the approach suggested in this paper does not rely on the likelihood estimation of marginal structural models, nor of nested structural models requiring the estimation of conditional covariate densities. Similar to TMLE, our approach is based on combining nested conditional mean outcomes with propensity score estimation. Different to TMLE, however, we base estimation on the efficient influence function, which does not iteratively update the nested outcomes. In addition, we also consider weighted treatment effect estimation as a function of baseline covariates. As we estimate the plug-in parameters by machine learning as recently also considered in  \cite{Tranetal2019}, we formally show that our approach fits the double machine learning framework of \cite{Chetal2018} and discuss regularity conditions under which $\sqrt{n}$-consistency is attained.

\cite{LewisSyrgkanis2020} propose an alternative DML estimator of dynamic treatment effects. It is based on residualizing or debiasing the outcome and the treatment by purging the effects of observed confounders using machine learning and regressing the debiased outcome on the debiased treatment in a specific period. This approach may also be applied to continuous (rather than discrete) treatments, but in contrast to our method assumes partial linearity in the outcome model. Finally, \cite{VivianoBradic2021} suggest a further doubly robust method that can be combined with machine learning, but replaces weighting by the inverse of the propensity scores (as applied in our paper) by a dynamic version of covariate balancing as discussed in \cite{Zubizarreta2015} and \cite{AtheyImbensWager2018}.

This paper proceeds as follows. Section \ref{effdef} introduces the concepts of dynamic treatment effects in the potential outcome framework, 
presents the identifying assumptions and discusses identification. Section \ref{section:CrossFitting} proposes an estimation procedure based on double machine learning and shows $\sqrt{n}$-consistency and asymptotic normality under specific conditions. Section \ref{section:weighted} extends the procedure to the evaluation of weighted dynamic treatment effects.
Section \ref{Sim} provides a simulation study. Section \ref{Application} presents an empirical application to data from Job Corps, an educational program for disadvantaged youth. Section \ref{conclusion} concludes.

\section{Definition of dynamic treatment effects and identification}\label{effdef}

We are interested in the causal effect of a sequence of discretely distributed treatments and will for the sake of simplicity focus on the case of two sequential treatments in the subsequent discussion. To this end, denote by $D_t$ and $Y_t$ the treatment (e.g a training program) and the outcome (e.g. employment) in period $T=t$.  Therefore, $D_1$ and $D_2$ are the treatments in the first and second periods, respectively, and may take values $d_1,d_2$ $\in$ $\{0,1,...,Q\}$, with $0$ indicating non-treatment and $1,...,Q$ the different treatment choices (where $Q$ denotes the number of non-zero treatments).  Let $Y_2$ denote the outcome of interest measured in the second period after the realization of treatment sequence $D_1$ and $D_2$.\footnote{We do not consider the evaluation of treatment effects on outcomes in the first period, as this corresponds to the conventional static treatment framework as for instance considered in \cite{Chetal2018}.} To define the dynamic treatment effects of interest, we make use of the potential outcome framework, see for instance  \cite{Rubin74}. We denote by $\underline{d}_2$ a specific treatment sequence $(d_1,d_2)$ with $d_1,d_2$ $\in$ $\{0,1,...,Q\},$ then $\underline{D}_2 \equiv (D_1,D_2)$ and let $Y_2(\underline{d}_2)$ denotes the potential outcome hypothetically realized when the treatments are set to that sequence $\underline{d}_2$. We also define $\{0,1,...,Q\}^2 = \{0,1,...,Q\} \times \{0,1,...,Q\}$.

We aim at evaluating the average treatment effect (ATE) of two distinct treatment sequences in the population,
\begin{eqnarray}
\Delta(\underline{d}_2,\underline{d}^*_2)=E[Y_2(\underline{d}_2)-Y_2(\underline{d}^*_2)],
\end{eqnarray}
with $\underline{d}_2\neq\underline{d}^*_2$ such that the sequences differ either in $d_1$ or in both $\underline{d}_2$.\footnote{In the case of $\underline{d}_2,\underline{d}^*_2$  sharing the same $d_1$ but differing in terms $d_2$, the identification problem collapses to the standard case with one treatment period (namely $T=2$) under the condition that $D_1=d_1$. The case of a single treatment period also prevails when considering the effects on $Y_1$, i.e.\ the outcome in period $T=1$, which only permits assessing the effect of $D_1$. In either case, the standard double machine learning (DML) framework for single treatment periods can be applied as e.g.\ outlined in \cite{Bellonietal2017}, such that we do not consider these scenarios in this paper.} Examples are the evaluation of a sequence of two binary treatments vs. no treatment, e.g.\ $\underline{d}_2= (1,1)$ and $\underline{d}^*_2=(0,0)$, or the effect of the first treatment when holding the second treatment constant,  $\underline{d}_2=(1,d_2)$ and $\underline{d}^*_2=(0,d_2)$, with $d_2$ $\in$ $\{0,1\}$. The latter parameter is known as the controlled direct effect in causal mediation analysis, see for instance \cite{Pearl01}, assessing the net effect of the first treatment when setting the second treatment to be $D_2=d_2$ for everyone.\footnote{From the perspective of causal mediation analysis, our paper complements the study of \cite{Farbmacheretal2020}, who apply DML to the estimation of so-called natural direct and indirect effects. In the latter case, $D_2$ is not prescribed to have the same value $d_2$ for everyone, but corresponds to the potential value $D_2(d_1)$, i.e.\ the hypothetical treatment state of $D_2$ that would be `naturally chosen' (i.e.\ without prescription) as a consequence of  $D_1=d_1$.} Throughout the paper we assume that stable unit treatment value assumption (SUTVA, \cite{Rubin80}) holds such that $\Pr(\underline{D}_2 = \underline{d}_2  \implies Y_2 = Y_2(\underline{d}_2))=1.$  This rules out interaction effects, general equilibrium effects and implicitly assumes that treatments are uniquely defined.

\begin{figure}[!htp]
\centering \caption{\label{figuresetup}  Causal paths under sequential conditional independence}\bigskip
\includegraphics[width=0.75\textwidth]{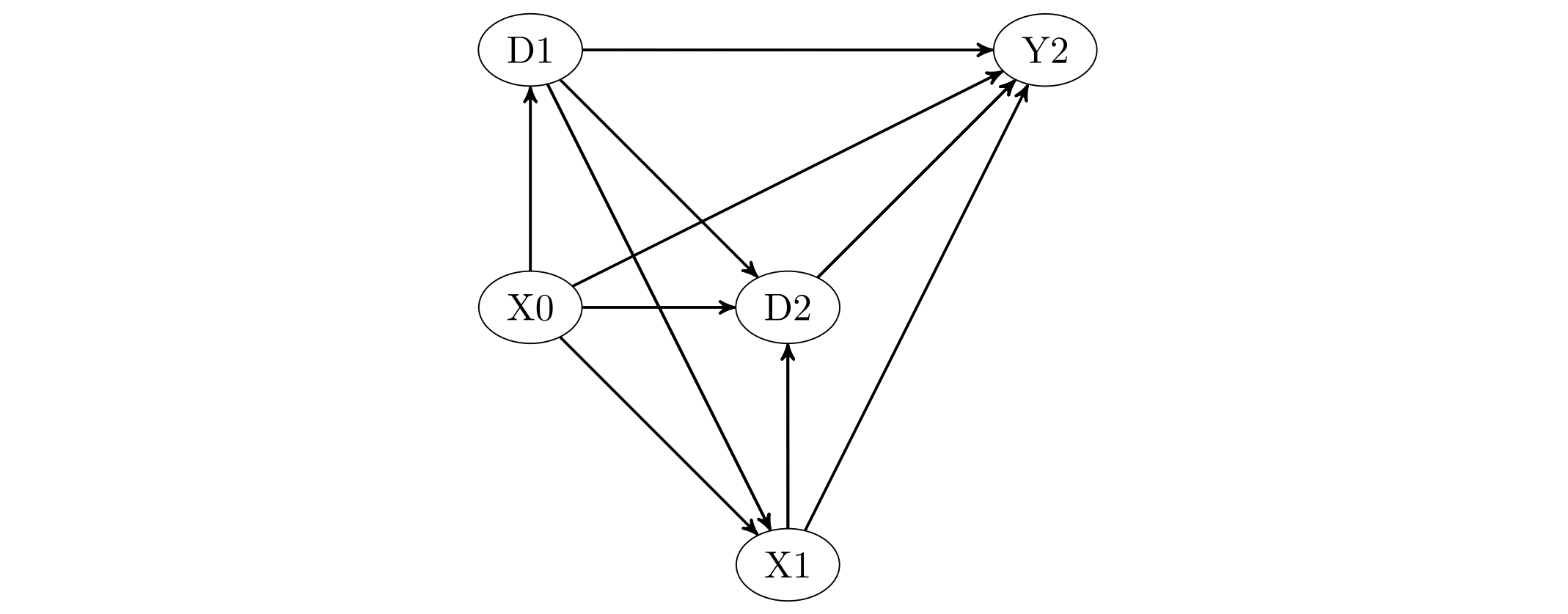}
\end{figure}



Identification relies on a sequential conditional independence assumption, requiring that the treatment in each period is conditionally independent of the potential outcomes, conditional on previous treatments and (histories of) observed covariates measured prior to treatment, which might include past outcomes, too. Let to this end $X_t$ denote the observed characteristics in period $T=t$. $X_0$ consists of pre-treatment characteristics measured prior to the first treatment $D_1$, while $X_1$ (which may contain $Y_1$) is measured prior to $D_2$, but may be influenced by $D_1$ as well as $X_0$. Covariates in a particular period may therefore be affected by previous covariates and treatments, implying that confounding may be dynamic in the sense that identification relies on time varying observables rather than on baseline covariates alone. Figure \ref{figuresetup} provides a graphical illustration using a directed acyclic graph, with arrows representing causal effects. Each of $D_1$, $D_2$, and $Y_2$  might be causally affected by distinct and statistically independent sets of unobservables not displayed in Figure \ref{figuresetup}, but none of these unobservables may jointly affect $D_1$ and $Y_2$ given $X_0$ or $D_2$ and $Y_2$ given $D_1$, $X_0$, and $X_1$.

Formally, the first assumption invokes conditional independence of the treatment in the first period $D_1$ and the potential outcomes $Y_2(\underline{d}_2)$ given $X_0$ as commonly invoked in the treatment evaluation literature, see e.g.\ \cite{Imbens03}. It rules out unobserved confounders jointly affecting $D_1$ and $Y_2(\underline{d}_2)$ conditional on $X_0$.
\vspace{5pt}\newline
\textbf{Assumption 1 (conditional independence of the first treatment):}\newline
$Y_2(\underline{d}_2)  \bot D_1 | X_0$, for $\underline{d}_2$ $\in$ $\{0,1,...,Q\}^2$.\vspace{5pt}\newline
where `$\bot$' denotes statistical independence.
\vspace{5pt}\newline
The second assumption invokes conditional independence of the second treatment $D_2$ given the first treatment $D_1$ and the (history of) covariates $X_0$ and $X_1$, which we denote by $\underline{X}_1=(X_0,X_1)$ to ease notation. It rules out unobserved confounders jointly affecting $D_2$ and $Y_2(\underline{d}_2)$ conditional on $D_1$ and $\underline{X}_1$.
\vspace{5pt}\newline
\textbf{Assumption 2 (conditional independence of the second treatment):}\newline
$Y_2(\underline{d}_2)  \bot D_2 | D_1, X_0, X_1$, for $\underline{d}_2$ $\in$ $\{0,1,...,Q\}^2$.  \vspace{5pt}\newline
\vspace{5pt}\newline
The third assumption imposes common support, meaning that the treatment in each period is not a deterministic function of the respective observables in the conditioning set, which rules out conditional treatment probabilities (or propensity scores) of $0$ or $1$. This implies that conditional on each value of the observables occuring in the population, subjects with distinct treatment assignments $\{0,1,...,Q\}$ exist.
\vspace{5pt}\newline
\textbf{Assumption 3 (common support):}\newline
$\Pr(D_1=d_1|  X_0)>0$, $\Pr(D_2=d_2| D_1, \underline{X}_1)>0$ for $d_1, d_2$ $\in$ $\{0,1...,Q\}$.\vspace{5pt}\newline

To ease notation, we henceforth denote the propensity scores by $p^{d_1}(X_0)=\Pr(D_1=d_1|X_0)$ and $p^{d_2}(D_1,\underline{X}_1)=\Pr(D_2=d_2|D_1,\underline{X}_1)$. Furthermore, we denote the conditional mean outcome in the second period by   $\mu^{Y_2}(\underline{D}_2,\underline{X}_1)=E[Y_2|\underline{D}_2,X_0,X_1]$ and the nested conditional mean outcome in the first period by
\begin{eqnarray}\label{nu}
 \nu^{Y_2}(\underline{D}_2,X_0)&=&\int E[Y_2|\underline{D}_2,X_0,X_1=x_1] dF_{X_1=x_1|D_1,X_0},
\end{eqnarray}

 where $F_{X_1=x_1|D_1,X_0}$ denotes the conditional distribution function of $X_1$ given $(D_1,X_0)$ at value $x_1$. 
For a fixed vector of treatments $\underline{D}_2=\underline{d}_2$ the quantity $\nu^{Y_2}(\underline{d}_2,X_0)$ is equal to $E[E[Y_2|\underline{D}_2=\underline{d}_2,X_0,X_1]|D_1=d_1,X_0]$, and this suggests that it can be obtained by a sequential estimation of nested conditional means.
 This is the approach followed in this paper, as it avoids the estimation of conditional covariate distributions, which might be cumbersome if covariates are high dimensional.

As for instance discussed in  \cite{Tranetal2019}, Assumptions 1-3 permit identifying the mean potential outcome $E[Y(\underline{d}_2)]$ based on the following expression:
\begin{eqnarray}
E[Y(\underline{d}_2)]&=&E[\psi^{\underline{d}_2}],\textrm{ where}\notag\\
\psi^{\underline{d}_2} &=&  \frac{I\{D_1=d_1\} \cdot I\{D_2=d_2\} \cdot [Y_2-\mu^{Y_2}(\underline{d}_2,\underline{X}_1)]}{p^{d_1}(X_0)\cdot p^{d_2}(d_1,\underline{X}_1)} \notag\\
& + & \frac{I\{D_1=d_1\}\cdot  [\mu^{Y_2}(\underline{d}_2,\underline{X}_1)-\nu^{Y_2}(\underline{d}_2,X_0)]}{p^{d_1}(X_0)}+\nu^{Y_2}(\underline{d}_2,X_0).  \label{1}
\end{eqnarray}

This follows from the fact that $\psi^{\underline{d}_2}-E[Y(\underline{d}_2)]$, which corresponds to the efficient score function of dynamic treatment effects as discussed in \cite{Robins1999}, has a zero mean property: $E[\psi^{\underline{d}_2}-E[Y(\underline{d}_2)]]=0$.

\section{Estimation of the counterfactual with K-fold Cross-Fitting}\label{estimsec}
\label{section:CrossFitting}
We subsequently propose an estimation strategy for the counterfactual  $E[Y(\underline{d}_2)]$ with $\underline{d}_2 \in \{0,1,...,Q\}^2$ and show its $\sqrt{n}$-consistency under specific regularity conditions. Define
\begin{eqnarray}
\psi^{\underline{d}_2}(W, \eta, \Psi^{\underline{d}_2}_{0}) &=&  \frac{I\{D_1=d_1\} \cdot I\{D_2=d_2\} \cdot [Y_2-\mu^{Y_2}(\underline{d}_2,\underline{X}_1)]}{p^{d_1}(X_0)\cdot p^{d_2}(d_1,\underline{X}_1)} \notag\\
& + & \frac{I\{D_1=d_1\}\cdot  [\mu^{Y_2}(\underline{d}_2,\underline{X}_1)-\nu^{Y_2}(\underline{d}_2,X_0)]}{p^{d_1}(X_0)} \notag\\
& + &\nu^{Y_2}(\underline{d}_2,X_0) - \Psi^{\underline{d}_2}_{0},
\end{eqnarray}
where $\mathcal{W} = \{W_i|1\leq i \leq N\}$ with $W_i = (Y_{2i}, D_{1i}, D_{2i}, X_{0i},X_{1i})$ for all $ i $ denotes the set of observations and $I\{\cdot\}$ denotes the indicator function.
The true nuisance parameters are denoted by $\eta_0=(p_0^{d_1}(X_0), p_0^{d_2}(D_1,\underline{X}_1),\mu_0^{Y_2}(\underline{D}_2,\underline{X}_1), \nu_0^{Y_2}(\underline{D}_2,X_0))$, their estimates by \newline
$\hat{\eta}=(\hat{p}^{d_1}(X_0), \hat{p}^{d_2}(D_1,\underline{X}_1), \hat{\mu}^{Y_2}(\underline{D}_2,\underline{X}_1), \hat{\nu}^{Y_2}(\underline{D}_2,X_0))$. Let $\Psi^{\underline{d}_2}_{0}=E[Y(\underline{d}_2)]$ denotes the true counterfactual.

We suggest estimating the $\Psi^{\underline{d}_2}_{0}$ using the following algorithm that combines orthogonal score estimation and sample splitting. Further below we will outline the conditions under which this estimation strategy leads to $\sqrt{n}$-consistent estimates for the counterfactual. \vspace{8pt}\newline
\textbf{Algorithm 1: Estimation of $E[Y(\underline{d}_2)]$ }
\begin{enumerate}
	\item Split $\mathcal{W}$ in $ K $ subsamples. For each subsample $ k $, let $n_k$ denote its size, $\mathcal{W}_k$ the set of observations in the sample and $\mathcal{W}_k^{C}$ the complement set of all observations not in $k$.
	\item For each $k$, use  $\mathcal{W}_k^{C}$ to estimate the model parameters of $p^{d_1}(X_0)$ and $p^{d_2}(d_1,\underline{X}_1)$. Split  $\mathcal{W}_k^{C}$ into 2 non-overlapping subsamples and estimate the model parameters of the conditional mean $\mu^{Y_2}(\underline{d}_2,\underline{X}_1)$ and the nested conditional mean $\nu^{Y_2}(\underline{d}_2,X_0)$   in the distinct subsamples.  Predict the models among $\mathcal{W}_k$, where the predictions are denoted by $\hat{p}_k^{d_1}(X_0)$, $\hat{p}_k^{d_2}(D_1,\underline{X}_1)$, $\hat{\mu}_k^{Y_2}(\underline{d}_2,\underline{X}_1)$, $\hat \nu_k^{Y_2}(\underline{d}_2,X_0)$.
	\item For each $ k $, obtain an estimate of the moment condition for each observation $i$ in $\mathcal{W}_k$, denoted by $\hat \psi^{\underline{d}_2}_{i,k}$ :
	\begin{eqnarray}
	\hat \psi^{\underline{d}_2}_{i,k}&=& \frac{I\{D_{1i}=d_1\} \cdot I\{D_{2i}=d_2\} \cdot [Y_{2i}-\hat\mu_k^{Y_2}(\underline{d}_2,\underline{X}_{1i})]}{\hat p_k^{d_1}(X_{0i})\cdot \hat p_k^{d_2}(d_1,\underline{X}_{1i})} \notag\\
	& + &\frac{I\{D_{1i}=d_1\}\cdot  [\hat \mu_k^{Y_2}(\underline{d}_2,\underline{X}_{1i})-\hat \nu_k^{Y_2}(\underline{d}_2,X_{0i})]}{\hat p_k^{d_1}(X_{0i})} +\hat \nu_k^{Y_2}(\underline{d}_2,X_{0i}).\notag
	\end{eqnarray}
	\item Average the estimated scores $\hat \psi^{\underline{d}_2}_{i,k}$ over all observations across all $ K $ subsamples to obtain an estimate of  $\Psi^{\underline{d}_2}$ in the total sample, denoted by $\hat \Psi^{\underline{d}_2}=1/n \sum_{k=1}^{K}  \sum_{i=1}^{n_k} \hat \psi^{\underline{d}_2}_{i,k}$.
\end{enumerate}
As a remark concerning step 2 of the algorithm, it may appear non-standard to estimate $\mu^{Y_2}(\underline{d}_2,\underline{X}_1)$ and $\nu^{Y_2}(\underline{d}_2,X_0)$ in distinct subsamples. This approach aims at avoiding correlations between both estimation steps and thus, overfitting bias, because the estimate of $\mu^{Y_2}(\underline{d}_2,\underline{X}_1)$ is used as a plug-in parameter for estimating $\nu^{Y_2}(\underline{d}_2,X_0)$. 

In order to achieve $\sqrt{n}$-consistency for counterfactual estimation, we make the following assumption on the prediction quality of the machine learners when estimating the nuisance parameters. Closely following  \cite{Chetal2018}, we introduce some further notation. Let $(\delta_n)_{n=1}^{\infty}$ and $(\Delta_n)_{n=1}^{\infty}$ denote sequences of positive constants with $\lim_{n\rightarrow \infty} \delta_n = 0 $ and $\lim_{n\rightarrow \infty} \Delta_n = 0.$ Furthermore, let $c, \epsilon, C$ and $q$ be positive constants such that $q>2,$ and let $K \geq 2$ be a fixed integer. Also, for any random vector $Z = (Z_1,...,Z_l)$, let $\left\| Z \right\|_{q} = \max_{1\leq j \leq l}\left\| Z_l \right\|_{q},$ where
$\left \| Z_l \right\|_{q}  =  \left( E\left[ \left| Z_l \right|^q \right] \right)^{\frac{1}{q}}$.
In order to ease notation, we assume that $n/K$ is an integer. For the sake of brevity we omit the dependence of probability $\Pr_P,$ expectation $E_P(\cdot),$ and norm $\left\| \cdot  \right\|_{P,q}$ on the probability measure $P.$
\vspace{5pt}\newline
\textbf{Assumption 4 (regularity conditions and quality of plug-in parameter estimates):}  \newline
For all probability laws $P \in \mathcal{P}$
 the following conditions hold for the random vector $( Y_2 ,D_1,D_2,X_0,X_1)$ for all $d_1,d_2 \in \{0,1,...,Q\}$:
\begin{enumerate}
\item[(a)] $  \left\| Y_2 \right\|_{q} \leq C,$

$\left\|E[Y_2^2| D_1 = d_1, D_2 = d_2, \underline{X}_1 ] \right\|_{\infty} \leq C^2$,

\item[(b)]





$\Pr(\epsilon \leq p_0^{d_1} (X_0) \leq 1-\epsilon) = 1,$


$\Pr(\epsilon \leq p_0^{d_2} (d_1,\underline{X}_1) \leq 1-\epsilon) = 1,$


\item[(c)]
$\left\| Y_2-\mu_0^{Y_2}(\underline{d}_2,\underline{X}_1)  \right\|_{2} = E_{ } \Big[\left(Y_2-\mu_0^{Y_2}(\underline{d}_2,\underline{X}_1) \right)^2 \Big]^{\frac{1}{2}} \geq c$

\item[(d)] Given a random subset $I$ of $[n]$ of size $n_k=n/K,$ the nuisance parameter estimator $\hat \eta_0 = \hat \eta_0((W_i)_{i \in I^C})$ satisfies the following conditions. With $P$-probability no less than $1-\Delta_n:$
\begin{eqnarray}
\left\|  \hat \eta_0 - \eta_0 \right\|_{q} &\leq& C, \notag \\
\left\|  \hat \eta_0 - \eta_0 \right\|_{2} &\leq& \delta_n, \notag \\
\left\| \hat  p_0^{d_1}(X_0)-1/2\right\|_{\infty}  &\leq& 1/2-\epsilon, \notag\\
 \left\| \hat  p_0^{d_2}(D_1,\underline{X}_1)-1/2\right\|_{\infty} &\leq & 1/2-\epsilon, \notag \\
\left\|  \hat \mu_0^{Y_2}(\underline{D}_2,\underline{X}_1)-\mu^{Y_2}_0(\underline{D}_2,\underline{X}_1)\right\|_{2} \times \left\| \hat  p_0^{d_1}(X_0)-p^{d_1}_0(X_0)\right\|_{2}  &\leq & \delta^{}_n n^{-1/2}, \notag \\
\left\|  \hat \mu_0^{Y_2}(\underline{D}_2,\underline{X}_1)-\mu^{Y_2}_0(\underline{D}_2,\underline{X}_1)\right\|_{2} \times \left\| \hat  p_0^{d_2}(D_1,\underline{X}_1)-p^{d_2}_0(D_1,\underline{X}_1)\right\|_{2} &\leq & \delta^{}_n n^{-1/2},\notag \\
\left\|  \hat \nu_0^{Y_2}(\underline{D}_2,X_0)-\nu^{Y_2}_0(\underline{D}_2,X_0)\right\|_{2} \times \left\|  \hat p_0^{d_1}(X_0)-p^{d_1}_0(X_0)\right\|_{2} &\leq & \delta^{}_n n^{-1/2}.\notag
\end{eqnarray}
\end{enumerate}

The only non-primitive condition is the condition (d). It puts restrictions on the quality of the nuisance parameter estimators. Condition (a) states that the distribution of the outcome does not have unbounded moments. (b) refines the common support condition such that the propensity scores are bounded away from $0$and $1$. Finally, (c) states that the covariates $\underline{X}_1$ do not perfectly predict the conditional mean outcome.

%

For demonstrating the $\sqrt{n}$-consistency of our estimator of the mean potential outcome, we show that it satisfies the requirements of the DML framework in \cite{Chetal2018}  by first verifying linearity and Neyman orthogonality of the score (see Appendix \ref{Neyman}). Then, as  $ \psi^{\underline{d}_2}(W, \eta, \Psi^{\underline{d}_2}_{0}) $ is smooth in $ (\eta, \Psi^{\underline{d}_2}_{0}) $,  it is sufficient that the  plug-in estimators converge with a rate at least as fast as $ n^{-1/4} $ for achieving $n^{-1/2} $-convergence for the estimation of $ \hat{ \Psi}^{\underline{d}_2}$ as postulated in Theorem 1. This convergence rate of $ n^{-1/4} $ has been shown to be achieved by many commonly used machine learners under specific conditions, such as lasso, random forests, boosting and neural nets, see for instance \cite{Bellonietal2014}, \cite{LuoSpindler2016}, \cite{WagerAthey2018}, and \cite{FarrellLiangMisra2018}.\vspace{5pt}\newline
\textbf{Theorem 1}\\
Under  Assumptions 1-4, it holds for estimating $E[Y(\underline{d}_2)]$ based on Algorithm 1:  \\
$\sqrt{n} \Big(\hat \Psi^{\underline{d}_2} -  \Psi^{\underline{d}_2}_{0} \Big) \rightarrow N(0,\sigma_{\psi^{\underline{d}_2}})$,  where $\sigma_{\psi^{\underline{d}_2}}= E[(\psi^{\underline{d}_2}-\Psi^{\underline{d}_2}_{0})^2]$. \\	
The proof of Theorem 1 is provided in Appendix \ref{Neyman}.
\vspace{5pt}\newline

\section{Evaluation of weighted dynamic treatment effects}\label{section:weighted}

\cite{LechnerMiquel2010} show that under our assumptions, one may identify treatment effects for specific subgroups that are defined as a function of the distribution of the baseline covariates $X_0$. To this end, let $S$ denote a binary indicator for belonging to the subgroup of interest that satisfies  $S \bot Y_2(\underline{d}_2)|X_0$, as $S$ may be selective in $X_0$, but not w.r.t.\ the post-treatment covariates $X_1$ after controlling for $X_0$. Furthermore, denote by $g(X_0)=\Pr(S=1|X_0)$ the probability of being in that group conditional on $X_0$. Interesting examples for such subgroups are the treated or non-treated populations in the first period, obtained by defining $S=I\{ D_1 = d_1\}$ with $d_1$ $\in$ $\{0,1,...,Q\}$. Mean potential outcomes conditional on $S=1$ are identified based on reweighting by $g(X_0)$, see e.g.\ \cite{Hirano+00} who use this approach for weighted ATE evaluation based on IPW. That is,
\begin{eqnarray}
E[Y_2(\underline{d}_2)|S=1]&=&E\Bigg[\frac{S\cdot Y_2(\underline{d}_2)}{\Pr(S=1)}\Bigg]=E\Bigg[\frac{g(X_0)}{\Pr(S=1)}\cdot E[Y_2(\underline{d}_2)|X_0]\Bigg]\label{weighted}\\
&=&E\Bigg[\frac{S}{\Pr(S=1)}\cdot E[Y_2(\underline{d}_2)|X_0]\Bigg],\notag
\end{eqnarray}
where the first equality follows from basic probability theory and the remaining ones from the fact that $S \bot Y_2(\underline{d}_2)|X_0$ and the law of iterated expectations. This suggests the following identification approach:
\begin{eqnarray}
E[Y_2(\underline{d}_2)|S=1]&=&E[\psi^{\underline{d}_2,S=1}],\textrm{ where}\notag\\
\psi^{\underline{d}_2,S=1} &=&  \frac{g(X_0)}{\Pr(S=1)}\cdot \frac{I\{D_1=d_1\} \cdot I\{D_2=d_2\} \cdot [Y_2-\mu^{Y_2}(\underline{d}_2,\underline{X}_1)]}{p^{d_1}(X_0)\cdot p^{d_2}(d_1,\underline{X}_1)} \label{wcond}\\
& + & \frac{g(X_0)}{\Pr(S=1)}\cdot \frac{I\{D_1=d_1\}\cdot  [\mu^{Y_2}(\underline{d}_2,\underline{X}_1)-\nu^{Y_2}(\underline{d}_2,X_0)]}{p^{d_1}(X_0)}+\frac{S}{\Pr(S=1)}\cdot \nu^{Y_2}(\underline{d}_2,X_0). \label{6} \notag
\end{eqnarray}
Note that the term $\frac{S}{\Pr(S=1)}\cdot \nu^{Y_2}(\underline{d}_2,X_0)$ in \eqref{wcond} corresponds to $\frac{S}{\Pr(S=1)}\cdot E[Y_2(\underline{d}_2)|X_0]$ in \eqref{weighted}. Appendix \ref{weightedscore} shows that the moment condition $E[\psi^{\underline{d}_2,S=1}-E[Y_2(\underline{d}_2)|S=1]]=0$ holds, such that $E[\psi^{\underline{d}_2,S=1}]$ identifies the weighted mean potential outcome, and proves Neyman orthogonality. It demonstrates that DML is $\sqrt{n}$-consistent and asymptotically normal under Assumption 5 below. The latter formalizes the rate restrictions on the plug-in estimates, which now also contain an estimate of $g(X_0)$ denoted by $\hat{g}(X_0)$. To this end, Algorithm 1 outlined in Section~\ref{estimsec} is applied to estimate $E[Y_2(\underline{d}_2)|S=1]$ by using modified moment conditions in steps 3 and 4.

More specifically, the previously used $\hat \psi^{\underline{d}_2}_{i,k}$ computed in some subsample $k$ is replaced by
\begin{eqnarray}
\hat \psi^{\underline{d}_2,S=1}_{i,k}&=&  \hat{g}_k(X_{0i})\cdot \frac{I\{D_{1i}=d_1\} \cdot I\{D_{2i}=d_2\} \cdot [Y_{2i}-\hat\mu_k^{Y_2}(\underline{d}_2,\underline{X}_{1i})]}{\hat p_k^{d_1}(X_{0i})\cdot \hat p_k^{d_2}(d_1,\underline{X}_{1i})} \notag\\
& + &\hat{g}_k(X_{0i})\cdot\frac{I\{D_{1i}=d_1\}\cdot  [\hat \mu_k^{Y_2}(\underline{d}_2,\underline{X}_{1i})-\hat \nu_k^{Y_2}(\underline{d}_2,X_{0i})]}{\hat p_k^{d_1}(X_{0i})} +S_i\cdot\hat \nu_k^{Y_2}(\underline{d}_2,X_{0i}).
\end{eqnarray}
In step 4, the estimated scores $\hat \psi^{\underline{d}_2,S=1}_{i,k}$ are averaged over all observations across all $ K $ subsamples and divided by an estimate of $\Pr(S=1)$ to obtain an estimate of $\Psi^{\underline{d}_2,S=1}_0=E[Y_2(\underline{d}_2)|S=1]$ based on $\hat \Psi^{\underline{d}_2,S=1}=\Big[\sum_{k=1}^{K}  \sum_{i=1}^{n_k} \hat \psi^{\underline{d}_2}_{i,k}\Big]\Big/\Big[\sum_{k=1}^{K}  \sum_{i=1}^{n_k}S_i\Big]$.



The following assumption refines the conditions of Assumption 4 such that asymptotic normality holds for the DML estimator based on (\ref{wcond}).\vspace{5pt}\newline
\textbf{Assumption 5 (regularity conditions and quality of plug-in parameter estimates):}  \newline
For all probability laws $P \in \mathcal{P}$
 the following conditions hold for the random vector $( Y_2 ,D_1,D_2,X_0,X_1, S)$ for all $d_1,d_2 \in \{0,1,...,Q\}$:
\begin{enumerate}
\item[(a)] $  \left\| Y_2 \right\|_{q} \leq C,$

$\left\|E[Y_2^2| D_1 = d_1, D_2 = d_2, \underline{X}_1 ] \right\|_{\infty} \leq C^2$,

\item[(b)]


$\Pr(\epsilon \leq p_0^{d_1} (X_0) \leq 1-\epsilon) = 1$

$\Pr(\epsilon \leq p_0^{d_2} (d_1,\underline{X}_1) \leq 1-\epsilon) = 1$
\item[(c)]
$\left\| Y_2-\mu_0^{Y_2}(\underline{d}_2,\underline{X}_1)  \right\|_{2} = E_{ } \Big[\left(Y_2-\mu_0^{Y_2}(\underline{d}_2,\underline{X}_1) \right)^2 \Big]^{\frac{1}{2}} \geq c$

\item[(d)] Given a random subset $I$ of $[n]$ of size $n_k=N/K,$ the nuisance parameter estimator $\hat \chi_0 = \hat \chi_0((W_i)_{i \in I^C})$ satisfies the following conditions. With $P$-probability no less than $1-\Delta_n:$
\begin{eqnarray}
\left\|  \hat \chi_0 - \chi_0 \right\|_{q} &\leq& C, \notag \\
\left\|  \hat \chi_0 - \chi_0 \right\|_{2} &\leq& \delta_n, \notag \\
 \left\| \hat  g(X_0)-1/2\right\|_{\infty}  &\leq& 1/2-\epsilon, \notag\\
\left\| \hat  p_0^{d_1}(X_0)-1/2\right\|_{\infty}  &\leq& 1/2-\epsilon, \notag\\
 \left\| \hat  p_0^{d_2}(D_1,\underline{X}_1)-1/2\right\|_{\infty} &\leq & 1/2-\epsilon, \notag \\
\left\|  \hat \chi_0^{Y_2}(\underline{D}_2,\underline{X}_1)-\mu^{Y_2}_0(\underline{D}_2,\underline{X}_1)\right\|_{2} \times \left\| \hat  p_0^{d_1}(X_0)-p^{d_1}_0(X_0)\right\|_{2}  &\leq & \delta^{}_n n^{-1/2}, \notag \\
\left\|  \hat \mu_0^{Y_2}(\underline{D}_2,\underline{X}_1)-\mu^{Y_2}_0(\underline{D}_2,\underline{X}_1)\right\|_{2} \times \left\| \hat  p_0^{d_2}(D_1,\underline{X}_1)-p^{d_2}_0(D_1,\underline{X}_1)\right\|_{2} &\leq & \delta^{}_n n^{-1/2},\notag \\
\left\|  \hat \nu_0^{Y_2}(\underline{D}_2,X_0)-\nu^{Y_2}_0(\underline{D}_2,X_0)\right\|_{2} \times \left\|  \hat p_0^{d_1}(X_0)-p^{d_1}_0(X_0)\right\|_{2} &\leq & \delta^{}_n n^{-1/2}.\notag \\
\left\|  \hat\mu^{Y_2}(\underline{D}_2,\underline{X}_1)-\mu^{Y_2}_0(\underline{D}_2,\underline{X}_1)\right\|_{2} \times \left\| \hat  g(X_0)-g_0(X_0)\right\|_{2}  &\leq & \delta^{}_n n^{-1/2}, \notag \\
\left\| \hat \nu^{Y_2}(\underline{D}_2,X_0)-\nu^{Y_2}_0(\underline{D}_2,X_0)\right\|_{2} \times \left\| \hat g(X_0)-g_0(X_0)\right\|_{2} &\leq & \delta^{}_n n^{-1/2}.\notag
\end{eqnarray}
\end{enumerate}

Assumption 5 can be satisfied if the plug-in estimator $\hat g(X_0)$ converges to its' true value $g_0(X_0)$ with rate $ n^{-1/4}$ just like the estimators of the other nuisance terms. Then, the average treatment effect in the subgroup, denoted by
\begin{eqnarray}
\Delta(\underline{d}_2,\underline{d}^*_2, S=1)=E[Y_2(\underline{d}_2)-Y_2(\underline{d}^*_2)|S=1],
\end{eqnarray}
is $\sqrt{n}$-consistently estimated, as postulated in Theorem 2.\vspace{5pt}\newline
\textbf{Theorem 2}\\
Under  Assumptions 1-3 and 5, it holds for estimating $E[Y_2(\underline{d}_2)|S=1]$ based on Algorithm 1:  \\
$\sqrt{n} \Big(\hat \Psi^{\underline{d}_2,S=1} -  \Psi^{\underline{d}_2,S=1}_{0} \Big) \rightarrow N(0,\sigma_{\psi^{\underline{d}_2,S=1}})$,  where $\sigma_{\psi^{\underline{d}_2,S=1}}= E[(\psi^{\underline{d}_2,S=1}-\Psi^{\underline{d}_2,S=1}_{0})^2]$. \\	
The proof of Theorem 2 is provided in Appendix \ref{weightedscore}.

\section{Simulation study}\label{Sim}

This section provides a simulation study to investigate the finite sample behavior of our double machine learning method for dynamic treatment effects based on the following data generating process:
\begin{eqnarray*}
    Y_2 &=& D_1 + D_2 + X_0'\beta_{X_0}  + X_1'\beta_{X_1}  + U ,\\
	D_1 &=& I\{X_0'\beta_{X_0}+ V>0\},\\
	D_2 &=& I\{ 0.3 D_1 + X_0'\beta_{X_0} + X_1'\beta_{X_1}  + W>0\},\\
	X_0 &\sim& N(0,\Sigma_0),\quad 	X_1 \sim N(0,\Sigma_1),\\
	U, V, W &\sim& N(0,1), \textrm{ independently of each other}.
\end{eqnarray*}
Outcome $Y_2$ is a function of the observed variables $D_1,D_2,X_0,X_1,$ and the unobserved scalar $U$. 
The treatment effects of both $D_1$ and $D_2$ are equal to 1. $D_1$ is a function of $X_0$ and the unobserved scalar $V$. $D_2$ is a function of both pre- and post-treatment covariates $X_0$ and $X_1$, the first treatment $D_1$, and the unobservable scalar $W$. Both $X_0$ and $X_1$ are vectors of covariates of dimension $p$, drawn from a multivariate normal distribution with zero mean and covariance matrices $\Sigma_0$ and $\Sigma_1$, respectively. $U, V, W$ are random and standard normally distributed. We consider two sample sizes of $n=2500$ and $10000$, running $1000$ simulations for the smaller and $250$ simulations for the larger sample sizes.

In our simulations, we set $p$, the number of covariates in $X_1$ and $X_0$, respectively, to 50 or 100. $\Sigma_0$ and $\Sigma_1$ are defined based on setting the covariance of the $i$th and $j$th covariate in $X_0$ or $X_1$ to $0.5^{|i-j|}$. 
The coefficients $\beta_{X_0}$  and $\beta_{X_1}$  gauge the impacts of the covariates on $Y_2$, $D_2$, and $D_1$, respectively, and thus, the magnitude of confounding. The $i$th element in the coefficient vectors $\beta_{X_0}$ and $\beta_{X_1}$ is set to $0.4/i^4$ for $i=1,...,p$, implying a quadratic decay of covariate importance in terms of confounding. 
As reported in Table~\ref{tab:confounding_appendix}, this specification implies that the $R^2$ statistic based on linearly predicting $Y_2$ by $\underline{X}_1$  ranges from 36 to 41\%, depending on the number of covariates and the sample size. Furthermore, the \cite{Nagelkerke1991} pseudo-$R^2$ when predicting $D_1$ by $X_0$  and $D_2$ by $D_1,\underline{X}_1$ based on probit models ranges from 13 to 17\% and 26 to 33\%, respectively. These figures point to a substantial level of confounding as it may be reasonably encountered in empirical applications.

\begin{table}[htbp]
	\centering\normalsize
	\caption{Confounding based on $\beta_{X_0}=\beta_{X_1}=0.4/i^4$}
	\label{tab:confounding_appendix}
	\begin{adjustbox}{max width=\textwidth}
		\begin{tabular}{ll|ccc}
			\hline \hline
			number of & sample& pseudo-$R^2$ (\%) & pseudo-$R^2$ (\%)& $R^2$ (\%)  \\
			covariates & size & $\hat p^{d_1}(X_{0})$  & $\hat p^{d_2}(d_1,\underline{X}_{1})$ & $\hat\mu^{Y_2}(\underline{X}_{1})$  \\
			\hline
			50  & 2500   &15&29&38 \\
			50  & 10000  &17&33&41 \\
			100 & 2500   &13&26&36 \\
			100 & 10000  &14&27&37 \\
			\hline	
		\end{tabular}
	\end{adjustbox}
\end{table}


We investigate the performance of ATE estimation when comparing the sequences of obtaining both treatments $(\underline{d}_2=(d_1=1,d_2=1))$ vs.\ no treatment $(\underline{d}^*_2=(d_1=0,d_2=0))$ in the total population based on Theorem 1 and in the treated in the first period based on Theorem 2. The nuisance parameters, i.e.\ the linear and probit specifications of the outcome and treatment equations, are estimated by lasso regressions using the default options of the \textit{SuperLearner} package provided by \cite{vanderLaanetal2007} for the statistical software \textsf{R}. 3-fold cross-fitting is used for the estimation of the treatment effects. We drop observations whose products of estimated treatment propensity scores in the first and second period, $\hat{p}^{d_1}(X_0)\cdot \hat{p}^{d_2}(D_1,\underline{X}_1)$, are close to zero, namely smaller than a trimming threshold of $0.01$ (or 1\%). This avoids an explosion of the propensity score-based weights and thus of the variance when estimating the mean potential outcomes by the sample analogue of identification result \eqref{6}, where the product of the propensity scores enters the denominator for reweighing the outcome. Our estimation procedure is available in the \textit{causalweight} package for \textsf{R} by \citet{BodoryHuber2018}.

\begin{table}[htbp]
\centering\normalsize
\caption{Simulation results based on $\beta_{X_0}=\beta_{X_1}=0.4/i^4$}
\label{tab:sim_appendix}
\begin{adjustbox}{max width=\textwidth}
\begin{tabular}{cc|cccccc}
\hline\hline
covar-   & sample& true   & absolute     & standard    & average     & RMSE  & coverage  \\ 
 iates  & size  & effect  & bias      & deviation & SE   &       & in \% \\ 
\hline
 & & \multicolumn{6}{c}{ATE: $\hat{\Delta}(\underline{d}_2,\underline{d}^*_2)$ (all) }\\
\hline
50  & 2500   &  2     & 0.027  & 0.07  & 0.049  & 0.075 & 80 \\ 
50  & 10000  &  2     & 0.007  & 0.035 & 0.024  & 0.036 & 82 \\
100 & 2500   &  2     & 0.04   & 0.072 & 0.048  & 0.083 & 74  \\
100 & 10000  &  2     & 0.011  & 0.035 & 0.024  & 0.037 & 78  \\
\hline\hline
 & & \multicolumn{6}{c}{ATE on selected: $\hat{\Delta}(\underline{d}_2,\underline{d}^*_2,S=1)$  } \\
\hline
50  & 2500   &  2     & 0.027  & 0.076    & 0.06     & 0.081 & 85 \\
50  & 10000  &  2     & 0.006  & 0.037    & 0.029    & 0.038 & 90 \\
100 & 2500   &  2     & 0.042  & 0.079    & 0.06     & 0.089 & 81 \\
100 & 10000  &  2     & 0.01   & 0.039    & 0.029    & 0.04  & 88 \\
\hline	
\end{tabular}
\end{adjustbox}
\caption*{\scriptsize Notes: SE and RMSE denote the standard error and the root mean squared error, respectively. Coverage is based on 95\% confidence intervals. 
}
\end{table}

Table~\ref{tab:sim_appendix} presents the main findings when estimating the ATE in the total population, $\hat{\Delta}(\underline{d}_2,\underline{d}^*_2)$, and among the subgroup of treated in the first period, $\hat{\Delta}(\underline{d}_2,\underline{d}^*_2,S=1)$.  Irrespective of the number of covariates, the absolute biases go to zero as the sample size increases. Furthermore, the standard deviations and root mean squared errors (RMSE) of the ATE estimators are roughly cut by half when quadrupling the sample size, as implied by $\sqrt{n}$-consistency. The levels of the standard deviations and RMSEs are somewhat higher for $\hat{\Delta}(\underline{d}_2,\underline{d}^*_2,S=1)$ than for  $\hat{\Delta}(\underline{d}_2,\underline{d}^*_2)$, which comes from the additional weighting step due to targeting the treated subpopulation with $S=1$.  We also observe that the average standard errors (average SE)  based on the asymptotic variance approximations appear to converge at $\sqrt{n}$-rate, however, they underestimate the true standard deviations. This results in under-coverage of the true effects when constructing 95\% confidence intervals based on those standard errors, an issue that decreases in the sample size.

\section{Empirical application}\label{Application}

We apply our double machine learning approach to evaluate the effects of training sequences provided by the Job Corps program on employment. Job Corps is the largest U.S.\ program offering vocational training and academic classroom instruction for disadvantaged individuals aged 16 to 24. It is financed by the U.S.\ Department of Labor and currently has about 50,000 participants every year. 
Besides vocational credentials, students may obtain a high school diploma or equivalent qualifications. Individuals meeting specific low-income requirements can participate in Jobs Corps without any costs.

A range of studies analyzes the impact of Job Corps based on an experimental study with randomized access to the program between November 1994 and February 1996. In particular, \citet{ScBuGl01} and \citet{ScBuMc2008} discuss in detail the study design and report the average effects of random program assignment on a broad range of outcomes. Their findings suggest that Job Corps increases educational attainment, reduces criminal activity, and increases employment and earnings, at least for some years after the program. \citet{Floresetal2012} assess the impact of a continuously defined treatment, namely the length of exposure to academic and vocational instruction on earnings and find positive effects. As the length of the treatment is (in contrast to program assignment) not random, they impose a conditional independence assumption and control for baseline characteristics at Job Corps assignment. \citet{ColangeloLee2020} suggest double machine learning-based estimation of continuous treatment effects and apply it to assess the employment effects of Job Corps.  In contrast to these contributions on continuous treatment doses of Job Corps, we consider discrete sequences of multiple treatments and also control for post-treatment confounders rather than baseline covariates only.

Several contributions assess specific causal mechanisms of the program. \citet{FlFl09} find a positive direct effect of program assignment on earnings when controlling for work experience which they assume to be conditionally independent given observed covariates. Also \citet{Huber2012} imposes a conditional independence assumption and estimates a positive direct health effect when controlling for the mediator employment. Using a partial identification approach permitting mediator endogeneity, \citet{FlFl10} compute bounds on the direct and indirect effects of Job Corps assignment on employment and earnings mediated by obtaining a GED, high school degree, or vocational degree. Under their strongest set of assumptions, the results point to a positive direct effect net of obtaining a degree. \citet{FrHu17} use an instrumental variable strategy based on two instruments to disentangle the earnings effect of being enrolled in Job Corps into an indirect effect via hours worked and a direct effect, likely related to a change in human capital. Their results point to the existence of an indirect rather than a direct mechanism. Even though our framework of analyzing sequences of treatments is in terms of statistical issues somewhat related to the evaluation of causal mechanisms, it relies on distinct identifying assumptions than the previously mentioned studies, which e.g.\ do not consider controlling for post-treatment confounders.


Our sample consists of 11313 individuals with completed follow-up interviews four years after randomization, out of which 6828 and 4485 were randomized in and out of Job Corps, respectively. We exploit the sequential structure of academic education and vocational training in the program to define dynamic treatment states. Since most of the education and training activities were taken in the first two years, we focus on the latter when generating a sequence of binary treatments for each observation. 
The treatment states in our application can take four different values: $d_1$,$d_2$,$d_1^*$,$d_2^*$ $\in \{0,1,2,3\}$. State 0 refers to no instruction offered due to being randomized out of Job Corps (control group), 1 to no instruction despite being randomized in (never takers in the denomination of \citet{Angrist+96}), 2 to academic education among program participants, and 3 to vocational training among program participants. If individuals participate in both academic education and vocational training in a specific year, we assign the code of the treatment that was attended to a larger extent in terms of completed hours.

\begin{table}[htbp]
\centering\normalsize
\caption{Sequences of binary treatments}
\label{tab:treat}
\begin{adjustbox}{max width=\textwidth}
\begin{tabular}{l|rrrr}
\hline\hline
\multicolumn{3}{c}{dynamic treatments} & in Job Corps & observations \\	
\hline
code & year 1 & year 2 & &\\
\hline
00 & no educ/train & no educ/train & no  & 4485 \\
11 & no educ/train & no educ/train & yes & 320  \\
12 & no educ/train & acad educ     & yes & 43  \\
13 & no educ/train & voc train     & yes & 42 \\
21 & acad educ     & no educ/train & yes & 1328 \\
22 & acad educ     & acad educ     & yes & 341 \\
23 & acad educ     & voc train     & yes & 183 \\
31 & voc train     & no educ/train & yes & 1279 \\
32 & voc train     & acad educ     & yes & 109 \\
33 & voc train     & voc train     & yes & 573 \\
missings & & & & 2610 \\
\hline	
\end{tabular}
\end{adjustbox}
\caption*{\scriptsize Notes: \textit{no educ/train} means not participating in any Job Corps program related to education or training measures. \textit{acad educ} and \textit{voc train} stand for academic education and vocational training, respectively, offered by Job Corps.}
\end{table}

Table~\ref{tab:treat} reports various sequences of treatments in the data along with the corresponding number of observations. For instance, the treatment sequence 00 refers to those 4485 control group members that were randomized out  and did not participate in any education activities offered by Job Corps. Furthermore, 320 individuals assigned to Job Corps do not participate in any form of education either, as indicated by the sequence 11. We also note that for 2610 out of the 11313 individuals, information on the treatment sequences is missing. The literature explains the missing values by a random skip logic error, due to which asking questions about treatment participation was randomly omitted for a subset of survey participants, see page J.5 in \citet{ScEtAl2003}. In our analysis, we drop the control group with treatment sequence 00, but make use of it in a placebo test outlined further below. Furthermore, for several potential comparisons of treatment sequences, small sample issues and/or problems of a lack of common support in propensity scores (and thus, covariates) arise. For this reason, we confine our evaluation to comparing treatment sequence 33 (vocational training in both years) to either 22 (academic education in both years), 21 (academic education in the first year), or 11 (no participation in either year).

\begin{table}[htbp]
\centering\normalsize
\caption{Mean outcome conditional on treatment sequence}
\label{tab:out}
\begin{adjustbox}{max width=\textwidth}
\begin{tabular}{l|rr}
\hline\hline
treatment code & \multicolumn{2}{c}{employment}  \\	
& mean & missings \\
\hline
 00 &       0.78 &         46 \\
 11 &       0.78 &          7 \\
 21 &       0.78 &         17 \\
 22 &       0.82 &          1 \\
 33 &       0.89 &          3 \\
 missings & 0.77 &         37 \\
\hline	
\end{tabular}
\end{adjustbox}
\caption*{\scriptsize Notes: The first column 1 provides the codes of the treatment sequences, see Table~\ref{tab:treat}. The second column gives the average employment per sequence, the third one the number of missing observations.}
\end{table}

Our outcome variable is a binary employment indicator measured four years after randomization. Table~\ref{tab:out} reports the mean outcome across various treatment sequences, which ranges from 77 to 89 percent. It also provides the sequence-specific numbers of cases with missing outcomes that are dropped from the analysis, which appear quite low. We aim at estimating the ATE of treatment sequences $\underline{d}_2$ vs.\ $\underline{d}^*_2$ among individuals whose treatment in the first year corresponds to the first-year-treatment of either $\underline{d}_2$ or $\underline{d}^*_2$. An alternative would be to assess the ATE in the total sample randomized into Job Corps (which would thus  also include individuals with different first-year treatments than the ones evaluated), but this proved to be problematic due to lacking common support in terms of treatment propensity scores. 


We make use of a large set of potential control variables that also include covariates which have been identified as important confounders in several articles assessing the sensitivity of program evaluations to the inclusion and omission of such confounders in observational labor market studies. \citet{BiFiOsPa14}, for instance, conclude that imposing conditional independence assumptions requires the availability of rich data on employment and benefit histories, and socio-economic characteristics. \citet{LeWu13} point to the importance of factors like health, caseworker assessments, regional information, timing of unemployment and program start, pre-treatment outcomes, job search behavior, and labor market histories. In line with these findings, our covariates comprise information about socio-economic characteristics, pre-treatment labor market histories, education and training, job search activities, welfare receipt, health, crime, and how one learnt about the existence of Job Corps. Table~\ref{tab:xvar} in the Appendix~\ref{xvar} reports more details on these features, including variable descriptions and distributions across treatment sequences.

We condition on observed characteristics $X_t$ in periods $t \in \{0,1\}$. $X_0$ denotes control variables measured at baseline prior to the first treatment $D_1$, whereas $X_1$ is observed one year after randomization but prior to the second treatment $D_2$. Table~\ref{tab:x} provides the number and types of variables assigned to $X_0$ and $X_1$. Our raw data include 1188 characteristics. After some data manipulations based on generating dummies for values of categorical variables and missing items in dummy or categorical variables, we end up with all in all 2336 regressors. Missing observations in numerical variables were replaced by the mean values of the non-missing items. Furthermore, we standardized numerical covariates to have a zero mean and a standard deviation of 0.5.

\begin{table}[htbp]
\centering\normalsize
\caption{Number of covariates}
\label{tab:x}
\begin{adjustbox}{max width=\textwidth}
\begin{tabular}{l|rr}
\hline\hline
raw variables  & $X_0$ & $X_1$ \\
\hline
 dummy       & 295 &  575 \\
 categorical &  53 &   13 \\
 numeric     &  26 &  226 \\
 total       & 374 &  814 \\
\hline
processed variables & $X_0$ & $X_1$ \\
\hline
 dummy       & 884 & 1200 \\
 numeric     &  26 &  226 \\
 total       & 910 & 1426 \\
\hline
\end{tabular}
\end{adjustbox}
\caption*{\scriptsize Note: $X_0$ and $X_1$ denote regressors measured prior to the first and second periods, respectively.}
\end{table}


We estimate $\Delta(\underline{d}_2,\underline{d}^*_2, S=1)$, with $S=1$ if the first treatment corresponds to either the first treatment in  $\underline{d}_2$ or $\underline{d}^*_2$, based on 3-fold cross-fitting and the random forest (see \citet{Breiman2001}) as machine learner of the nuisance parameters. To this end, we use the \textit{SuperLearner} package with default options provided by \cite{vanderLaanetal2007} for the statistical software \textsf{R}. Our motivation for choosing the random forest is that it is (in the spirit of kernel regression) a nonparametric estimator that does not impose functional form assumptions (like linearity) on the conditional outcome or treatment models. As in our simulation study, we drop observations whose products of propensity scores in the first and second period are smaller than $0.01$ to impose common support in our sample
and avoid an explosion in the propensity score-based weights.
For a visual assessment of the common support, Appendix~\ref{overlap} provides plots with the propensity score distributions across all treatment sequences considered in this application. In general, common support is rather decent for the first period propensity scores $\hat{p}^{d_1}(X_0)$, while the overlap is weaker for the scores in the second period $\hat{p}^{d_2}(D_1,\underline{X}_1)$, especially at the boundaries of the distributions.

\begin{table}[htbp]
\centering\normalsize
\caption{Effect estimates with a trimming threshold of 0.01}
\label{tab:res}
\begin{adjustbox}{max width=\textwidth}
\begin{tabular}{ll|cccccc}
\hline
\hline
$\underline{d}_2$ & $\underline{d}^*_2$ & $\hat{E}[Y_2(\underline{d}^*_2)|S=1]$ &  $\hat{\Delta}(\underline{d}_2,\underline{d}^*_2, S=1)$ & SE & p-value & observations & trimmed \\
\hline
    33    & 22    & 0.77  & 0.08 & 0.07 & 0.23 & 3783 & 512 \\
    33    & 21    & 0.82  & 0.04 & 0.03 & 0.15 & 3783 &  43 \\
    33    & 11    & 0.81  & 0.08 & 0.03 & 0.02 & 2346 &  27 \\
\hline
\end{tabular}
\end{adjustbox}
\caption*{\scriptsize Notes: $\underline{d}_2$ and $\underline{d}^*_2$ indicate the treatment sequences under treatment and non-treatment, respectively. $\hat{E}[Y_2(\underline{d}^*_2)|S=1]$ denotes the mean potential outcome under non-treatment conditional on $S=1$, where $S$ is an indicator for the first treatment corresponding to either the first treatment in  $\underline{d}_2$ or $\underline{d}^*_2$. $\hat{\Delta}(\underline{d}_2,\underline{d}^*_2, S=1)$ provides the ATE estimate, SE the standard error. The last column gives the number of observations dropped according to the trimming rule $p^{d_1}(X_0) \cdot p^{d_2}(D_1,\underline{X}_1)<0.01$.}
\end{table}


Table~\ref{tab:res} presents the results for our three different comparisons of treatment sequences. As displayed in the first and second rows, we find no statistically significant increase in employment when attending two years of vocational training rather than one or two years of academic classroom training, respectively. Even though the point estimate $\hat{\Delta}(\underline{d}_2,\underline{d}^*_2, S=1)$ suggests an increase of 8 and 4 percentage points in the employment probability (starting from a counterfactual probability of 77\% and 82\%, respectively), the p-values are beyond any conventional level of statistical significance. For the comparison of vocational training to no training in either year presented in the third row, however, the effect of 8 percentage points is statistically significant at the 5\% level. We therefore conclude that vocational training appears to increase the employment probability 4 years after randomization into Job Corps, while it is less clear whether it performs relatively better than academic classroom training. The results are qualitatively similar when increasing the trimming threshold for the products of the propensity scores to $0.03$, see Table~\ref{tab:rob03}. However, the p-value of the effect of vocational vs.\ no training is now somewhat higher (6\%), while the effect of vocational training (in both periods) vs.\ classroom training in the first period only is borderline significant at the 10\% level.

\begin{table}[htbp]
	\centering\normalsize
	\caption{Effect estimates with a trimming threshold of 0.03}
	\label{tab:rob03}
	\begin{adjustbox}{max width=\textwidth}
		\begin{tabular}{ll|cccccc}
			\hline
			\hline
			$\underline{d}_2$ & $\underline{d}^*_2$ & $\hat{E}[Y_2(\underline{d}^*_2)|S=1]$ &  $\hat{\Delta}(\underline{d}_2,\underline{d}^*_2, S=1)$ & SE & p-value & observations & trimmed \\
			\hline
			33    & 22    & 0.79  & 0.05 & 0.06 & 0.4  & 3783 & 1932 \\
			33    & 21    & 0.82  & 0.05 & 0.03 & 0.1  & 3783 &  578 \\
			33    & 11    & 0.81  & 0.07 & 0.04 & 0.06 & 2346 &  157 \\
			\hline
		\end{tabular}
	\end{adjustbox}
	\caption*{\scriptsize Notes: $\underline{d}_2$ and $\underline{d}^*_2$ indicate the treatment sequences under treatment and non-treatment, respectively. $\hat{E}[Y_2(\underline{d}^*_2)|S=1]$ denotes the mean potential outcome under non-treatment conditional on $S=1$, where $S$ is an indicator for the first treatment corresponding to either the first treatment in  $\underline{d}_2$ or $\underline{d}^*_2$. $\hat{\Delta}(\underline{d}_2,\underline{d}^*_2, S=1)$ provides the ATE estimate, SE the standard error. The last column gives the number of observations dropped according to the trimming rule $p^{d_1}(X_0) \cdot p^{d_2}(D_1,\underline{X}_1)<0.03$.}
\end{table}

%
%

To partially assess the validity of the conditional independence assumptions imposed in this application, we conduct a placebo test based on comparing the outcomes of two control groups as for instance discussed in \citet{AthImb17}. The first control group are the never takers, i.e.\ those randomized into Job Corps who never attended any form of instruction with treatment sequence 11. The second control group are those randomized out and thus without access to Job Corps instruction with treatment sequence 00. We estimate the pseudo-treatment effect of Job Corps on the employment outcome using the  double machine learning approach for assessing static (rather than dynamic) treatments as for instance discussed in \cite{Chetal2018}. To this end, we consider sequence 11 as pseudo-treatment and sequence 00 as non-treatment and control for the baseline covariates $X_0$ based on the random forest as machine learner of the nuisance parameters. As neither group attended any training, the true ATE is equal to zero. As shown in Table~\ref{tab:pseu}, the estimated ATE is indeed approximately zero with a p-value of 93\%. This provides some statistical support for the satisfaction of the conditional independence assumption, at least w.r.t.\ the baseline covariates $X_0$.

\begin{table}[htbp]
\centering\normalsize
\caption{Placebo test with a trimming threshold of 0.01}
\label{tab:pseu}
\begin{adjustbox}{max width=\textwidth}
\begin{tabular}{ccccc}
\hline
\hline
 ATE estimate  & SE & p-value & observations & trimmed \\
\hline
 -0.00 & 0.02 & 0.93 & 4752 & 196 \\
\hline
\end{tabular}
\end{adjustbox}
\caption*{\scriptsize Notes: The ATE estimate provides the pseudo-treatment effect when comparing the employment outcomes of never takers (treatment sequence $11$) and those randomized out (treatment sequence $00$) conditional on baseline covariates $X_0$. The last column states the number of observations dropped according to the trimming rule: $p(X_0)<0.01$.}
\end{table}

\section{Conclusion}\label{conclusion}

In this paper, we combined dynamic treatment evaluation with double machine learning under sequential selection-on-observables assumptions which avoids adhoc pre-selection of control variables. This approach appears particularly fruitful in high-dimensional data with many potential control variables. We suggested estimators for the (weighted) average effects of sequences of treatments (with the so-called controlled direct effect being a special case) based on Neyman orthogonal score functions, sample splitting, and machine learning-based plug-in estimates of conditional mean outcomes and treatment propensity scores. We demonstrated the $\sqrt{n}$-consistency and asymptotic normality of the treatment effect estimators under specific regularity conditions and analyzed their finite sample behavior in a Monte Carlo simulation. Finally, we applied our method to the Job Corps data to analyze the effects of distinct sequences of educational programs and found positive employment effects for vocational training when compared to no program participation.

\bibliographystyle{econometrica}
\bibliography{research}

\bigskip

\renewcommand\appendix{\par
	\setcounter{section}{0}%
	\setcounter{subsection}{0}%
	\setcounter{table}{0}%
	\setcounter{figure}{0}%
	\renewcommand\thesection{\Alph{section}}%
	\renewcommand\thetable{\Alph{section}.\arabic{table}}}
\renewcommand\thefigure{\Alph{section}.\arabic{subsection}.\arabic{subsubsection}.\arabic{figure}}
\clearpage

\begin{appendix}
	
	\numberwithin{equation}{section}
	\noindent \textbf{\LARGE Appendices}

{\small

\section{Proofs}

\subsection{Proof of Theorem 1} \label{Neyman}

For the proof of Theorem 1 it is sufficient to check the conditions of Assumptions 3.1 and 3.2 from Theorem 3.1 and 3.2 and Corollary 3.2 from \cite{Chetal2018}. All bounds hold uniformly over $P \in \mathcal{P},$ where $\mathcal{P}$ is the set of all possible probability laws, and we omit $P$ for brevity.

Define the nuisance parameters to be the vector of functions $\eta=(p^{d_1}(X_0), p^{d_2}(D_1,\underline{X}_1), \mu^{Y_2}(\underline{D}_2,\underline{X}_1))$,
$ \nu^{Y_2}(D_1,X_0)$, with $p^{d_1}(X_0)=\Pr(D_1=d_1|X_0)$, $p^{d_2}(D_1,\underline{X}_1)=\Pr(D_2=d_2|D_1,X_0, X_1)$,   $\mu^{Y_2}(\underline{D}_2,\underline{X}_1)=E[Y_2|\underline{D}_2,X_0,X_1]$, and $\nu^{Y_2}(\underline{D}_2,X_0)=\int E[Y_2|\underline{d}_2,X_0,X_1=x_1] dF_{X_1=x_1|D_1,X_0}$, where $F_{X_1=x_1|D_1,X_0}$ denotes the conditional distribution function of $X_1$ at value $x_1$. The score function for the counterfactual $\Psi^{\underline{d}_2}_{0}=E[Y_2(\underline{d}_2)]$ is given by:
\begin{eqnarray}
\psi^{\underline{d}_2}(W, \eta, \Psi^{\underline{d}_2}_{0}) &=& \frac{I\{D_1=d_1\} \cdot I\{D_2=d_2\} \cdot [Y_2-\mu^{Y_2}(\underline{d}_2,\underline{X}_1)]}{p^{d_1}(X_0)\cdot p^{d_2}(d_1,\underline{X}_1)} \notag\\
& + & \frac{I\{D_1=d_1\}\cdot  [\mu^{Y_2}(\underline{d}_2,\underline{X}_1)-\nu^{Y_2}(\underline{d}_2,X_0)]}{p^{d_1}(X_0)} \notag\\
& + &\nu^{Y_2}(\underline{d}_2,X_0) - \Psi^{\underline{d}_2}_{0}. \notag
\end{eqnarray}

Let $\mathcal{T}_n$ be the set fo all $\eta=(p^{d_1}, p^{d_2},\mu^{Y_2}, \nu^{Y_2})$ consisting of $P$-square integrable functions $p^{d_1}, p^{d_2},\mu^{Y_2}$ and $\nu^{Y_2}$ such that
\begin{eqnarray} \label{Tn}
\left\|  \eta - \eta_0 \right\|_{q} &\leq& C,  \\
\left\|  \eta - \eta_0 \right\|_{2} &\leq& \delta_n, \notag \\
\left\|   p^{d_1}(X_0)-1/2\right\|_{\infty}  &\leq& 1/2-\epsilon, \notag\\
 \left\|   p^{d_2}(D_1,\underline{X}_1)-1/2)\right\|_{\infty} &\leq & 1/2-\epsilon, \notag \\
\left\|  \mu^{Y_2}(\underline{D}_2,\underline{X}_1)-\mu^{Y_2}_0(\underline{D}_2,\underline{X}_1)\right\|_{2} \times \left\|  p^{d_1}(X_0)-p^{d_1}_0(X_0)\right\|_{2}  &\leq & \delta^{}_n n^{-1/2}, \notag \\
\left\|  \mu^{Y_2}(\underline{D}_2,\underline{X}_1)-\mu^{Y_2}_0(\underline{D}_2,\underline{X}_1)\right\|_{2} \times \left\|  p^{d_2}(D_1,\underline{X}_1)-p^{d_2}_0(D_1,\underline{X}_1)\right\|_{2} &\leq & \delta^{}_n n^{-1/2},\notag \\
\left\|  \nu^{Y_2}(\underline{D}_2,X_0)-\nu^{Y_2}_0(\underline{D}_2,X_0)\right\|_{2} \times \left\|  p^{d_1}(X_0)-p^{d_1}_0(X_0)\right\|_{2} &\leq & \delta^{}_n n^{-1/2}.\notag
\end{eqnarray}
We furthermore replace the sequence $(\delta_n)_{n \geq 1}$ by $(\delta_n')_{n \geq 1},$ where $\delta_n' = C_{\epsilon} \max(\delta_n,n^{-1/2}),$ where $C_{\epsilon}$ is sufficiently large constant that only depends on $C$ and $\epsilon.$

\newpage
\textbf{Assumption 3.1:  Linear scores and Neyman orthogonality}
\vspace{5pt}\newline

\textbf{Assumption 3.1(a)}

\textbf{Moment Condition:} The moment condition $E\Big[\psi^{\underline{d}_2}(W, \eta_0, \Psi^{\underline{d}_2}_{0})\Big] = 0$ is satisfied, which follows from the law of iterated expectations:
\begin{eqnarray}
E\Big[\psi^{\underline{d}_2}(W, \eta_0,  \Psi^{\underline{d}_2}_{0})\Big] &=& E\Bigg[\overbrace{ E\Bigg[\frac{I\{D_1=d_1\} \cdot I\{D_2=d_2\} }{p^{d_1}_0(X_0)\cdot p^{d_2}_0(d_1,\underline{X}_1)} \cdot  [Y_2-\mu_0^{Y_2}(\underline{d}_2,\underline{X}_1)]\Bigg|X_0\Bigg]}^{=E[E[Y_2-\mu_0^{Y_2}(\underline{d}_2,\underline{X}_1)|\underline{D}_2=\underline{d}_2,\underline{X}_1]|D_1=d_1,X_0]=0} \Bigg] \notag\\
&+& \ E\Bigg[\overbrace{ E\Bigg[ \frac{I\{D_1=d_1\}\cdot [ \mu_0^{Y_2}(\underline{d}_2,\underline{X}_1)-\nu_0^{Y_2}(\underline{d}_2,X_0)]}{p_0^{d_1}(X_0)} \Bigg|   X_0 \Bigg]}^{=\int E\big[   \mu_0^{Y_2}(\underline{d}_2,\underline{x}_1)-\nu_0^{Y_2}(\underline{d}_2,x_0) \big| D_1=d_1,  \underline{X}_1=\underline{x}_1 \big] dF_{X_1=x_1|D_1=d_1,X_0}=0} \Bigg] \notag\\
&+&  \  E\big[  \nu_0^{Y_2}(\underline{d}_2,X_0) \big] \ \ - \ \ \Psi^{\underline{d}_2}_{0} \ \ = \ \ \Psi^{\underline{d}_2}_{0}\ \ - \ \ \Psi^{\underline{d}_2}_{0} \ \ =  0 \notag
\end{eqnarray}
To better see this result, note that
\begin{eqnarray}
&&E\Bigg[\frac{I\{D_1=d_1\} \cdot I\{D_2=d_2\} }{p_0^{d_1}(X_0)\cdot  p_0^{d_2}(d_1,\underline{X}_1)} \cdot  [Y_2- \mu_0^{Y_2}(\underline{d}_2,\underline{X}_1)]\Bigg|X_0\Bigg]\notag\\
&=&E\Bigg[\frac{I\{D_2=d_2\}  }{ p_0^{d_2}(d_1,\underline{X}_1)} \cdot  [Y_2- \mu_0^{Y_2}(\underline{d}_2,\underline{X}_1)]\Bigg|D_1=d_1,X_0\Bigg]\notag\\
&=&E\Bigg[ E\Bigg[\frac{I\{D_2=d_2\}  }{ p_0^{d_2}(d_1,\underline{X}_1)} \cdot  [Y_2- \mu_0^{Y_2}(\underline{d}_2,\underline{X}_1)]\Bigg|D_1=d_1,\underline{X}_1)\Bigg] \Bigg|D_1=d_1,X_0\Bigg]\notag\\
&=&E [ E[Y_2- \mu_0^{Y_2}(\underline{d}_2,\underline{X}_1) |\underline{D}_2=\underline{d}_2,\underline{X}_1 ]  |D_1=d_1,X_0 ]\notag \\
&=&E [ \mu^{Y_2}_0(\underline{d}_2,\underline{X}_1) -  \mu^{Y_2}_0(\underline{d}_2,\underline{X}_1)|D_1=d_1,X_0  ]=0, \notag
\end{eqnarray}
where the first and third equalities follow from basic probability theory and the second from the law of iterated expectations.
Furthermore,
\begin{eqnarray}
&&E\Bigg[ \frac{I\{D_1=d_1\} \cdot [ \mu_0^{Y_2}(\underline{d}_2,\underline{X}_1)- \nu_0^{Y_2}(\underline{d}_2,x_0)]}{p^{d_1}_0(x_0)} \Bigg|   X_0 =  x_0 \Bigg]\notag\\
&=&E\big[   \mu_0^{Y_2}(\underline{d}_2,\underline{X}_1)- \nu_0^{Y_2}(\underline{d}_2,x_0) \big| D_1=d_1,  X_0 = x_0 \big]\notag\\
&=& \int E\big[   \mu_0^{Y_2}(\underline{d}_2,\underline{x}_1)- \nu_0^{Y_2}(\underline{d}_2,x_0) \big| D_1=d_1,  \underline{X}_1=\underline{x}_1 \big] dF_{X_1=x_1|D_1=d_1,X_0=x_0}\notag\\
&=& \int E\big[  \mu_0^{Y_2}(\underline{d}_2,\underline{x}_1) \big| D_1=d_1,  \underline{X}_1=\underline{x}_1 \big] dF_{X_1=x_1|D_1=d_1,X_0=x_0}-\nu_0^{Y_2}(\underline{d}_2,x_0)\notag\\
&=& \nu^{Y_2}_0(\underline{d}_2,x_0)-\nu^{Y_2}_0(\underline{d}_2,x_0) =0.\notag
\end{eqnarray}
where the first equality follows from basic probability theory, the second from conditioning on and integrating over $X_1$, and the third from the fact that $ \nu_0^{Y_2}(\underline{d}_2,X_0)$ is not a function of $X_1$.

\textbf{Assumption 3.1(b)}

\textbf{Linearity:} The score $ \psi^{\underline{d}_2}(W, \eta_0, \Psi^{\underline{d}_2}_{0}) $ is linear in $\Psi^{\underline{d}_2}_{0}$ :
 $\psi^{\underline{d}_2}(W, \eta_0, \Psi^{\underline{d}_2}_{0}) = \psi^{\underline{d}_2}_a(W, \eta_0) \cdot\Psi^{\underline{d}_2}_0 + \psi^{\underline{d}_2}_b(W, \eta_0) $
with $\psi^{\underline{d}_2}_a(W, \eta_0) = -1$ and
\begin{eqnarray}
  \psi^{\underline{d}_2}_b(W, \eta_0) &=&\frac{I\{D_1=d_1\} \cdot I\{D_2=d_2\} \cdot [Y_2-\mu_0^{Y_2}(\underline{d}_2,\underline{X}_1)]}{p_0^{d_1}(X_0)\cdot p_0^{d_2}(d_1,\underline{X}_1)}\notag\\
 & + & \frac{I\{D_1=d_1\}\cdot  [\mu^{Y_2}(\underline{d}_2,\underline{X}_1)-\nu^{Y_2}(\underline{d}_2,X_0)]}{p^{d_1}(X_0)} + \nu^{Y_2}(\underline{d}_2,X_0). \notag
\end{eqnarray}

\textbf{Assumption 3.1(c)}

\textbf{Continuity:}
The expression for the second Gateaux derivative of a map $\eta \mapsto E[\psi^{\underline{d}_2}(W, \eta, \Psi^{\underline{d}_2})]$ is continuous.
\newpage

\textbf{Assumption 3.1(d)}

\textbf{Neyman Orthogonality}: For any $\eta \in \mathcal{T}_N,$ the Gateaux derivative in the direction $ \eta - \eta_0 =
(p^{d_1}(X_0)-p_0^{d_1}(X_0), p^{d_2}(D_1,\underline{X}_1)-p_0^{d_2}(D_1,\underline{X}_1),\mu^{Y_2}(\underline{d}_2,X_0)-\mu_0^{Y_2}(\underline{d}_2,X_0), \nu^{Y_2}(\underline{d}_2,X_0)-\nu_0^{Y_2}(\underline{d}_2,X_0)) $ is given by:
	\begin{align}
	&\partial E \big[\psi^{\underline{d}_2}(W, \eta, \Psi^{\underline{d}_2})\big] \big[\eta - \eta_0 \big]  = \notag\\
	& - E \Bigg[  \frac{I\{D_1=d_1\} \cdot I\{D_2=d_2\} \cdot [\mu^{Y_2}(\underline{d}_2,\underline{X}_1)-\mu_0^{Y_2}(\underline{d}_2,\underline{X}_1)]}{p_0^{d_1}(X_0)\cdot p_0^{d_2}(d_1,\underline{X}_1)}  \Bigg]  \tag{$*$}\\
	& + E \Bigg[  \frac{I\{D_1=d_1\}  \cdot [\mu^{Y_2}(\underline{d}_2,\underline{X}_1)-\mu_0^{Y_2}(\underline{d}_2,\underline{X}_1)]}{p_0^{d_1}(X_0)}  \Bigg]  \tag{$**$}\\
	&- \  E \Bigg[ \frac{\overbrace{I\{D_1=d_1\} \cdot I\{D_2=d_2\} \cdot  [Y_2-\mu_0^{Y_2}(\underline{d}_2,\underline{X}_1)]}^{E[\cdot|X_0]=E[E[Y_2-\mu_0^{Y_2}(\underline{d}_2,\underline{X}_1)|\underline{D}_2=\underline{d}_2,\underline{X}_1]|D_1=d_1,X_0]=0}}{ p_0^{d_1}(X_0)\cdot p_0^{d_2}(d_1,\underline{X}_1) }\cdot\frac{[p^{d_1}(X_0)-p_0^{d_1}(X_0)] }{p_0^{d_1}(X_0)} \Bigg] \notag\\
	& - \  E \Bigg[  \overbrace{\frac{I\{D_1=d_1\}\cdot  [\mu_0^{Y_2}(\underline{d}_2,\underline{X}_1)-\nu_0^{Y_2}(\underline{d}_2,X_0)]}{p_0^{d_1}(X_0)}}^{E[\cdot|X_0]=\int E\big[  \mu_0^{Y_2}(\underline{d}_2,\underline{x}_1)-\nu_0^{Y_2}(\underline{d}_2,x_0) \big| D_1=d_1,  \underline{X}_1=\underline{x}_1 \big] dF_{X_1=x_1|D_1=d_1,X_0}=0} \cdot \frac{[p^{d_1}(X_0)-p_0^{d_1}(X_0)] }{p_0^{d_1}(X_0)} \Bigg]  \notag \\
	&- \  E \Bigg[ \frac{\overbrace{I\{D_1=d_1\} \cdot I\{D_2=d_2\} \cdot  [Y_2-\mu_0^{Y_2}(\underline{d}_2,\underline{X}_1)]}^{E[\cdot|X_0]=E[E[Y_2-\mu_0^{Y_2}(\underline{d}_2,\underline{X}_1)|\underline{D}_2=\underline{d}_2,\underline{X}_1]|D_1=d_1,X_0]=0}}{ p_0^{d_1}(X_0)\cdot p_0^{d_2}(d_1,\underline{X}_1) }\cdot\frac{[p^{d_2}(d_1,\underline{X}_1)-p_0^{d_2}(d_1,\underline{X}_1)] }{p_0^{d_2}(d_1,\underline{X}_1)} \Bigg] \notag\\
	&\underbrace{ - \  E \Bigg[  \underbrace{\frac{I\{D_1=d_1\}  \cdot [\nu^{Y_2}(\underline{d}_2,X_0)-\nu_0^{Y_2}(\underline{d}_2,X_0)]}{p_0^{d_1}(X_0)}}_{ E[\cdot|X_0]=\frac{p_0^{d_1}(X_0)}{p_0^{d_1}(X_0)}\cdot[\nu^{Y_2}(\underline{d}_2,X_0)-\nu_0^{Y_2}(\underline{d}_2,X_0)]}   \Bigg]+  E[\nu^{Y_2}(\underline{d}_2,X_0)-\nu_0^{Y_2}(\underline{d}_2,X_0)]}_{=0} =0\notag
	\end{align}
	The Gateaux derivative is zero because expressions $(*)$ and $(**)$ cancel out. To see this, note that
	\begin{eqnarray}
	&&E\Bigg[ \frac{I\{D_1=d_1\} \cdot [\mu^{Y_2}(\underline{d}_2,\underline{X}_1)-\mu_0^{Y_2}(\underline{d}_2,\underline{X}_1)]}{p^{d_1}(x_0)} \Bigg|   X_0 =  x_0 \Bigg]\notag\\
	&=&E\big[  \mu^{Y_2}(\underline{d}_2,\underline{X}_1)-\mu_0^{Y_2}(\underline{d}_2,\underline{X}_1) \big| D_1=d_1,  X_0 = x_0 \big]\notag\\
	&=& \int E\big[  \mu^{Y_2}(\underline{d}_2,\underline{x}_1)-\mu_0^{Y_2}(\underline{d}_2,\underline{x}_1) \big| D_1=d_1,  \underline{X}_1=\underline{x}_1 \big] dF_{X_1=x_1|D_1=d_1,X_0=x_0}, \notag 
	\end{eqnarray}
	where the first equality follows from basic probability theory and the second from conditioning on and integrating over $X_1$. Furthermore,
	\begin{eqnarray}
	&&E\Bigg[ \frac{I\{D_1=d_1\}  \cdot I\{D_2=d_2\}  \cdot [\mu^{Y_2}(\underline{d}_2,\underline{X}_1)-\mu_0^{Y_2}(\underline{d}_2,\underline{X}_1)]}{p^{d_1}(x_0)  \cdot p_0^{d_2}(d_1,\underline{X}_1) } \Bigg|   X_0 =  x_0 \Bigg]\notag\\
	&=&E\Bigg[   \frac{I\{D_2=d_2\}  \cdot[\mu^{Y_2}(\underline{d}_2,\underline{X}_1)-\mu_0^{Y_2}(\underline{d}_2,\underline{X}_1)]} {p_0^{d_2}(d_1,\underline{X}_1) } \Bigg| D_1=d_1,  X_0 = x_0 \Bigg]\notag\\
	&=& \int E\Bigg[  \frac{I\{D_2=d_2\}  \cdot [\mu^{Y_2}(\underline{d}_2,\underline{x}_1)-\mu_0^{Y_2}(\underline{d}_2,\underline{X}_1)]}{p_0^{d_2}(d_1,\underline{x}_1) }  \Bigg| D_1=d_1,  \underline{x}_1=\underline{x}_1 \Bigg] dF_{X_1=x_1|D_1=d_1,X_0=x_0}\notag\\
	&=& \int E\big[  \mu^{Y_2}(\underline{d}_2,\underline{x}_1)-\mu_0^{Y_2}(\underline{d}_2,\underline{x}_1) \big| D_1=d_1, D_2=d_2,  \underline{X}_1=\underline{x}_1 \big] dF_{X_1=x_1|D_1=d_1,X_0=x_0}\notag\\
	&=& \int E\big[  \mu^{Y_2}(\underline{d}_2,\underline{x}_1)-\mu_0^{Y_2}(\underline{d}_2,\underline{x}_1) \big| D_1=d_1,   \underline{X}_1=\underline{x}_1 \big] dF_{X_1=x_1|D_1=d_1,X_0=x_0},\notag
	\end{eqnarray}
	where the first equality follows from basic probability theory, the second from conditioning on and integrating over $X_1$, the third from basic probability theory, and the fourth from simplification as $\mu^{Y_2}(\underline{d}_2,\underline{x}_1)=E[Y_2|D_1=d_1,D_2=d_2,\underline{X}_1=\underline{x}_1]$ is already conditional on $D_2=d_2$.
\begin{align}
&\partial E \big[\psi^{\underline{d}_2}(W, \eta, \Psi^{\underline{d}_2})\big] \big[\eta - \eta_0 \big] = 0 \notag
\end{align}
proving that the score function is orthogonal.

\textbf{Assumption 3.1(e)}

\textbf{Singular values of $E[\psi^{\underline{d}_2}_a(W;\eta_0)]$ are bounded:}
Holds trivially, because $\psi^{\underline{d}_2}_a(W;\eta_0) = -1.$

\bigskip

\textbf{Assumption 3.2:  Score regularity and quality of nuisance parameter estimators}

\textbf{Assumption 3.2(a)}

This assumption directly follows from the construction of the set $\mathcal{T}_n$ and the regularity conditions (Assumption 4).

\textbf{Assumption 3.2(b)} 

\textbf{Bound for $m_n$:}
\begin{eqnarray}
\left\|  \mu^{Y_2}_0(\underline{D}_2,\underline{X}_1)  \right\|_{q} &=& \left( E\left[ \left| \mu_0^{Y_2}(\underline{D}_2,\underline{X}_1) \right|^q \right] \right)^{\frac{1}{q}} \notag \\
&=&  \left( \sum_{\underline{d}_2 \in \{0,1,...,Q\}^2} E\left[ \left| \mu_0^{Y_2}(\underline{d}_2,\underline{X}_1) \right|^q \Pr  _{P}(\underline{D}_2=\underline{d}_2|\underline{X}_1)  \right]  \right)^{\frac{1}{q}} \notag \\
&\geq& \epsilon^{2/q} \left( \sum_{\underline{d}_2 \in \{0,1,...,Q\}^2} E\left[ \left| \mu_0^{Y_2}(\underline{d}_2,\underline{X}_1) \right|^q \right] \right)^{\frac{1}{q}} \notag \\
&\geq& \epsilon^{2/q} \left( \max_{\underline{d}_2 \in \{0,1,...,Q\}^2} E\left[ \left| \mu_0^{Y_2}(\underline{d}_2,\underline{X}_1) \right|^q \right] \right)^{\frac{1}{q}} \notag \\
&=& \epsilon^{2/q} \left( \max_{\underline{d}_2 \in \{0,1,...,Q\}^2} \left\|  \mu_0^{Y_2}(\underline{d}_2,\underline{X}_1)  \right\|_{q}\right), \notag
\end{eqnarray}
where the first equality follows from definition, the second from the law of total probability, and the third line from the fact that $\Pr(\underline{D}_2 = \underline{d}_2|\underline{X}_1) = p_0^{d_1} (X_0) \cdot p_0^{d_2} (d_1,\underline{X}_1) \geq \epsilon^2.$  Similarly, we obtain
\begin{equation}
\left\|  \nu^{Y_2}_0(\underline{D}_2,X_0)  \right\|_{q} \geq \epsilon^{2/q} \left( \max_{\underline{d}_2 \in \{0,1,...,Q\}^2} \left\|  \nu_0^{Y_2}(\underline{d}_2,X_0)  \right\|_{q}\right). \notag
\end{equation}

Notice that by Jensen's inequality $\left\|  \mu^{Y_2}_0(\underline{D}_2,\underline{X}_1)  \right\|_{q} \leq \left\|  Y_2  \right\|_{q}$ and
$\left\|  \nu^{Y_2}_0(\underline{D}_2,X_0)  \right\|_{q} \leq \left\|  Y_2  \right\|_{q}$ and hence
$\left\|  \mu^{Y_2}_0(\underline{d}_2,\underline{X}_1)  \right\|_{q} \leq C/\epsilon^{2/q}$ and
$\left\|  \nu^{Y_2}_0(\underline{d}_2,X_0)  \right\|_{q} \leq C/\epsilon^{2/q},$
by conditions (\ref{Tn}). Similarly, for any $\eta \in \mathcal{T}_N:$
$\left\|  \mu^{Y_2}(\underline{d}_2,\underline{X}_1) - \mu^{Y_2}_0(\underline{d}_2,\underline{X}_1)  \right\|_{q} \leq C/\epsilon^{2/q}$ and
$\left\|    \nu^{Y_2}(\underline{d}_2,X_0) - \nu^{Y_2}_0(\underline{d}_2,X_0)  \right\|_{q} \leq C/\epsilon^{2/q},$ because
$\left\|  \mu^{Y_2}(\underline{D}_2,\underline{X}_1) - \mu^{Y_2}_0(\underline{D}_2,\underline{X}_1)  \right\|_{q} \leq C$ and
$\left\|    \nu^{Y_2}(\underline{D}_2,X_0) - \nu^{Y_2}_0(\underline{D}_2,X_0)  \right\|_{q} \leq C.$

Consider
\begin{eqnarray}
E\Big[ \psi^{\underline{d}_2}(W, \eta, \Psi_{0}^{\underline{d}_2})\Big] &=& E\Bigg[ \underbrace{ \frac{ I\{D_1=d_1\} \cdot I\{D_2=d_2\}}{p^{d_1}(X_0)\cdot p^{d_2}(d_1,\underline{X}_1)} \cdot Y_2 }_{=I_1}  \notag\\
& + & \underbrace{ \frac{I\{D_1=d_1\}}{p^{d_1}(X_0)} \cdot \bigg(1- \frac{I\{D_2=d_2\}}{p^{d_2}(d_1,\underline{X}_1) } \bigg) \cdot \mu^{Y_2}(\underline{d}_2,\underline{X}_1)  }_{=I_2}  \notag\\
& + & \underbrace{  \bigg(1-   \frac{I\{D_1=d_1\}}{p^{d_1}(X_0)}\bigg)  \nu^{Y_2}(\underline{d}_2,X_0)}_{=I_3} - \Psi^{\underline{d}_2}_{0}  \Bigg]  \notag
\end{eqnarray}
and thus
\begin{eqnarray}
\left\| \psi^{\underline{d}_2}(W, \eta, \Psi_{0}^{\underline{d}_2}) \right\|_{q} &\leq&  \left\| I_1 \right\|_{q}  + \left\| I_2 \right\|_{q}  + \left\| I_3 \right\|_{q} +  \left\| \Psi^{\underline{d}_2}_{0}  \right\|_{q} \notag \\
&\leq& \frac{1}{\epsilon^2}  \left\| Y_2 \right\|_{q} \notag + \frac{1- \epsilon}{\epsilon^2} \left\|  \mu^{Y_2}(\underline{d}_2,\underline{X}_1)  \right\|_{q} + \\
&+& \frac{1-\epsilon}{\epsilon} \left\|  \nu^{Y_2}(\underline{d}_2,X_0)  \right\|_{q}   +  | \Psi^{\underline{d}_2}_{0} | \notag \\
&\leq& C \left( \frac{1}{\epsilon^2} + \frac{2(1-\epsilon)}{\epsilon^{2/q}} \left(\frac{1}{\epsilon^2} + \frac{1}{\epsilon} \right) + \frac{1}{\epsilon} \right),  \notag
\end{eqnarray}
because of triangular inequality and because the following set of inequalities hold:
\begin{eqnarray} \label{32b}
\left\|  \mu^{Y_2}(\underline{d}_2,\underline{X}_1)  \right\|_{q} &\leq& \left\|  \mu^{Y_2}(\underline{d}_2,\underline{X}_1) -  \mu^{Y_2}_0(\underline{d}_2,\underline{X}_1)  \right\|_{q} + \left\|  \mu^{Y_2}_0(\underline{d}_2,\underline{X}_1)  \right\|_{q} \leq 2C/\epsilon^{2/q},  \\
\left\|  \nu^{Y_2}(\underline{d}_2,X_0)  \right\|_{q} &\leq& \left\|  \nu^{Y_2}(\underline{d}_2,X_0) -  \nu^{Y_2}_0(\underline{d}_2,X_0)  \right\|_{q} + \left\|  \nu^{Y_2}_0(\underline{d}_2,X_0)  \right\|_{q} \leq 2C/\epsilon^{2/q}, \notag \\
|\Psi^{\underline{d}_2}_{0} | &=& |E[ \nu^{Y_2}_0(\underline{d}_2,X_0)] | \leq  E_{ } \Big[\left|  \nu^{Y_2}_0(\underline{d}_2,X_0)  \right|^1 \Big]^{\frac{1}{1}} =  \left\| \nu^{Y_2}_0(\underline{d}_2,X_0) \right\|_{1}  \notag \\
&\leq&  \left\| \nu^{Y_2}_0(\underline{d}_2,X_0) \right\|_{2} \leq  \left\| Y_2 \right\|_{2}/\epsilon^{2/2} \overbrace{ \leq}^{q > 2}  \left\| Y_2 \right\|_{q}/\epsilon \leq C /\epsilon.  \notag
\end{eqnarray}
which gives the upper bound on $m_n$ in Assumption 3.2(b) of \cite{Chetal2018}.

\textbf{Bound for $m'_n$:}

Notice that
$$\Big(E[ |\psi_a^{\underline{d}_2}(W, \eta) |^q] \Big)^{1/q}=1$$
and this gives the upper bound on $m'_n$ in Assumption 3.2(b).

\textbf{Assumption 3.2(c)}

In the following, we omit arguments for the sake of brevity and use $p^{d_1}= p^{d_1}(X_0),p^{d_2}= p^{d_2}(d_1,\underline{X}_1),\nu^{Y_2} = \nu^{Y_2}(\underline{d}_2,X_0),  \mu^{Y_2} = \mu^{Y_2}(\underline{d}_2,\underline{X}_1)$ and similarly for $p^{d_1}_0,p^{d_2}_0,\nu^{Y_2}_0,\mu^{Y_2}_0.$

\textbf{Bound for $r_n$:}

For any $\eta = (p^{d_1}, p^{d_2},\mu^{Y_2}, \nu^{Y_2})$ we have
$$ \Big| E\Big( \psi_a^{\underline{d}_2}(W, \eta) - \psi_a^{\underline{d}_2}(W, \eta_0) \Big) \Big| = |1-1| = 0 \leq \delta'_N,$$
and thus we have the bound on $r_n$ from Assumption 3.2(c).

\textbf{Bound for $r'_n$:}
\begin{eqnarray}\label{rprimeN}
&& \left\|   \psi^{\underline{d}_2}(W, \eta, \Psi_{0}^{\underline{d}_2}) - \psi^{\underline{d}_2}(W, \eta_0, \Psi_{0}^{\underline{d}_2})  \right\|_{2} \leq  \left\|    I\{D_1=d_1\} \cdot I\{D_2=d_2\} \cdot Y \cdot \left( \frac{1}{p^{d_1} p^{d_2}} - \frac{1}{p_0^{d_1} p_0^{d_2}} \right) \right\|_{2}   \\
&+&  \left\|    I\{D_1=d_1\} \cdot I\{D_2=d_2\} \left( \frac{\mu^{Y_2}}{p^{d_1} p^{d_2}} - \frac{\mu^{Y_2}_0}{p_0^{d_1} p_0^{d_2}} \right) \right\|_{2}  + \left\|    I\{D_1=d_1\}  \left( \frac{\mu^{Y_2}}{p^{d_1}} - \frac{\mu^{Y_2}_0}{p_0^{d_1}} \right) \right\|_{2} \notag \\
&+& \left\|    I\{D_1=d_1\}  \left( \frac{\nu^{Y_2}}{p^{d_1}} - \frac{\nu^{Y_2}_0}{p_0^{d_1}} \right) \right\|_{2} + \left\| \nu^{Y_2} - \nu^{Y_2}_0 \right\|_{2} \notag \\
&\leq&  \left\| Y \cdot \left( \frac{1}{p^{d_1} p^{d_2}} - \frac{1}{p_0^{d_1} p_0^{d_2}} \right) \right\|_{2} +   \left\| \frac{\mu^{Y_2}}{p^{d_1} p^{d_2}} - \frac{\mu^{Y_2}_0}{p_0^{d_1} p_0^{d_2}}  \right\|_{2} +  \left\|  \frac{\mu^{Y_2}}{p^{d_1}} - \frac{\mu^{Y_2}_0}{p_0^{d_1}}  \right\|_{2} +  \left\|   \frac{\nu^{Y_2}}{p^{d_1}} - \frac{\nu^{Y_2}_0}{p_0^{d_1}}  \right\|_{2} + \left\| \nu^{Y_2} - \nu^{Y_2}_0 \right\|_{2} \notag \\
&\leq&
 \frac{C}{\epsilon^4} \delta_n \left(1 + \frac{1}{\epsilon} \right) + \delta_n \left(  \frac{1}{\epsilon^5} + C + \frac{C}{\epsilon} \right) +\delta_n \left( \frac{1}{\epsilon^3} + \frac{C}{\epsilon^2} \right) + \delta_n \left( \frac{1}{\epsilon^3} + \frac{C}{\epsilon^2} \right)+  \frac{\delta_n}{\epsilon}
 \leq \delta_n' \notag
\end{eqnarray}
as long as $C_\epsilon$ in the definition of $\delta_n'$ is sufficiently large.  This gives the bound on $r'_n$ from Assumption 3.2(c).
Here we made use of the fact that $\left\| \mu^{Y_2} - \mu^{Y_2}_0 \right\|_{2} = \left\| \mu^{Y_2}(\underline{d}_2,\underline{X}_1) - \mu^{Y_2}_0(\underline{d}_2,\underline{X}_1) \right\|_{2} \leq \delta_n/\epsilon,$ $\left\| \nu^{Y_2} - \nu^{Y_2}_0 \right\|_{2} = \left\| \nu^{Y_2}(\underline{d}_2,X_0) - \nu^{Y_2}_0(\underline{d}_2,X_0) \right\|_{2} \leq \delta_n/\epsilon$ and $\left\| p^{d_2} - p^{d_2}_0 \right\|_{2} = \left\|p^{d_2} (d_1,X_0) - p^{d_2} _0(d_1,X_0) \right\|_{2} \leq \delta_n/\epsilon$ using similar steps as in Assumption 3.1(b).

The last inequality in (\ref{rprimeN}) is satisfied because we can bound the first term by
\begin{eqnarray*}
&&  \left\| Y \cdot \left( \frac{1}{p^{d_1} p^{d_2}} - \frac{1}{p_0^{d_1} p_0^{d_2}} \right) \right\|_{2}  \leq   C  \left\|  \frac{1}{p^{d_1} p^{d_2}} - \frac{1}{p_0^{d_1} p_0^{d_2}}  \right\|_{2} \leq \frac{C}{\epsilon^4} \left\| p_0^{d_1} p_0^{d_2} - p^{d_1} p^{d_2}  \right\|_{2} \\
&=& \frac{C}{\epsilon^4} \left\| p_0^{d_1} p_0^{d_2} - p^{d_1} p^{d_2} +  p_0^{d_1}  p^{d_2}  -  p_0^{d_1}  p^{d_2}  \right\|_{2} \leq \frac{C}{\epsilon^4}\left( \left\| p_0^{d_1} (p_0^{d_2} - p^{d_2} ) \right\|_{2} + \left\| p_0^{d_2} (p_0^{d_1} - p^{d_1})  \right\|_{2} \right) \\
&\leq&  \frac{C}{\epsilon^4}\left( \left\| p_0^{d_2} - p^{d_2}  \right\|_{2} + \left\| p_0^{d_1} - p^{d_1}  \right\|_{2} \right) \leq \frac{C}{\epsilon^4} \delta_n \left(1 + \frac{1}{\epsilon} \right),
\end{eqnarray*}
where the first inequality follows from the second inequality in Assumption 4(a). The second term in (\ref{rprimeN}) is bounded by
\begin{eqnarray*}
&& \left\|  \frac{\mu^{Y_2}}{p^{d_1} p^{d_2}} - \frac{\mu^{Y_2}_0}{p_0^{d_1} p_0^{d_2}} \right\|_{2} \leq \frac{1}{\epsilon^4} \left\| p_0^{d_1} p_0^{d_2} \mu^{Y_2} -  p^{d_1} p^{d_2} \mu^{Y_2}_0 \right\|_{2} =  \frac{1}{\epsilon^4}\left\|  p_0^{d_1} p_0^{d_2} \mu^{Y_2} -  p^{d_1} p^{d_2}  \mu^{Y_2}_0 + p_0^{d_1} p_0^{d_2} \mu^{Y_2}_0 - p_0^{d_1} p_0^{d_2} \mu^{Y_2}_0 \right\|_{2} \\
&\leq&  \frac{1}{\epsilon^4} \left( \left\| p_0^{d_1} p_0^{d_2}  (\mu^{Y_2} - \mu^{Y_2}_0) \right\|_{2} + \left\| \mu^{Y_2}_0 ( p_0^{d_1} p_0^{d_2}  - p^{d_1} p^{d_2}  ) \right\|_{2} \right) \leq   \frac{1}{\epsilon^4} \left( \left\| \mu^{Y_2} - \mu^{Y_2}_0 \right\|_{2} + C  \left\| p_0^{d_1} p_0^{d_2}  - p^{d_1} p^{d_2} \right\|_{2} \right) \\
&\leq& \frac{1}{\epsilon^4} \left( \frac{\delta_n}{\epsilon} + C \left\| p_0^{d_1} p_0^{d_2}  - p^{d_1} p^{d_2} \right\|_{2} \right) \leq \delta_n \left(  \frac{1}{\epsilon^5} + C + \frac{C}{\epsilon} \right) ,
\end{eqnarray*}
where the third inequality follows from $E[Y^2_2| D_1 =d_1, D_2=d_2, \underline{X_1}] \geq (E[Y_2| D_1 =d_1, D_2=d_2, \underline{X_1}])^2= \mu^2_0(\underline{d}_2, \underline{X_1}) $ by the conditional Jensen's inequality and therefore $\left\| \mu^{Y_2}_0 (\underline{d}_2,\underline{X}_1) \right\|_\infty \leq C^2.$

For the third term we get
\begin{eqnarray*}
&& \left\|  \frac{\mu^{Y_2}}{p^{d_1}} - \frac{\mu^{Y_2}_0}{p_0^{d_1}}  \right\|_{2} = \frac{1}{\epsilon^2}  \left\| p_0^{d_1} \mu^{Y_2}-  p^{d_1} \mu^{Y_2}_0  \right\|_{2} =  \frac{1}{\epsilon^2} \left\| p_0^{d_1} \mu^{Y_2}-  p^{d_1} \mu^{Y_2}_0 + p_0^{d_1} \mu^{Y_2}_0 - p_0^{d_1} \mu^{Y_2}_0   \right\|_{2} \\
&\leq&\frac{1}{\epsilon^2} \left( \left\|  p_0^{d_1}(\mu^{Y_2} - \mu^{Y_2}_0) \right\|_{2} +  \left\| \mu^{Y_2}_0 (p_0^{d_1} - p^{d_1}) \right\|_{2} \right) \leq \frac{1}{\epsilon^2} \left( \left\| \mu^{Y_2} - \mu^{Y_2}_0 \right\|_{2} + C \left\| p_0^{d_1} - p^{d_1} \right\|_{2} \right) \leq \delta_n \left( \frac{1}{\epsilon^3} + \frac{C}{\epsilon^2} \right),
\end{eqnarray*}
and similarly, for the fourth term we obtain
\begin{eqnarray*}
&& \left\|  \frac{\nu^{Y_2}}{p^{d_1}} - \frac{\nu^{Y_2}_0}{p_0^{d_1}}  \right\|_{2}  \leq \delta_n \left( \frac{1}{\epsilon^3} + \frac{C}{\epsilon^2} \right),
\end{eqnarray*}
where we used Jensen's inequality twice to get $\left\| \nu^{Y_2}_0 (\underline{d}_2,X_0) \right\|_\infty \leq C^2$.

\textbf{Bound for $\lambda'_n$:}

Now consider


\begin{equation}
f(r) := E[\psi(W;\Psi_0^{\underline{d}_2},\eta + r(\eta-\eta_0)]. \notag
\end{equation}
 For any $r \in (0,1):$
\begin{eqnarray} \label{secondGatDer}
\frac{\partial^2 f(r)}{\partial r^2}&=& E\Bigg[  I\{D_1=d_1\} \cdot I\{D_2=d_2\} (-2) \frac{(\mu^{Y_2}-\mu^{Y_2}_0)(p^{d_1} - p^{d_1}_0)}{\left(p^{d_1}_0 + r(p^{d_1} -p^{d_1}_0)\right)^2\left(p^{d_2}_0 + r(p^{d_2} -p^{d_2}_0)\right)}  \Bigg]  \\
&+& E\Bigg[  I\{D_1=d_1\} \cdot I\{D_2=d_2\} (-2) \frac{(\mu^{Y_2}-\mu^{Y_2}_0)(p^{d_2} - p^{d_2}_0)}{\left(p^{d_1}_0 + r(p^{d_1} -p^{d_1}_0)\right)\left(p^{d_2}_0 + r(p^{d_2} -p^{d_2}_0)\right)^2}  \Bigg]  \notag \\
&+& E\Bigg[  I\{D_1=d_1\} \cdot I\{D_2=d_2\} 2 \frac{(Y_2 - \mu^{Y_2}_0 - r(\mu^{Y_2}-\mu^{Y_2}_0) )(p^{d_1} - p^{d_1}_0)^2}{\left(p^{d_1}_0 + r(p^{d_1} -p^{d_1}_0)\right)^3\left(p^{d_2}_0 + r(p^{d_2} -p^{d_2}_0)\right)}  \Bigg]  \notag \\
&+& E\Bigg[  I\{D_1=d_1\} \cdot I\{D_2=d_2\} 2 \frac{(Y_2 - \mu^{Y_2}_0 - r(\mu^{Y_2}-\mu^{Y_2}_0) )(p^{d_2} - p^{d_2}_0)^2}{\left(p^{d_1}_0 + r(p^{d_1} -p^{d_1}_0)\right)\left(p^{d_2}_0 + r(p^{d_2} -p^{d_2}_0)\right)^3}  \Bigg]  \notag \\
&+& E\Bigg[  I\{D_1=d_1\} \cdot I\{D_2=d_2\} 2 \frac{(Y_2 - \mu^{Y_2}_0 - r(\mu^{Y_2}-\mu^{Y_2}_0) )(p^{d_1} - p^{d_1}_0)(p^{d_2} - p^{d_2}_0)}{\left(p^{d_1}_0 + r(p^{d_1} -p^{d_1}_0)\right)^2\left(p^{d_2}_0 + r(p^{d_2} -p^{d_2}_0)\right)^2}  \Bigg]  \notag \\
&+& E\Bigg[  I\{D_1=d_1\} (-2) \frac{(\mu^{Y_2}-\mu^{Y_2}_0)(p^{d_1} - p^{d_1}_0)}{\left(p^{d_1}_0 + r(p^{d_1} -p^{d_1}_0)\right)^2}  \Bigg] + E\Bigg[  I\{D_1=d_1\} 2 \frac{(\nu^{Y_2}-\nu^{Y_2}_0)(p^{d_1} - p^{d_1}_0)}{\left(p^{d_1}_0 + r(p^{d_1} -p^{d_1}_0)\right)^2}  \Bigg]  \notag \\
&+& E\Bigg[  I\{D_1=d_1\} 2 \frac{r(\mu^{Y_2}-\mu^{Y_2}_0)(p^{d_1} - p^{d_1}_0)^2}{\left(p^{d_1}_0 + r(p^{d_1} -p^{d_1}_0)\right)^3}  \Bigg] + E\Bigg[  I\{D_1=d_1\} 2 \frac{(r(\nu^{Y_2}-\nu^{Y_2}_0)(p^{d_1} - p^{d_1}_0)^2}{\left(p^{d_1}_0 + r(p^{d_1} -p^{d_1}_0)\right)^3}  \Bigg] \notag \\
&+& E\Bigg[  I\{D_1=d_1\} 2 \frac{(\mu^{Y_2}_0 -  \nu^{Y_2}_0  )(p^{d_1} - p^{d_1}_0)^2}{\left(p^{d_1}_0 + r(p^{d_1} -p^{d_1}_0)\right)^3}  \Bigg]   \notag
\end{eqnarray}

Note that because
\begin{eqnarray}
E[Y_2-\mu_0^{Y_2}(\underline{d}_2,\underline{X}_1)|D_1=d_1,D_2=d_2,\underline{X}_1]  &= & 0, \notag \\
|p^{d_1} - p^{d_1}_0|  \leq  2, \ \  \ \ \ \ |p^{d_2} - p^{d_2}_0|  &\leq & 2 \notag \\
\left\| \mu^{Y_2}_0 \right\|_{q} \leq  \left\| Y_2 \right\|_{q}/\epsilon^{1/q} &\leq & C/\epsilon^{2/q} \notag \\
\left\| \nu^{Y_2}_0 \right\|_{q} \leq  \left\| Y_2 \right\|_{q}/\epsilon^{1/q} &\leq & C/\epsilon^{2/q} \notag \\
\left\|  \mu^{Y_2}-\mu^{Y_2}_0\right\|_{2} \times \left\|  p^{d_1}-p^{d_1}_0\right\|_{2}  &\leq & \delta^{}_n n^{-1/2}/\epsilon, \notag \\
\left\|  \mu^{Y_2}-\mu^{Y_2}_0\right\|_{2} \times \left\|  p^{d_2}-p^{d_2}_0\right\|_{2} &\leq & \delta^{}_n n^{-1/2}/\epsilon^2,\notag \\
\left\|  \nu^{Y_2}-\nu^{Y_2}_0\right\|_{2} \times \left\|  p^{d_1}-p^{d_1}_0\right\|_{2} &\leq & \delta^{}_n n^{-1/2}/\epsilon.\notag
\end{eqnarray}
we get that for some constant $C_{\epsilon}''$ that only depends on $C$ and $\epsilon$
\begin{equation}
\left|\frac{\partial^2 f(r)}{\partial r^2} \right| \leq C_{\epsilon}'' \delta_n n^{-1/2} \leq \delta_n' n^{-1/2} \notag
\end{equation}
and this gives the upper bound on $\lambda'_n$ in Assumption 3.2(c) of \cite{Chetal2018} as long as $C_{\epsilon} \geq C_{\epsilon}''$. We used the following inequalities
\begin{eqnarray}
\left\| \mu^{Y_2}-\mu^{Y_2}_0\right\|_{2} &=& \left\|\mu^{Y_2}(\underline{d}_2,\underline{X}_1)-\mu^{Y_2}_0(\underline{d}_2,\underline{X}_1)\right\|_{2} \leq  \left\|\mu^{Y_2}(\underline{D}_2,\underline{X}_1)-\mu^{Y_2}_0(\underline{D}_2,\underline{X}_1)\right\|_{2}/\epsilon \notag \\
\left\| \nu^{Y_2}-\nu^{Y_2}_0\right\|_{2} &=& \left\|\nu^{Y_2}(\underline{d}_2,X_0)-\nu^{Y_2}_0(\underline{d}_2,X_0)\right\|_{2} \leq  \left\|\nu^{Y_2}(\underline{D}_2,X_0)-\nu^{Y_2}_0(\underline{D}_2,X_0)\right\|_{2}/\epsilon^2 \notag \\
\left\| p^{d_2}-p^{d_2}_0\right\|_{2} &=& \left\|p^{d_2}(d_1,X_0)-p^{d_2}_0(d_1,X_0)\right\|_{2} \leq  \left\|p^{d_2}(D_1,X_0)-p^{d_2}_0(D_1,X_0)\right\|_{2}/\epsilon, \notag
\end{eqnarray}
and these can be shown using similar steps as in Assumption 3.1(b).

To verify that $\left|\frac{\partial^2 f(r)}{\partial r^2} \right|  \leq C_{\epsilon}'' \delta_n n^{-1/2}$ holds, note that by the triangular inequality it is sufficient to bound the absolute value of each of the ten terms in (\ref{secondGatDer}) separately. We illustrate it for the first, third, and last terms.  For the first term:
\begin{eqnarray}
&& \left| E\Bigg[  I\{D_1=d_1\} \cdot I\{D_2=d_2\} (-2) \frac{(\mu^{Y_2}-\mu^{Y_2}_0)(p^{d_1} - p^{d_1}_0)}{\left(p^{d_1}_0 + r(p^{d_1} -p^{d_1}_0)\right)^2\left(p^{d_2}_0 + r(p^{d_2} -p^{d_2}_0)\right)}  \Bigg] \right| \notag \\
&\leq& 2 \left|  E\Bigg[  \frac{(\mu^{Y_2}-\mu^{Y_2}_0)(p^{d_1} - p^{d_1}_0)}{\left(p^{d_1}_0 + r(p^{d_1} -p^{d_1}_0)\right)^2(p^{d_2}_0 + r(p^{d_2} -p^{d_2}_0)}  \Bigg] \right| \notag \\
&\leq&  \frac{2}{ \epsilon^3} \left| E\Bigg[  (\mu^{Y_2}-\mu^{Y_2}_0)(p^{d_1} - p^{d_1}_0)  \Bigg] \right| \leq  \frac{2}{\epsilon^3} \frac{\delta^{}_N}{\epsilon} n^{-1/2}.  \notag
\end{eqnarray}
For the second inequality we used the fact that for $i \in \{1,2\}: 1 \geq p^{d_i}_0 + r(p^{d_i} -p^{d_i}_0) =  (1-r)p^{d_i}_0 + r p^{d_i} \geq (1-r)\epsilon + r \epsilon = \epsilon$ and in the third Holder's inequality. For the third term, we get
\begin{eqnarray}
&& \left| E\Bigg[  I\{D_1=d_1\} \cdot I\{D_2=d_2\} 2 \frac{(Y_2 - \mu^{Y_2}_0 - r(\mu^{Y_2}-\mu^{Y_2}_0) )(p^{d_1} - p^{d_1}_0)^2}{\left(p^{d_1}_0 + r(p^{d_1} -p^{d_1}_0)\right)^3\left(p^{d_2}_0 + r(p^{d_2} -p^{d_2}_0)\right)}  \Bigg] \right|  \notag \\
&\leq& \frac{2}{\epsilon^4} \left| E\Bigg[  I\{D_1=d_1\} \cdot I\{D_2=d_2\}(Y_2 - \mu^{Y_2}_0 - r(\mu^{Y_2}-\mu^{Y_2}_0) )(p^{d_1} - p^{d_1}_0)^2 \Bigg] \right|  \notag \\
&\leq&  \frac{8}{\epsilon^4} \left| E\Bigg[ I\{D_1=d_1\} \cdot I\{D_2=d_2\}(Y_2 - \mu^{Y_2}_0) \Bigg] \right|  +  \frac{2}{\epsilon^4}  \left| E\Bigg[r(\mu^{Y_2}-\mu^{Y_2}_0) (p^{d_1} - p^{d_1}_0)^2 \Bigg] \right| \notag \\
&\leq& \frac{2 \cdot 2}{\epsilon^4}  \left| E\Bigg[1\cdot(\mu^{Y_2}-\mu^{Y_2}_0) )(p^{d_1} - p^{d_1}_0) \Bigg] \right| \leq \frac{4}{\epsilon^4} \frac{\delta^{}_N}{\epsilon} n^{-1/2}, \notag
\end{eqnarray}
where in addition we made use of conditions (\ref{Tn}).

For the last term, we have
\begin{eqnarray}
&& E\Bigg[  I\{D_1=d_1\} 2 \frac{(\mu^{Y_2}_0 -  \nu^{Y_2}_0  )(p^{d_1} - p^{d_1}_0)^2}{\left(p^{d_1}_0 + r(p^{d_1} -p^{d_1}_0)\right)^3}  \Bigg] \notag \\
&=& E\Bigg[ \overbrace{ I\{D_1=d_1\} \frac{(\mu^{Y_2}_0 -  \nu^{Y_2}_0 )}{p^{d_1}_0} }^{\int E\big[  \mu_0^{Y_2}(\underline{d}_2,\underline{x}_1)-\nu_0^{Y_2}(\underline{d}_2,x_0) \big| D_1=d_1,  \underline{X}_1=\underline{x}_1 \big] dF_{X_1=x_1|D_1=d_1,X_0=x_0}=0} \cdot \frac{2p^{d_1}_0(p^{d_1} - p^{d_1}_0)^2}{\left(p^{d_1}_0 + r(p^{d_1} -p^{d_1}_0)\right)^3}  \Bigg]  = 0. \notag
\end{eqnarray}

The remaining terms in (\ref{secondGatDer}) are bounded similarly.

\textbf{Assumption 3.2(d)} 
\begin{eqnarray}
E\Big[ (\psi^{\underline{d}_2}(W, \eta_0, \Psi_{0}^{\underline{d}_2}) )^2\Big] &=& E\Bigg[ \Bigg( \underbrace{ \frac{ I\{D_1=d_1\} \cdot I\{D_2=d_2\} \cdot [Y_2-\mu_0^{Y_2}(\underline{d}_2,\underline{X}_1)]}{p_0^{d_1}(X_0)\cdot p_0^{d_2}(d_1,\underline{X}_1)} }_{=I_1}  \notag\\
& + & \underbrace{ \frac{I\{D_1=d_1\}\cdot  [\mu_0^{Y_2}(\underline{d}_2,\underline{X}_1)-\nu_0^{Y_2}(\underline{d}_2,X_0)]}{p_0^{d_1}(X_0)} }_{=I_2}  \notag\\
& + & \underbrace{ \nu_0^{Y_2}(\underline{d}_2,X_0) - \Psi^{\underline{d}_2}_{0}}_{=I_3} \Bigg)^2 \Bigg]  \notag\\
& = & E[I_1^2 + I_2^2 + I_3^2] \geq E[I^2_1]\notag\\
& = & E\Bigg[  I\{D_1=d_1\} \cdot I\{D_2=d_2\} \cdot \Bigg( \frac{ [Y_2-\mu_0^{Y_2}(\underline{d}_2,\underline{X}_1)]}{p_0^{d_1}(X_0)\cdot p_0^{d_2}(d_1,\underline{X}_1)} \Bigg)^2 \Bigg] \notag\\
& \geq & \epsilon^2 E\Bigg[ \Bigg( \frac{ [Y_2-\mu_0^{Y_2}(\underline{d}_2,\underline{X}_1)]}{p_0^{d_1}(X_0)\cdot p_0^{d_2}(d_1,\underline{X}_1)} \Bigg)^2 \Bigg] \notag\\
&\geq& \frac{\epsilon^2 c^2}{(1-\epsilon)^4} > 0. \notag
\end{eqnarray}
because $\Pr(\underline{D}_2 = \underline{d}_2|\underline{X}_1)= p_0^{d_1} (X_0) \cdot p_0^{d_2} (d_1,\underline{X}_1) \geq \epsilon^2, \  p_0^{d_1}(X_0) \leq 1-\epsilon$ and $p_0^{d_2}(d_1,\underline{X}_1) \leq 1-\epsilon.$



\noindent where the second equality follows from
\begin{eqnarray}
E\Big[ I_1 \cdot I_2\Big] &=& E\Bigg[  \overbrace{\frac{ I\{D_1=d_1\} \cdot I\{D_2=d_2\}}{ (p_0^{d_1}(X_0))^2\cdot p_0^{d_2}(d_1,\underline{X}_1)}  \cdot [Y_2-\mu_0^{Y_2}(\underline{d}_2,\underline{X}_1)]}^{E[\cdot|X_0]=E[E[Y_2-\mu_0^{Y_2}(\underline{d}_2,\underline{X}_1)|\underline{D}_2=\underline{d}_2,\underline{X}_1]|D_1=d_1,X_0]=0}   \cdot  [\mu_0^{Y_2}(\underline{d}_2,\underline{X}_1)-\nu_0^{Y_2}(\underline{d}_2,X_0)] \Bigg], \notag\\
E\Big[ I_2 \cdot I_3\Big] &=& E\Bigg[ \overbrace{  \frac{ I\{D_1=d_1\}}{p_0^{d_1}(X_0)} \cdot  [\mu_0^{Y_2}(\underline{d}_2,\underline{X}_1)-\nu_0^{Y_2}(\underline{d}_2,X_0)]}^{E[\cdot|\underline{X}_0]=\int E\big[  \mu_0^{Y_2}(\underline{d}_2,\underline{x}_1)-\nu_0^{Y_2}(\underline{d}_2,x_0) \big| D_1=d_1,  \underline{X}_1=\underline{x}_1 \big] dF_{X_1=x_1|D_1=d_1,X_0=x_0}=0}  \cdot [ \nu_0^{Y_2}(\underline{d}_2,X_0) - \Psi^{\underline{d}_2}_{0}] \Bigg],\notag\\
E\Big[ I_1 \cdot I_3\Big] &=& E\Bigg[  \overbrace{\frac{ I\{D_1=d_1\} \cdot I\{D_2=d_2\}}{ p_0^{d_1}(X_0)\cdot p_0^{d_2}(d_1,\underline{X}_1)}  \cdot [Y_2-\mu_0^{Y_2}(\underline{d}_2,\underline{X}_1)]}^{E[\cdot|X_0]=E[E[Y_2-\mu_0^{Y_2}(\underline{d}_2,\underline{X}_1)|\underline{D}_2=\underline{d}_2,\underline{X}_1]|D_1=d_1,X_0]=0}   \cdot  [\nu_0^{Y_2}(\underline{d}_2,X_0) - \Psi^{\underline{d}_2}_{0}] \Bigg]. \notag
\end{eqnarray}

\newpage
\subsection{Proof of Theorem 2} \label{weightedscore}
The proof follows in a similar manner than the one of Theorem 1 (Section \ref{Neyman}). All bounds hold uniformly over all probability laws $P \in \mathcal{P}$ where $\mathcal{P}$ is the set of all possible probability laws, and we omit $P$ for brevity.


Denote by $S$ a binary indicator for being selected into the target population and by $g(X_0)=\Pr(S=1|X_0)$ the selection probability as a function of $X_0$.
Define the nuisance parameter to be $\chi=(g,\eta)=(g(X_0),p^{d_1}(X_0), p^{d_2}(D_1,\underline{X}_1), \mu^{Y_2}(\underline{D}_2,\underline{X}_1),\nu^{Y_2}(\underline{D}_2,X_0))$. The score function for the weighted counterfactual $\Psi^{\underline{d}_2, S=1}_{0}=E[Y_2(\underline{d}_2)|S=1]=E[g(X_0)\cdot Y_2(\underline{d}_2)/\Pr(S=1)]$ is given by:
\begin{eqnarray} \label{score2}
\psi^{\underline{d}_2, S=1}(W, \chi, \Psi^{\underline{d}_2,S=1}_{0}) &=& \frac{g(X_0)}{\Pr(S=1)}\cdot\frac{I\{D_1=d_1\} \cdot I\{D_2=d_2\} \cdot [Y_2-\mu^{Y_2}(\underline{d}_2,\underline{X}_1)]}{p^{d_1}(X_0)\cdot p^{d_2}(d_1,\underline{X}_1)} \notag\\
& + & \frac{g(X_0)}{\Pr(S=1)}\cdot\frac{I\{D_1=d_1\}\cdot  [\mu^{Y_2}(\underline{d}_2,\underline{X}_1)-\nu^{Y_2}(\underline{d}_2,X_0)]}{p^{d_1}(X_0)} \notag\\
& + &\frac{S}{\Pr(S=1)}\cdot\nu^{Y_2}(\underline{d}_2,X_0) - \Psi^{\underline{d}_2, S=1}_{0}. \notag
\end{eqnarray}

Let $\mathcal{T}^*_n$ be the set fo all $\chi= (g,\eta)=(g, p^{d_1}, p^{d_2},\mu^{Y_2}, \nu^{Y_2})$ consisting of $P$-square integrable functions $g, p^{d_1}, p^{d_2},\mu^{Y_2}$ and $\nu^{Y_2}$ such that
\begin{eqnarray}\label{Tnstar}
\left\|  \chi - \chi_0 \right\|_{q} &\leq& C,  \\
\left\|  \chi - \chi_0 \right\|_{2} &\leq& \delta_n, \notag \\
 \left\| g(X_0)-1/2\right\|_{\infty}  &\leq& 1/2-\epsilon, \notag\\
\left\|   p^{d_1}(X_0)-1/2\right\|_{\infty}  &\leq& 1/2-\epsilon, \notag\\
 \left\|   p^{d_2}(D_1,\underline{X}_1)-1/2)\right\|_{\infty} &\leq & 1/2-\epsilon, \notag \\
\left\|  \mu^{Y_2}(\underline{D}_2,\underline{X}_1)-\mu^{Y_2}_0(\underline{D}_2,\underline{X}_1)\right\|_{2} \times \left\|  p^{d_1}(X_0)-p^{d_1}_0(X_0)\right\|_{2}  &\leq & \delta^{}_n n^{-1/2}, \notag \\
\left\|  \mu^{Y_2}(\underline{D}_2,\underline{X}_1)-\mu^{Y_2}_0(\underline{D}_2,\underline{X}_1)\right\|_{2} \times \left\|  p^{d_2}(D_1,\underline{X}_1)-p^{d_2}_0(D_1,\underline{X}_1)\right\|_{2} &\leq & \delta^{}_n n^{-1/2},\notag \\
\left\|  \nu^{Y_2}(\underline{D}_2,X_0)-\nu^{Y_2}_0(\underline{D}_2,X_0)\right\|_{2} \times \left\|  p^{d_1}(X_0)-p^{d_1}_0(X_0)\right\|_{2} &\leq & \delta^{}_n n^{-1/2}.\notag \\
\left\|  \mu^{Y_2}(\underline{D}_2,\underline{X}_1)-\mu^{Y_2}_0(\underline{D}_2,\underline{X}_1)\right\|_{2} \times \left\|  g(X_0)-g_0(X_0)\right\|_{2}  &\leq & \delta^{}_n n^{-1/2}, \notag \\
\left\|  \nu^{Y_2}(\underline{D}_2,X_0)-\nu^{Y_2}_0(\underline{D}_2,X_0)\right\|_{2} \times \left\|  g(X_0)-g_0(X_0)\right\|_{2} &\leq & \delta^{}_n n^{-1/2}.\notag
\end{eqnarray}
We furthermore replace the sequence $(\delta_n)_{n \geq 1}$ by $(\delta_n')_{n \geq 1},$ where $\delta_n' = C_{\epsilon} \max(\delta_n,n^{-1/2}),$ where $C_{\epsilon}$ is sufficiently large constant that only depends on $C$ and $\epsilon.$

\newpage
\textbf{Assumption 3.1:  Linear scores and Neyman orthogonality}
\vspace{5pt}

\textbf{Assumption 3.1(a)}
\textbf{Moment Condition:}
 The moment condition $E\Big[\psi^{\underline{d}_2,S=1}(W, \chi_0, \Psi^{\underline{d}_2,S=1}_{0})\Big] =0$ holds by the law of iterated expectations:
\begin{eqnarray}
&&E\Big[\psi^{\underline{d}_2,S=1}(W,  \chi_0, \Psi^{\underline{d}_2,S=1}_{0})\Big]\notag\\
&=& E\Bigg[ \frac{g(X_0)}{\Pr(S=1)}\cdot \overbrace{ E\Bigg[\frac{I\{D_1=d_1\} \cdot I\{D_2=d_2\} }{ p_0^{d_1}(X_0)\cdot  p_0^{d_2}(d_1,\underline{X}_1)} \cdot [Y_2- \mu_0^{Y_2}(\underline{d}_2,\underline{X}_1)]\Bigg|\underline{X}_1\Bigg]}^{=E[E[Y_2- \mu_0^{Y_2}(\underline{d}_2,\underline{X}_1)|\underline{D}_2=\underline{d}_2,\underline{X}_1]|D_1=d_1,X_0]=0} \Bigg]\notag\\
&+& \ E\Bigg[\frac{g(X_0)}{\Pr(S=1)}\cdot \overbrace{ E\Bigg[ \frac{I\{D_1=d_1\}\cdot [ \mu_0^{Y_2}(\underline{d}_2,\underline{X}_1)- \nu_0^{Y_2}(\underline{d}_2,X_0)]}{ p_0^{d_1}(X_0)} \Bigg|   X_0 \Bigg]}^{= \int E\Big[   \mu_0^{Y_2}(\underline{d}_2,\underline{x}_1)- \nu_0^{Y_2}(\underline{d}_2,x_0) \big| D_1=d_1,  \underline{X}_1=\underline{x}_1 \big] dF_{X_1=x_1|D_1=d_1,X_0=x_0}=0} \Bigg] \notag\\
&+&  \  E\Bigg[ \frac{S}{\Pr(S=1)}\cdot  \nu_0^{Y_2}(\underline{d}_2,X_0) \Bigg]  -  \Psi^{\underline{d}_2,S=1}_{0} \ \ = \ \  \Psi^{\underline{d}_2,S=1}_{0}\ \ - \ \  \Psi^{\underline{d}_2,S=1}_{0} \ \
=  0. \notag
\end{eqnarray}

\textbf{Assumption 3.1(b)}

\textbf{Linearity:} The score $ \psi^{\underline{d}_2, S=1}(W, \chi_0, \Psi^{\underline{d}_2, S=1}_{0}) $ is linear in $\Psi^{\underline{d}_2, S=1}_{0}$:
$\psi^{\underline{d}_2,S=1}(W, \chi_0, \Psi^{\underline{d}_2,S=1}_{0}) = \psi^{\underline{d}_2,S=1}_a(W, \chi_0) \cdot\Psi^{\underline{d}_2,S=1}_0 + \psi^{\underline{d}_2,S=1}_b(W, \chi_0) $
with $\psi^{\underline{d}_2,S=1}_a(W, \chi_0) = -1$ and
\begin{eqnarray}
\psi^{\underline{d}_2,S=1}_b(W, \chi_0) &=&\frac{g(X_0)}{\Pr(S=1)}\cdot\frac{I\{D_1=d_1\} \cdot I\{D_2=d_2\} \cdot [Y_2-\mu_0^{Y_2}(\underline{d}_2,\underline{X}_1)]}{p_0^{d_1}(X_0)\cdot p_0^{d_2}(d_1,\underline{X}_1)}\notag\\
& + & \frac{g(X_0)}{\Pr(S=1)}\cdot\frac{I\{D_1=d_1\}\cdot  [\mu^{Y_2}(\underline{d}_2,\underline{X}_1)-\nu^{Y_2}(\underline{d}_2,X_0)]}{p^{d_1}(X_0)} \notag\\
& + &  \frac{S}{\Pr(S=1)}\cdot\nu^{Y_2}(\underline{d}_2,X_0). \notag
\end{eqnarray}

\textbf{Assumption 3.1(c)}
\textbf{Continuity:}
We may observe that the expression for the second Gateaux derivative of a map $\chi \mapsto E[\psi^{\underline{d}_2,S=1}(W, \chi, \Psi^{\underline{d}_2,S=1})]$, given in (\ref{secondGatDer_b}), is continuous.

\textbf{Assumption 3.1(d)}

\textbf{Neyman Orthogonality}: For any $\eta \in \mathcal{T}^*_N,$ the Gateaux derivative in the direction  $ \chi - \chi_0 = (g(X_0)-g_0(X_0),p^{d_1}(X_0)-p_0^{d_1}(X_0), p^{d_2}(D_1,\underline{X}_1)-p_0^{d_2}(D_1,\underline{X}_1),\mu^{Y_2}(\underline{d}_2,\underline{X}_1)-\mu_0^{Y_2}(\underline{d}_2,\underline{X}_1), \nu^{Y_2}(\underline{d}_2,X_0)-\nu_0^{Y_2}(\underline{d}_2,X_0))$  is given by
\begin{align}
&\partial E \big[\psi^{\underline{d}_2,S=1}(W, \chi,\Psi^{\underline{d}_2,S=1}_{0})\big] \big[\chi - \chi_0 \big]  = \notag\\
& - E \Bigg[\frac{g_0(X_0)}{\Pr(S=1)}\cdot  \frac{I\{D_1=d_1\} \cdot I\{D_2=d_2\} \cdot [\mu^{Y_2}(\underline{d}_2,\underline{X}_1)-\mu_0^{Y_2}(\underline{d}_2,\underline{X}_1)]}{p_0^{d_1}(X_0)\cdot p_0^{d_2}(d_1,\underline{X}_1)}  \Bigg]  \tag{$*$}\\
& + E \Bigg[  \frac{g_0(X_0)}{\Pr(S=1)}\cdot  \frac{I\{D_1=d_1\}  \cdot [\mu^{Y_2}(\underline{d}_2,\underline{X}_1)-\mu_0^{Y_2}(\underline{d}_2,\underline{X}_1)]}{p_0^{d_1}(X_0)}  \Bigg]  \tag{$**$}\\
&- \  E \Bigg[ \frac{g_0(X_0)}{\Pr(S=1)}\cdot \frac{\overbrace{I\{D_1=d_1\} \cdot I\{D_2=d_2\} \cdot  [Y_2-\mu_0^{Y_2}(\underline{d}_2,\underline{X}_1)]}^{E[\cdot|X_0]=0}}{ p_0^{d_1}(X_0)\cdot p_0^{d_2}(d_1,\underline{X}_1) }\cdot\frac{[p^{d_1}(X_0)-p_0^{d_1}(X_0)] }{p_0^{d_1}(X_0)} \Bigg] \notag\\
& - \  E \Bigg[ \frac{g_0(X_0)}{\Pr(S=1)}\cdot  \overbrace{\frac{I\{D_1=d_1\}\cdot  [\mu_0^{Y_2}(\underline{d}_2,\underline{X}_1)-\nu_0^{Y_2}(\underline{d}_2,X_0)]}{p_0^{d_1}(X_0)}}^{E[\cdot|X_0]=0} \cdot \frac{[p^{d_1}(X_0)-p_0^{d_1}(X_0)] }{p_0^{d_1}(X_0)} \Bigg]  \notag \\
&- \  E \Bigg[ \frac{g_0(X_0)}{\Pr(S=1)}\cdot \frac{\overbrace{I\{D_1=d_1\} \cdot I\{D_2=d_2\} \cdot  [Y_2-\mu_0^{Y_2}(\underline{d}_2,\underline{X}_1)]}^{E[\cdot|X_0]=0}}{ p_0^{d_1}(X_0)\cdot p_0^{d_2}(d_1,\underline{X}_1) }\cdot\frac{[p^{d_2}(d_1,\underline{X}_1)-p_0^{d_2}(d_1,\underline{X}_1)] }{p_0^{d_2}(d_1,\underline{X}_1)} \Bigg] \notag\\
& - \  E \Bigg[  \frac{g_0(X_0)}{\Pr(S=1)}\cdot  \underbrace{\frac{I\{D_1=d_1\}  \cdot [\nu^{Y_2}(\underline{d}_2,X_0)-\nu_0^{Y_2}(\underline{d}_2,X_0)]}{p_0^{d_1}(X_0)}}_{ E[\cdot|X_0]=\frac{p_0^{d_1}(X_0)}{p_0^{d_1}(X_0)}\cdot[\nu^{Y_2}(\underline{d}_2,X_0)-\nu_0^{Y_2}(\underline{d}_2,X_0)]}   \Bigg] \tag{$***$}\\
& +  E\Big[  \overbrace{\frac{S}{\Pr(S=1)}}^{E[\cdot|X_0]=\frac{g_0(X_0)}{\Pr(S=1)}}\cdot [\nu^{Y_2}(\underline{d}_2,X_0)-\nu_0^{Y_2}(\underline{d}_2,X_0)]\big] \tag{$****$}\\
&+E \Bigg[ \frac{\overbrace{I\{D_1=d_1\} \cdot I\{D_2=d_2\} \cdot [Y_2-\mu_0^{Y_2}(\underline{d}_2,\underline{X}_1)]}^{E[\cdot| X_0]=0}}{p_0^{d_1}(X_0)\cdot p_0^{d_2}(d_1,\underline{X}_1)}\cdot\frac{[g(X_0)-g_0(X_0)]}{\Pr(S=1)} \Bigg]\notag\\
& + E \Bigg[\underbrace{\frac{I\{D_1=d_1\}\cdot  [\mu_0^{Y_2}(\underline{d}_2,\underline{X}_1)-\nu_0^{Y_2}(\underline{d}_2,X_0)]}{p_0^{d_1}(X_0)}}_{E[\cdot|X_0]=0} \cdot\frac{[g(X_0)-g_0(X_0)]}{\Pr(S=1)} \Bigg] =0,\notag
\end{align}
where terms $(*)$ and $(**)$ as well as $(***)$ and $(****)$ cancel out.

\textbf{Assumption 3.1(e)}

\textbf{Singular values of $E[\psi^{\underline{d}_2,S=1}_a(W;\chi_0)]$ are bounded:}
Holds trivially, because $\psi^{\underline{d}_2,S=1}_a(W;\chi_0) = -1.$
\vspace{5pt}\newline
\newpage

\textbf{Assumption 3.2:  Score regularity and quality of nuisance parameter estimators}
\vspace{5pt}

\textbf{Assumption 3.2(a)}

This assumption directly follows from the defition of $\mathcal{T}^*_n$ and the regularity conditions (Assumption 5).

\textbf{Assumption 3.2(b)} 

\textbf{Bounds for $m_n$ :}

We start by rearranging the terms in the Neyman score function (\ref{score2})
\begin{eqnarray}
E\Big[ \psi^{\underline{d}_2,S=1}(W, \chi, \Psi_{0}^{\underline{d}_2,S=1})\Big] &=& E\Bigg[ \underbrace{ \frac{g(X_0)}{\Pr(S=1)}\cdot \frac{ I\{D_1=d_1\} \cdot I\{D_2=d_2\}}{p^{d_1}(X_0)\cdot p^{d_2}(d_1,\underline{X}_1)} \cdot Y_2 }_{=I_1}  \notag\\
& + & \underbrace{\frac{g(X_0)}{\Pr(S=1)}\cdot  \frac{I\{D_1=d_1\}}{p^{d_1}(X_0)} \cdot \bigg(1- \frac{I\{D_2=d_2\}}{p^{d_2}(d_1,\underline{X}_1) } \bigg) \cdot \mu^{Y_2}(\underline{d}_2,\underline{X}_1)  }_{=I_2}  \notag\\
& + & \underbrace{  \bigg( \frac{S}{\Pr(S=1)} -   \frac{g(X_0)}{\Pr(S=1)}\cdot \frac{I\{D_1=d_1\}}{p^{d_1}(X_0)}\bigg)  \nu^{Y_2}(\underline{d}_2,X_0)}_{=I_3} - \Psi^{\underline{d}_2,S=1}_{0}  \Bigg]  \notag
\end{eqnarray}
and then, Following the same steps as in (\ref{Neyman}), we get
\begin{eqnarray}
\left\| \psi^{\underline{d}_2,S=1}(W, \chi, \Psi_{0}^{\underline{d}_2,S=1}) \right\|_{q} &\leq&  \left\| I_1 \right\|_{q}  + \left\| I_2 \right\|_{q}  + \left\| I_3 \right\|_{q} +  \left\| \Psi^{\underline{d}_2,S=1}_{0}  \right\|_{q} \notag \\
&\leq& \frac{1}{\epsilon^3}  \left\| Y_2 \right\|_{q} \notag + \frac{1- \epsilon}{\epsilon^3} \left\|  \mu^{Y_2}(\underline{d}_2,\underline{X}_1)  \right\|_{q} + \\
&+& \frac{1-\epsilon}{\epsilon^2} \left\|  \nu^{Y_2}(\underline{d}_2,X_0)  \right\|_{q}   +  | \Psi^{\underline{d}_2,S=1}_{0} | \notag \\
&\leq& C \left( \frac{1}{\epsilon^3} + \frac{2(1-\epsilon)}{\epsilon^{2/q}} \left(\frac{1}{\epsilon^3} + \frac{1}{\epsilon^2} \right) + \frac{1}{\epsilon^2} \right)  \notag
\end{eqnarray}
because of triangular inequality and because the following set of inequalities hold (similarly to (\ref{32b})):
\begin{eqnarray*}
\left\|  \mu^{Y_2}(\underline{d}_2,\underline{X}_1)  \right\|_{q} &\leq& 2C/\epsilon^{2/q}, \ \  \left\|  \nu^{Y_2}(\underline{d}_2,X_0)  \right\|_{q} \leq 2C/\epsilon^{2/q}, \notag \\
|\Psi^{\underline{d}_2,S=1}_{0} | &=& \left|E\left[ \frac{S}{\Pr(S=1)} \nu^{Y_2}_0(\underline{d}_2,X_0) \right] \right| \leq  E_{ } \Big[\left|  \nu^{Y_2}_0(\underline{d}_2,X_0)  \right|^1 \Big]^{\frac{1}{1}} / \epsilon=  \left\| \nu^{Y_2}_0(\underline{d}_2,X_0) \right\|_{1}/ \epsilon  \notag \\
&\leq&  \left\|  \nu^{Y_2}_0(\underline{d}_2,X_0) \right\|_{2} / \epsilon \leq  \left\| Y_2 \right\|_{2}/\epsilon^{4/2} \overbrace{ \leq}^{q > 2}  \left\| Y_2 \right\|_{q}/\epsilon^{2} \leq C /\epsilon^{2}.  \notag
\end{eqnarray*}
which gives the upper bound on $m_n$ in Assumption 3.2(b) of \cite{Chetal2018}.


\textbf{Bounds for $m'_n$:}

Notice that
$$\Big(E[ |\psi_a^{\underline{d}_2,S=1}(W, \chi) |^q] \Big)^{1/q}=1$$
and this gives the upper bound on $m'_n$ in Assumption 3.2(b) of \cite{Chetal2018}.

\textbf{Assumption 3.2(c)}

\textbf{Bound for $r_n$:}

For any $\chi = (g, p^{d_1}, p^{d_2},\mu^{Y_2}, \nu^{Y_2})$ we have
$$ \Big| E\Big( \psi_a^{\underline{d}_2,S=1}(W, \chi) - \psi_a^{\underline{d}_2,S=1}(W, \chi_0) \Big) \Big| = |1-1| = 0 \leq \delta'_N,$$
and thus we have the bound on $r_n$ from Assumption 3.2(c) of \cite{Chetal2018}.

\textbf{Bound for $r'_n$:}
\begin{eqnarray*}
&& \left\|   \psi^{\underline{d}_2,S=1}(W, \chi, \Psi_{0}^{\underline{d}_2,S=1}) - \psi^{\underline{d}_2,S=1}(W, \chi_0, \Psi_{0}^{\underline{d}_2,S=1})  \right\|_{2} \leq
\left\|     \frac{I\{D_1=d_1\} \cdot I\{D_2=d_2\}}{\Pr(S=1)}  \cdot Y \cdot \left( \frac{g}{p^{d_1} p^{d_2}} - \frac{g_0}{p_0^{d_1} p_0^{d_2}} \right) \right\|_{2}  \notag \\
&+&  \left\|    \frac{I\{D_1=d_1\} \cdot I\{D_2=d_2\}}{\Pr(S=1)} \left( \frac{g \cdot \mu^{Y_2}}{p^{d_1} p^{d_2}} - \frac{g_0 \cdot \mu^{Y_2}_0}{p_0^{d_1} p_0^{d_2}} \right) \right\|_{2}  + \left\|     \frac{I\{D_1=d_1\}}{\Pr(S=1)}  \left( \frac{g \cdot \mu^{Y_2}}{p^{d_1}} - \frac{g_0 \cdot \mu^{Y_2}_0}{p_0^{d_1}} \right) \right\|_{2} \\ \notag
&+& \left\|     \frac{I\{D_1=d_1\}}{\Pr(S=1)} \left(   \frac{g \cdot \nu^{Y_2}}{p^{d_1}} - \frac{ g_0 \cdot  \nu^{Y_2}_0}{p_0^{d_1}} \right) \right\|_{2} + \left\| \frac{S}{\Pr(S=1)} (\nu^{Y_2} - \nu^{Y_2}_0) \right\|_{2} \\ \notag
&\leq& \frac{1}{\epsilon} \left\| Y \cdot \left( \frac{g}{p^{d_1} p^{d_2}} - \frac{g_0}{p_0^{d_1} p_0^{d_2}} \right) \right\|_{2} +   \frac{1}{\epsilon}\left\| \frac{g \cdot \mu^{Y_2}}{p^{d_1} p^{d_2}} - \frac{g_0 \cdot \mu^{Y_2}_0}{p_0^{d_1} p_0^{d_2}} \right\|_{2} + \frac{1}{\epsilon} \left\|  \frac{g \cdot \mu^{Y_2}}{p^{d_1}} - \frac{g_0 \cdot \mu^{Y_2}_0}{p_0^{d_1}} \right\|_{2} \\
&+& \frac{1}{\epsilon} \left\|   \frac{g \cdot \nu^{Y_2}}{p^{d_1}} - \frac{g_0 \cdot \nu^{Y_2}_0}{p_0^{d_1}} \right\|_{2} + \frac{1}{\epsilon}\left\| \nu^{Y_2} - \nu^{Y_2}_0 \right\|_{2} \\
&\leq& \frac{C}{\epsilon^5} \delta_n \left(2 + \frac{1}{\epsilon} \right) + \frac{\delta_n}{\epsilon^5}  \left( \frac{1}{\epsilon} + 2C +  \frac{C}{\epsilon} \right) + \frac{\delta_n}{\epsilon^3} \left( \frac{1}{\epsilon}+2C \right)+  \frac{\delta_n}{\epsilon^3} \left( \frac{1}{\epsilon}+2C \right) +   \frac{\delta_n}{\epsilon^{2}} \leq \delta_n'
\end{eqnarray*}
as long as $C_\epsilon$ in the definition of $\delta_n'$ is sufficiently large.  This gives the bound on $r'_n$ from Assumption 3.2(c) of \cite{Chetal2018}.

The last inequality holds because we can bound the first term by
\begin{eqnarray*}
&&  \left\| Y \cdot \left( \frac{g}{p^{d_1} p^{d_2}} - \frac{g_0}{p_0^{d_1} p_0^{d_2}} \right) \right\|_{2}  \leq    C \left\|  \frac{g}{p^{d_1} p^{d_2}} - \frac{g_0}{p_0^{d_1} p_0^{d_2}}  \right\|_{2} \leq \frac{C}{\epsilon^4} \left\| p_0^{d_1} p_0^{d_2} g -  p^{d_1} p^{d_2} g_0  \right\|_{2} \\
&=& \frac{C}{\epsilon^4} \left\|  p_0^{d_1} p_0^{d_2} g -  p^{d_1} p^{d_2} g_0  + p_0^{d_1} p_0^{d_2} g - p_0^{d_1} p_0^{d_2} g_0  \right\|_{2} \leq \frac{C}{\epsilon^4} \left( \left\| g - g_0 \right\|_{2} + 1 \cdot \left\| p_0^{d_1} p_0^{d_2}  - p^{d_1} p^{d_2} \right\|_{2} \right)   \\
&\leq&  \frac{C}{\epsilon^4}\left( \delta_n + 1 \cdot \delta_n \left(1 + \frac{1}{\epsilon} \right) \right) \leq \frac{C}{\epsilon^4} \delta_n \left(2 + \frac{1}{\epsilon} \right),
\end{eqnarray*}

the second term is bounded by
\begin{eqnarray*}
&& \left\|  \frac{g \cdot \mu^{Y_2}}{p^{d_1} p^{d_2}} - \frac{g_0 \cdot \mu^{Y_2}_0}{p_0^{d_1} p_0^{d_2}}  \right\|_{2} \leq \frac{1}{\epsilon^4} \left\|  p_0^{d_1} p_0^{d_2} g \cdot \mu^{Y_2} -   p^{d_1} p^{d_2} g_0 \cdot \mu^{Y_2}_0 \right\|_{2} \\
&=&  \frac{1}{\epsilon^4}\left\|   p_0^{d_1} p_0^{d_2} g \cdot \mu^{Y_2} -   p^{d_1} p^{d_2} g_0 \cdot \mu^{Y_2}_0 + p_0^{d_1} p_0^{d_2} g_0 \cdot \mu^{Y_2}_0 - p_0^{d_1} p_0^{d_2} g_0 \cdot \mu^{Y_2}_0 \right\|_{2} \\
&\leq&  \frac{1}{\epsilon^4} \left( \left\| p_0^{d_1} p_0^{d_2}  (g \cdot \mu^{Y_2} - g_0 \cdot \mu^{Y_2}_0) \right\|_{2} + \left\| g_0 \cdot \mu^{Y_2}_0 ( p_0^{d_1} p_0^{d_2}  - p^{d_1} p^{d_2}  ) \right\|_{2} \right) \\
&\leq&   \frac{1}{\epsilon^4} \left( \left\| g \cdot \mu^{Y_2} - g_0 \cdot \mu^{Y_2}_0 \right\|_{2} + 1 \cdot C  \left\| p_0^{d_1} p_0^{d_2}  - p^{d_1} p^{d_2} \right\|_{2} \right) \\
&\leq&\frac{1}{\epsilon^4} \left( \delta_n \left( C + \frac{1}{\epsilon}\right)   +  C \delta_n \left(1 + \frac{1}{\epsilon} \right) \right) = \frac{\delta_n}{\epsilon^4}  \left( \frac{1}{\epsilon} + 2C +  \frac{C}{\epsilon} \right)
\end{eqnarray*}
where $ \left\| g \cdot \mu^{Y_2} - g_0 \cdot \mu^{Y_2}_0 \right\|_{2} $ is bounded similarly as  $\left\| p_0^{d_1} \mu^{Y_2}-  p^{d_1} \mu^{Y_2}_0  \right\|_{2}$ and we also used bounds for $ \left\| p_0^{d_1} p_0^{d_2}  - p^{d_1} p^{d_2} \right\|_{2}$ derived in section (\ref{Neyman}).

while for the third term we get
\begin{eqnarray*}
&& \left\|  \frac{g \cdot \mu^{Y_2}}{p^{d_1}} - \frac{g_0 \cdot \mu^{Y_2}_0}{p_0^{d_1}}  \right\|_{2} = \frac{1}{\epsilon^2}  \left\| p_0^{d_1}g \cdot \mu^{Y_2}-  p^{d_1}g_0 \cdot \mu^{Y_2}_0  \right\|_{2} \\
&=&  \frac{1}{\epsilon^2} \left\| p_0^{d_1}g \cdot \mu^{Y_2}-  p^{d_1}g_0 \cdot \mu^{Y_2}_0 + p_0^{d_1}g_0 \cdot  \mu^{Y_2}_0 - p_0^{d_1}g_0 \cdot  \mu^{Y_2}_0   \right\|_{2} \\
&\leq&\frac{1}{\epsilon^2} \left( \left\|  p_0^{d_1}(g \cdot \mu^{Y_2} - g_0 \cdot \mu^{Y_2}_0) \right\|_{2} +  \left\| g_0 \cdot \mu^{Y_2}_0 (p_0^{d_1} - p^{d_1}) \right\|_{2} \right) \\
&\leq& \frac{1}{\epsilon^2} \left( \delta_n \left(C+\frac{1}{\epsilon} \right) +  C \left\| p_0^{d_1} - p^{d_1} \right\|_{2} \right) \leq  \frac{1}{\epsilon^2} \left(\delta_n \left(C+\frac{1}{\epsilon} \right)  +  C \delta_n \right) = \frac{\delta_n}{\epsilon^2} \left( \frac{1}{\epsilon}+2C \right).
\end{eqnarray*}
and similarly, for the fourth term we obtain
\begin{eqnarray*}
&& \left\|  \frac{\nu^{Y_2}}{p^{d_1}} - \frac{\nu^{Y_2}_0}{p_0^{d_1}}  \right\|_{2}  \leq \frac{\delta_n}{\epsilon^2} \left( \frac{1}{\epsilon}+2C \right).
\end{eqnarray*}

\textbf{Bound for $\lambda'_n$:}

Now consider
\begin{equation}
f(r) := E[\psi^{\underline{d}_2,S=1}(W;\Psi_0^{\underline{d}_2,S=1},\chi_0 + r(\chi-\chi_0)] \notag
\end{equation}
 For any $r \in (0,1):$
\begin{eqnarray} \label{secondGatDer_b}
\frac{\partial^2 f(r)}{\partial r^2} &=& E\Bigg[ \frac{I \{D_1 = d_1 \} }{\Pr(S=1)} \cdot \frac{ 2 (g-g_0) \Big( (\mu^{Y_2} - \mu^{Y_2}_0) - (\nu^{Y_2} - \nu^{Y_2}_0) \Big) }{p^{d_1}_0 + r(p^{d_1} -p^{d_1}_0)} \Bigg] \\  \notag  
&+& E\Bigg[ \frac{I \{D_1 = d_1 \} }{\Pr(S=1)} \cdot \frac{ (-2) \Big(g_0 + r(g-g_0) \Big) \Big( (\mu^{Y_2} - \mu^{Y_2}_0) - (\nu^{Y_2} - \nu^{Y_2}_0)\Big) (p^{d_1} -p^{d_1}_0)  }{\left(p^{d_1}_0 + r(p^{d_1} -p^{d_1}_0) \right)^2} \Bigg] \\  \notag 
&+& E\Bigg[ \frac{I \{D_1 = d_1 \} }{\Pr(S=1)} \cdot \frac{ 2 (g-g_0) \Big( (\mu^{Y_2}_0 + r(\mu^{Y_2} - \mu^{Y_2}_0)) - (\nu^{Y_2}_0 +  r(\nu^{Y_2} - \nu^{Y_2}_0)) \Big) }{\left( p^{d_1}_0 + r(p^{d_1} -p^{d_1}_0)\right)^2 } \Bigg] \\  \notag 
&+& E\Bigg[ \frac{I \{D_1 = d_1 \} }{\Pr(S=1)} \cdot \frac{ 2 \Big(g_0 + r(g-g_0) \Big) \Big( (\mu^{Y_2}_0 + r(\mu^{Y_2} - \mu^{Y_2}_0)) - (\nu^{Y_2}_0 +  r(\nu^{Y_2} - \nu^{Y_2}_0)) \Big) (p^{d_1} -p^{d_1}_0)^2 }{\left( p^{d_1}_0 + r(p^{d_1} -p^{d_1}_0)\right)^3 } \Bigg] \\  \notag 
&+& E\Bigg[ \frac{I \{D_1 = d_1 \} I \{D_2 = d_2 \} }{\Pr(S=1)} \cdot \frac{ (-2) (g-g_0) \Big(Y - (\mu^{Y_2}_0 + r(\mu^{Y_2}-\mu^{Y_2}_0)) \Big) (p^{d_1} -p^{d_1}_0)}{\left( p^{d_1}_0 + r(p^{d_1} -p^{d_1}_0)\right)^2 \left( p^{d_2}_0 + r(p^{d_2} -p^{d_2}_0)\right) } \Bigg] \\  \notag 
&+& E\Bigg[ \frac{I \{D_1 = d_1 \} I \{D_2 = d_2 \} }{\Pr(S=1)} \cdot \frac{ (-2) (g-g_0) \Big(Y - (\mu^{Y_2}_0 + r(\mu^{Y_2}-\mu^{Y_2}_0)) \Big) (p^{d_2} -p^{d_2}_0)}{\left( p^{d_1}_0 + r(p^{d_1} -p^{d_1}_0)\right) \left( p^{d_2}_0 + r(p^{d_2} -p^{d_2}_0)\right)^2 } \Bigg] \\  \notag 
&+& E\Bigg[ \frac{I \{D_1 = d_1 \} I \{D_2 = d_2 \} }{\Pr(S=1)} \cdot \frac{ (-2) (g-g_0) (\mu^{Y_2}-\mu^{Y_2}_0) }{\left( p^{d_1}_0 + r(p^{d_1} -p^{d_1}_0)\right) \left( p^{d_2}_0 + r(p^{d_2} -p^{d_2}_0)\right) } \Bigg] \\  \notag 
&+& E\Bigg[ \frac{I \{D_1 = d_1 \} I \{D_2 = d_2 \} }{\Pr(S=1)} \cdot \frac{ 2 \Big(g_0 + r(g-g_0) \Big) \Big(Y - (\mu^{Y_2}_0 + r(\mu^{Y_2}-\mu^{Y_2}_0)) \Big) (p^{d_1} -p^{d_1}_0)^2}{\left( p^{d_1}_0 + r(p^{d_1} -p^{d_1}_0)\right)^3 \left( p^{d_2}_0 + r(p^{d_2} -p^{d_2}_0)\right) } \Bigg] \\  \notag 
&+& E\Bigg[ \frac{I \{D_1 = d_1 \} I \{D_2 = d_2 \} }{\Pr(S=1)} \cdot \frac{ 2 \Big(g_0 + r(g-g_0) \Big) \Big(Y - (\mu^{Y_2}_0 + r(\mu^{Y_2}-\mu^{Y_2}_0)) \Big) (p^{d_2} -p^{d_2}_0)^2}{\left( p^{d_1}_0 + r(p^{d_1} -p^{d_1}_0)\right) \left( p^{d_2}_0 + r(p^{d_2} -p^{d_2}_0)\right)^3 } \Bigg] \\  \notag 
&+& E\Bigg[ \frac{I \{D_1 = d_1 \} I \{D_2 = d_2 \} }{\Pr(S=1)} \cdot \frac{ 2 \Big(g_0 + r(g-g_0) \Big) \Big(Y - (\mu^{Y_2}_0 + r(\mu^{Y_2}-\mu^{Y_2}_0)) \Big) (p^{d_1} -p^{d_1}_0)(p^{d_2} -p^{d_2}_0)}{\left( p^{d_1}_0 + r(p^{d_1} -p^{d_1}_0)\right)^2 \left( p^{d_2}_0 + r(p^{d_2} -p^{d_2}_0)\right)^2 } \Bigg] \\  \notag 
&+& E\Bigg[ \frac{I \{D_1 = d_1 \} I \{D_2 = d_2 \} }{\Pr(S=1)} \cdot \frac{ 2 \Big(g_0 + r(g-g_0) \Big) (\mu^{Y_2} - \mu^{Y_2}_0) (p^{d_1} -p^{d_1}_0)}{\left( p^{d_1}_0 + r(p^{d_1} -p^{d_1}_0)\right)^2 \left( p^{d_2}_0 + r(p^{d_2} -p^{d_2}_0)\right) } \Bigg] \\  \notag 
&+& E\Bigg[ \frac{I \{D_1 = d_1 \} I \{D_2 = d_2 \} }{\Pr(S=1)} \cdot \frac{ 2 \Big(g_0 + r(g-g_0) \Big) (\mu^{Y_2} - \mu^{Y_2}_0) (p^{d_2} -p^{d_2}_0)}{\left( p^{d_1}_0 + r(p^{d_1} -p^{d_1}_0)\right) \left( p^{d_2}_0 + r(p^{d_2} -p^{d_2}_0)\right)^2 } \Bigg] \\  \notag 
\end{eqnarray}

here we follow the same procedure as in bounding (\ref{secondGatDer}), with the only exception that now we have to make use of the last two inequalities from the regularity conditions  (\ref{Tnstar}).

As an example, consider the first term:

\begin{eqnarray*}
 \left| E\Bigg[ \frac{I \{D_1 = d_1 \} }{\Pr(S=1)} \cdot \frac{ 2 (g-g_0) \Big( (\mu^{Y_2} - \mu^{Y_2}_0) - (\nu^{Y_2} - \nu^{Y_2}_0) \Big) }{p^{d_1}_0 + r(p^{d_1} -p^{d_1}_0)} \Bigg] \right| &\leq&  \left|E\Bigg[ \frac{I \{D_1 = d_1 \} }{\Pr(S=1)} \cdot \frac{ 2 (g-g_0) (\mu^{Y_2} - \mu^{Y_2}_0) }{p^{d_1}_0 + r(p^{d_1} -p^{d_1}_0)} \Bigg]  \right| \\
&+&  \left|E\Bigg[ \frac{I \{D_1 = d_1 \} }{\Pr(S=1)} \cdot \frac{ 2 (g-g_0)  (\nu^{Y_2} - \nu^{Y_2}_0) }{p^{d_1}_0 + r(p^{d_1} -p^{d_1}_0)}\Bigg]  \right| \notag \\
&\leq& \frac{2}{\epsilon^2} \frac{\delta^{}_N}{\epsilon} n^{-1/2} +  \frac{2}{\epsilon^2} \frac{\delta^{}_N}{\epsilon} n^{-1/2}  =  4 \frac{\delta^{}_N}{\epsilon^{3}} n^{-1/2}.  \notag
\end{eqnarray*}
We may bound all the remaining terms similarly and get that for some constant $C_{\epsilon}''$ that only depends on $C$ and $\epsilon$
\begin{equation}
\left|\frac{\partial^2 f(r)}{\partial r^2} \right| \leq C_{\epsilon}'' \delta_n n^{-1/2} \leq \delta_n' n^{-1/2} \notag
\end{equation}
and this gives the upper bound on $\lambda'_n$ in Assumption 3.2(c) of \cite{Chetal2018} as long as $C_{\epsilon} \geq C_{\epsilon}''$.

\textbf{Assumption 3.2(d)}

\begin{eqnarray}
E\Big[ (\psi^{\underline{d}_2,S=1}(W, \chi_0, \Psi_{0}^{\underline{d}_2,S=1}) )^2\Big] &=& E\Bigg[ \Bigg( \underbrace{  \frac{g(X_0)}{\Pr(S=1)} \cdot \frac{ I\{D_1=d_1\} \cdot I\{D_2=d_2\} \cdot [Y_2-\mu_0^{Y_2}(\underline{d}_2,\underline{X}_1)]}{p_0^{d_1}(X_0)\cdot p_0^{d_2}(d_1,\underline{X}_1)} }_{=I_1}  \notag\\
& + & \underbrace{  \frac{g(X_0)}{\Pr(S=1)} \cdot\frac{I\{D_1=d_1\}\cdot  [\mu_0^{Y_2}(\underline{d}_2,\underline{X}_1)-\nu_0^{Y_2}(\underline{d}_2,X_0)]}{p_0^{d_1}(X_0)} }_{=I_2}  \notag\\
& + & \underbrace{  \frac{S}{\Pr(S=1)} \cdot \nu_0^{Y_2}(\underline{d}_2,X_0) - \Psi^{\underline{d}_2}_{0}}_{=I_3} \Bigg)^2 \Bigg]  \notag\\
& = & E[I_1^2 + I_2^2 + I_3^2] \geq E[I^2_1]\notag\\
& = & E\Bigg[  \frac{g(X_0)}{\Pr(S=1)} \cdot  I\{D_1=d_1\} \cdot I\{D_2=d_2\} \cdot \Bigg( \frac{ [Y_2-\mu_0^{Y_2}(\underline{d}_2,\underline{X}_1)]}{p_0^{d_1}(X_0)\cdot p_0^{d_2}(d_1,\underline{X}_1)} \Bigg)^2 \Bigg] \notag\\
& \geq & \frac{ \epsilon^3}{1-\epsilon} E\Bigg[ \Bigg( \frac{ [Y_2-\mu_0^{Y_2}(\underline{d}_2,\underline{X}_1)]}{p_0^{d_1}(X_0)\cdot p_0^{d_2}(d_1,\underline{X}_1)} \Bigg)^2 \Bigg] \notag\\
&\geq& \frac{\epsilon^3 c^2}{(1-\epsilon)^5} > 0. \notag
\end{eqnarray}
because $\Pr(\underline{D}_2 = \underline{d}_2|\underline{X}_1)= p_0^{d_1} (X_0) \cdot p_0^{d_2} (d_1,\underline{X}_1) \geq \epsilon^2$, $p_0^{d_2}(d_1,\underline{X}_1) \leq 1-\epsilon$ and $g(X_0) \geq  \epsilon.$



\noindent where the second equality follows from
\begin{eqnarray}
E\Big[  I_1 \cdot I_2\Big] &=& E\Bigg[ \left( \frac{g(X_0)}{\Pr(S=1)} \right)^2 \cdot \overbrace{\frac{ I\{D_1=d_1\} \cdot I\{D_2=d_2\}}{ (p_0^{d_1}(X_0))^2\cdot p_0^{d_2}(d_1,\underline{X}_1)}  \cdot [Y_2-\mu_0^{Y_2}(\underline{d}_2,\underline{X}_1)]}^{E[\cdot|X_0]=E[E[Y_2-\mu_0^{Y_2}(\underline{d}_2,\underline{X}_1)|\underline{D}_2=\underline{d}_2,\underline{X}_1]|D_1=d_1,X_0]=0}   \cdot  [\mu_0^{Y_2}(\underline{d}_2,\underline{X}_1)-\nu_0^{Y_2}(\underline{d}_2,X_0)] \Bigg], \notag\\
E\Big[ I_2 \cdot I_3\Big] &=& E\Bigg[  \frac{g(X_0) \cdot S}{\left( \Pr(S=1) \right)^2}  \cdot \overbrace{  \frac{ I\{D_1=d_1\}}{p_0^{d_1}(X_0)} \cdot  [\mu_0^{Y_2}(\underline{d}_2,\underline{X}_1)-\nu_0^{Y_2}(\underline{d}_2,X_0)]}^{E[\cdot|X_0]=\int E\big[  \mu_0^{Y_2}(\underline{d}_2,\underline{x}_1)-\nu_0^{Y_2}(\underline{d}_2,x_0) \big| D_1=d_1,  \underline{X}_1=\underline{x}_1 \big] dF_{X_1=x_1|D_1=d_1,X_0=x_0}=0}  \cdot [ \nu_0^{Y_2}(\underline{d}_2,X_0) - \Psi^{\underline{d}_2}_{0}] \Bigg],\notag\\
E\Big[ I_1 \cdot I_3\Big] &=& E\Bigg[ \frac{g(X_0) \cdot S}{\left( \Pr(S=1) \right)^2}  \cdot \overbrace{\frac{ I\{D_1=d_1\} \cdot I\{D_2=d_2\}}{ p_0^{d_1}(X_0)\cdot p_0^{d_2}(d_1,\underline{X}_1)}  \cdot [Y_2-\mu_0^{Y_2}(\underline{d}_2,\underline{X}_1)]}^{E[\cdot|X_0]=E[E[Y_2-\mu_0^{Y_2}(\underline{d}_2,\underline{X}_1)|\underline{D}_2=\underline{d}_2,\underline{X}_1]|D_1=d_1,X_0]=0}   \cdot  [\nu_0^{Y_2}(\underline{d}_2,X_0) - \Psi^{\underline{d}_2}_{0}] \Bigg]. \notag
\end{eqnarray}

} 

\newpage
\section{Covariates}\label{xvar}

Table~\ref{tab:xvar} provides information on the covariates in our empirical application across treatment sequences. The first two columns contain the name and a description of each variable. The variable type in the third column can take the value 1, 2, or 3, which stands for dummy, categorical, or continuous variable, respectively. The fourth column presents an indicator which is 0 for covariates observed prior to the first treatment ($X_0$) and 1 for covariates observed after the first treatment ($X_1$). The number of missing values is given in the fifth column, and columns 6-15 display the mean values of the covariates for the treatment sequences in Table~\ref{tab:treat}.

\begin{landscape}
	\tiny
\begin{longtable}{ll|rrr|rrrrrrrrrrr}
		\caption[Description of regressors and means across treatment sequences]{Description of regressors and means across treatment sequences}
		\label{tab:xvar} \\
		
		\hline \multicolumn{2}{c}{{Variable description}} & \multicolumn{3}{c}{{Statistics}} & \multicolumn{11}{c}{{Mean values across treatment sequences}} \\
		\hline name & description & type & x & miss & -1 & 00 & 11 & 12 & 13 & 21 & 22 & 23 & 31 & 32 & 33 \\ \hline
		\endfirsthead
		
		\multicolumn{16}{c}%
		{{\bfseries \tablename\ \thetable{} -- continued from previous page}} \\
		\hline name & description & type & x & miss & -1 & 00 & 11 & 12 & 13 & 21 & 22 & 23 & 31 & 32 & 33 \\ \hline
		\endhead
		
		\hline \multicolumn{16}{c}{{Continued on next page}} \\ \hline
		\endfoot
		jcmsa &                                      MSA CATEGORY &     2 &  0 &    219 &     1.88 &     1.91 &     1.98 &     2.09 &     1.95 &     1.86 &     1.86 &     1.77 &     1.95 &     1.86 &     1.90 \\
		age &                          AGE IN YEARS AT BASELINE &     3 &  0 &    219 &    19.08 &    18.78 &    19.04 &    18.44 &    18.98 &    18.46 &    18.54 &    18.79 &    19.01 &    18.55 &    19.32 \\
		RACE\_ETH &                                 RACE OR ETHNICITY &     2 &  0 &    222 &     2.04 &     2.05 &     2.00 &     2.05 &     2.05 &     2.10 &     2.22 &     2.30 &     1.96 &     2.15 &     2.02 \\
		NTV\_LANG &                                   NATIVE LANGUAGE &     2 &  0 &    236 &     1.18 &     1.18 &     1.19 &     1.09 &     1.20 &     1.17 &     1.33 &     1.37 &     1.14 &     1.32 &     1.20 \\
		HH14 &           HEAD OF HOUSE WHEN SAMPLE MEMBER WAS 14 &     2 &  0 &    306 &     2.46 &     2.43 &     2.37 &     2.37 &     2.39 &     2.56 &     2.41 &     2.45 &     2.36 &     2.37 &     2.31 \\
		WELF\_KID &                 FAMILY ON WELFARE WHEN GROWING UP &     2 &  0 &    905 &     2.10 &     2.08 &     2.08 &     2.40 &     2.08 &     2.15 &     2.08 &     2.02 &     1.98 &     2.41 &     1.93 \\
		HGC\_MOTH &                    HIGHEST GRADE MOTHER COMPLETED &     3 &  0 &   2243 &    11.47 &    11.55 &    11.29 &    10.92 &    11.39 &    11.60 &    10.53 &    10.88 &    11.75 &    10.89 &    11.53 \\
		M\_WORK14 &           MOTHER WORKED WHEN SAMPLE MEMBER WAS 14 &     1 &  0 &   1805 &     0.72 &     0.74 &     0.70 &     0.61 &     0.81 &     0.77 &     0.70 &     0.68 &     0.74 &     0.73 &     0.70 \\
\end{longtable}
\end{landscape}

\newpage
\section{Propensity score plots}\label{overlap}

The following figures display the overlap of the sequential treatment propensity scores across treatment states in the empirical application by means of kernel density plots. Each figure is divided into four windows. The upper windows show the first and second period propensity scores $\hat{p}^{d_1}(X_0)$ and $\hat{p}^{d_2}(d_1,\underline{X}_1)$ under the treatment sequence. The bottom windows display the overlap for the corresponding first and second period propensity scores under the control sequence.

\begin{figure}[htbp]
\caption{Support for treatment sequences 33 vs.\ 22 with a trimming threshold of 0.01}
\label{fig:support1}
\includegraphics[width=0.5\textwidth]{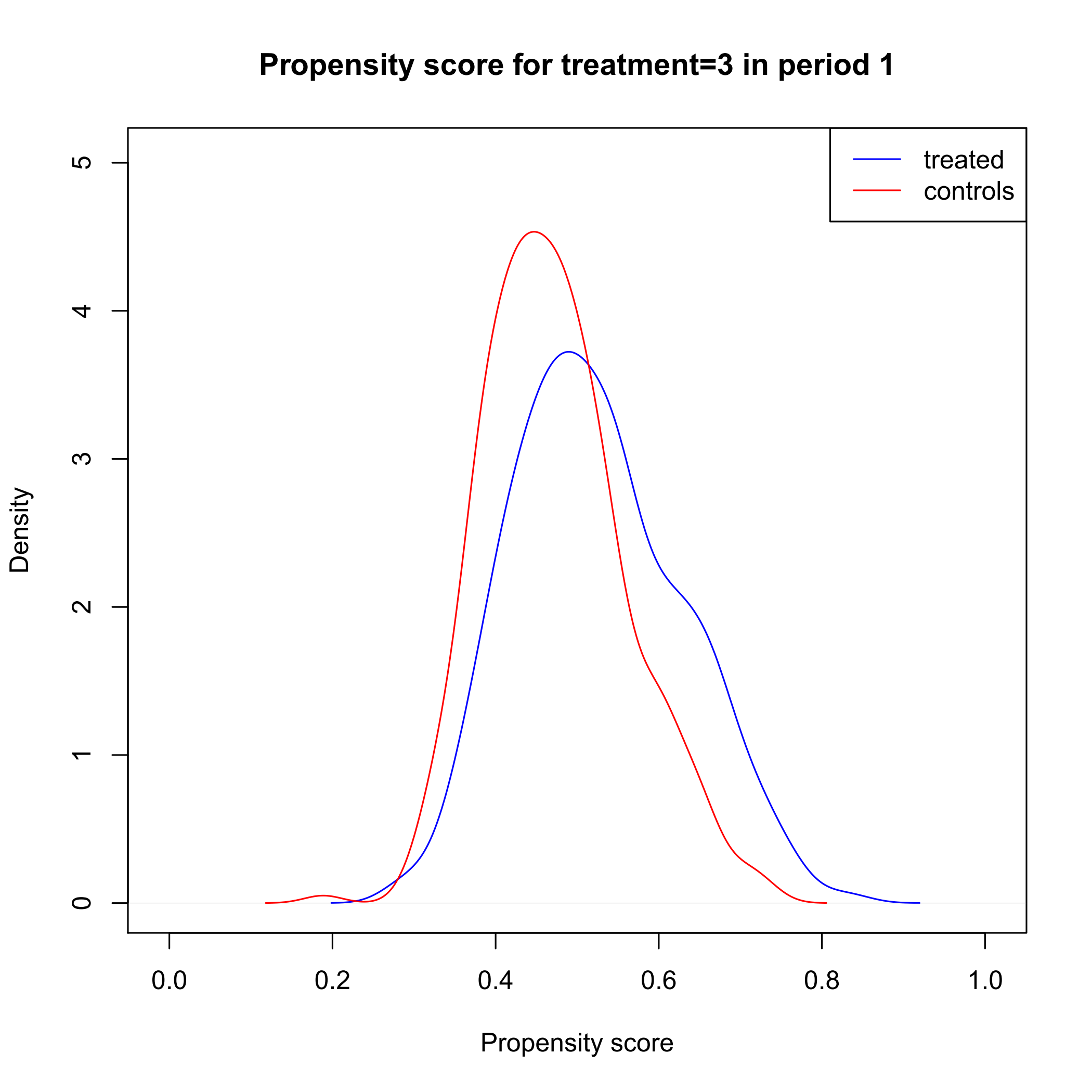}
\includegraphics[width=0.5\textwidth]{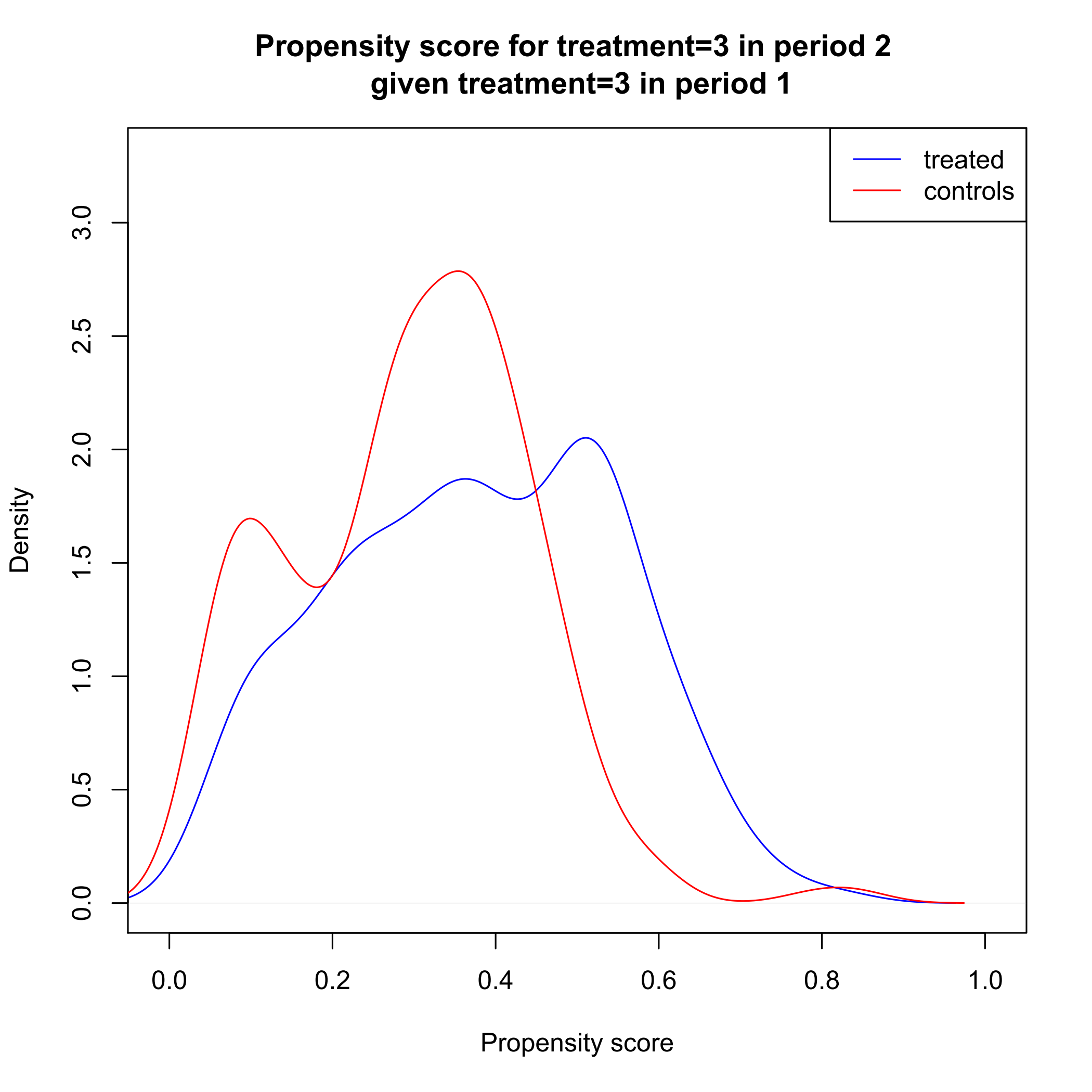}
\includegraphics[width=0.5\textwidth]{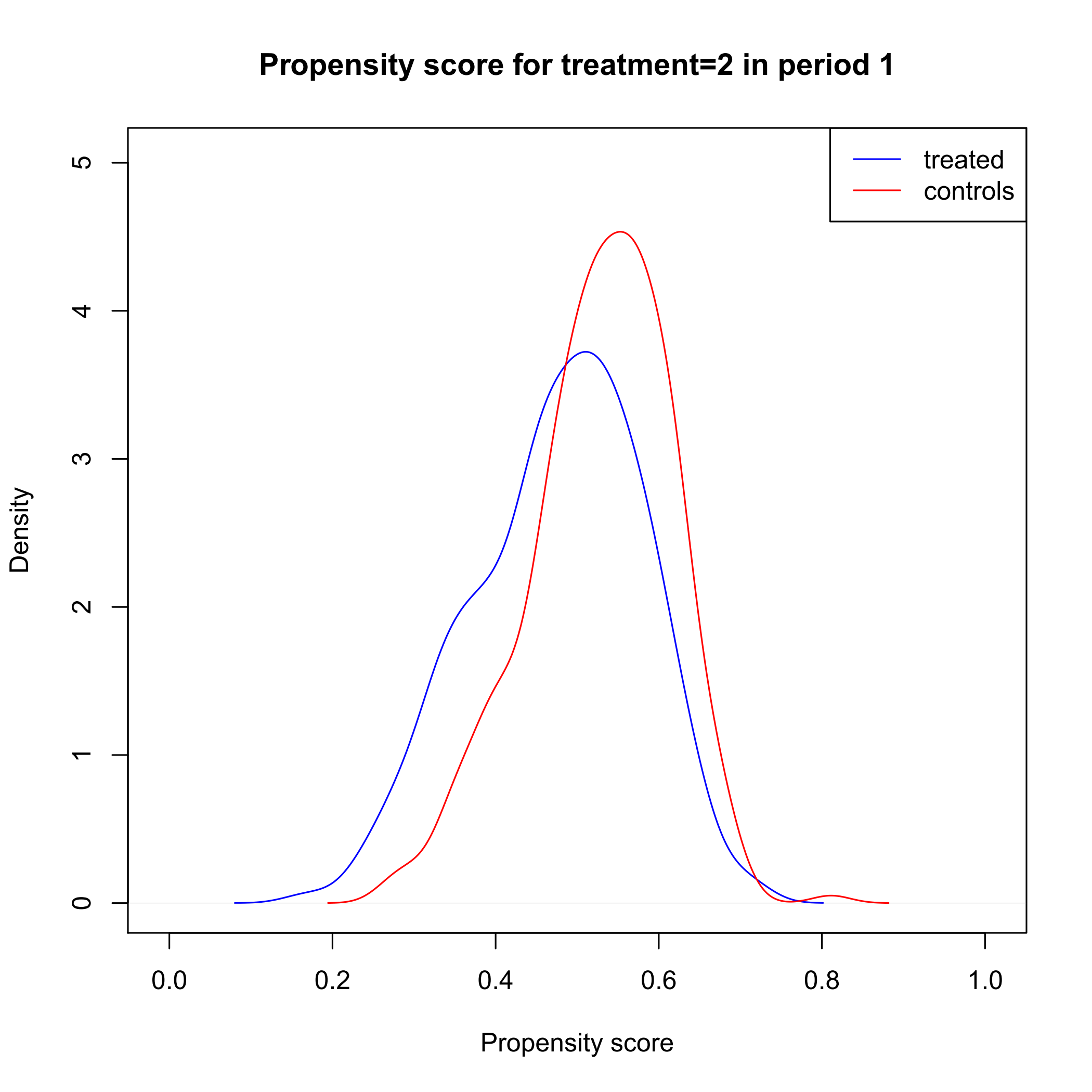}
\includegraphics[width=0.5\textwidth]{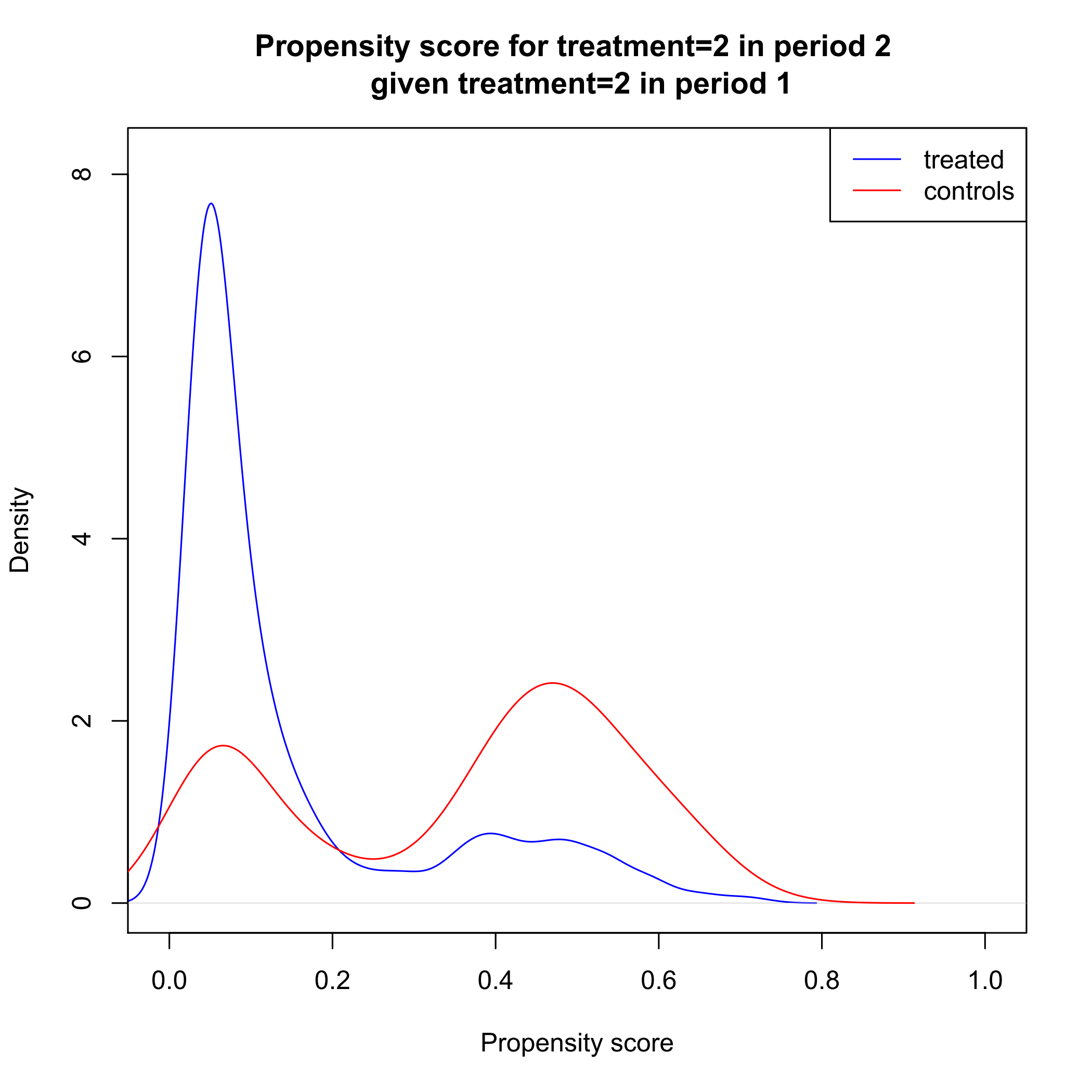}
\end{figure}

\newpage
\begin{figure}[htbp]
\caption{Support for treatment sequences 33 vs.\ 21 with a trimming threshold of 0.01}
\label{fig:support1}
\includegraphics[width=0.5\textwidth]{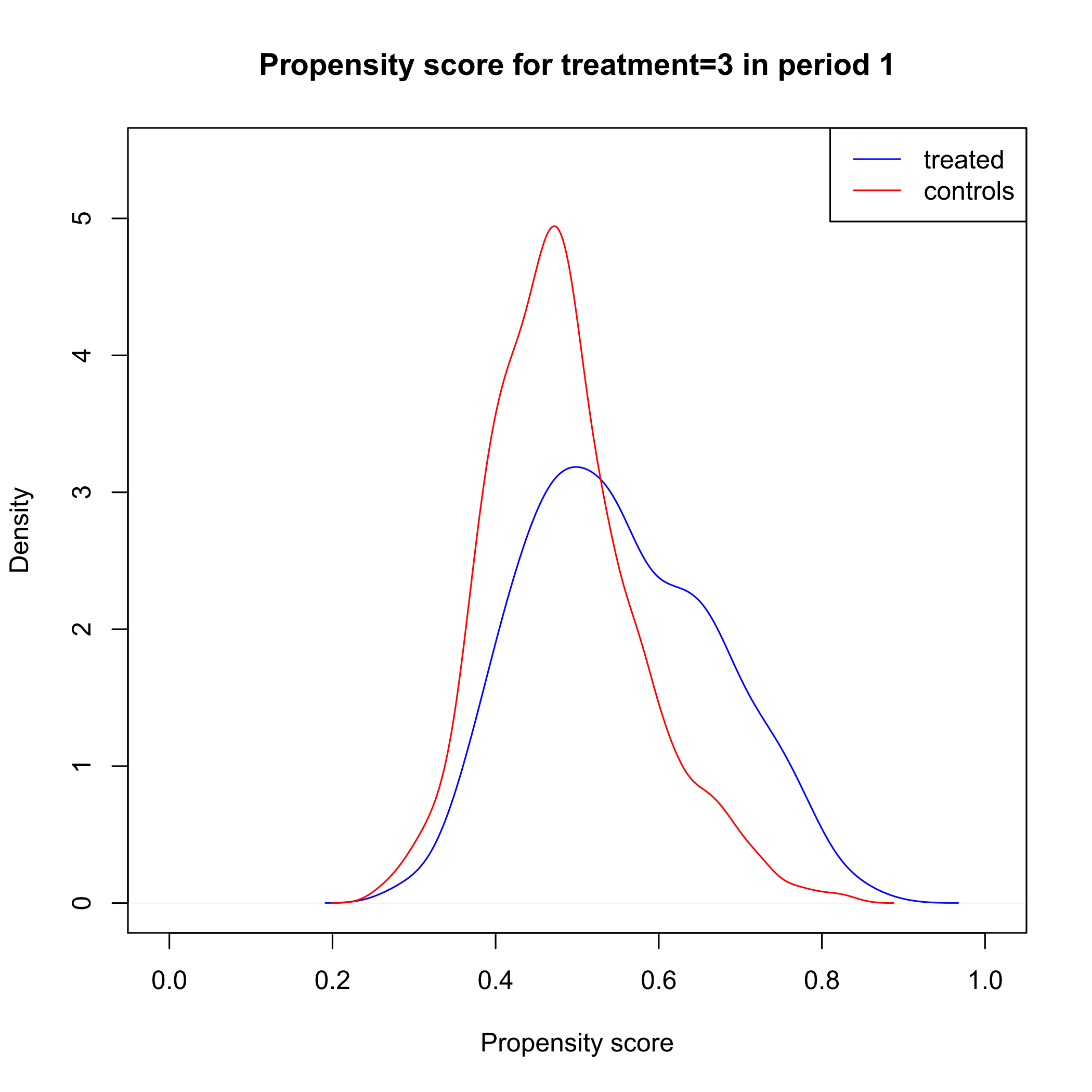}
\includegraphics[width=0.5\textwidth]{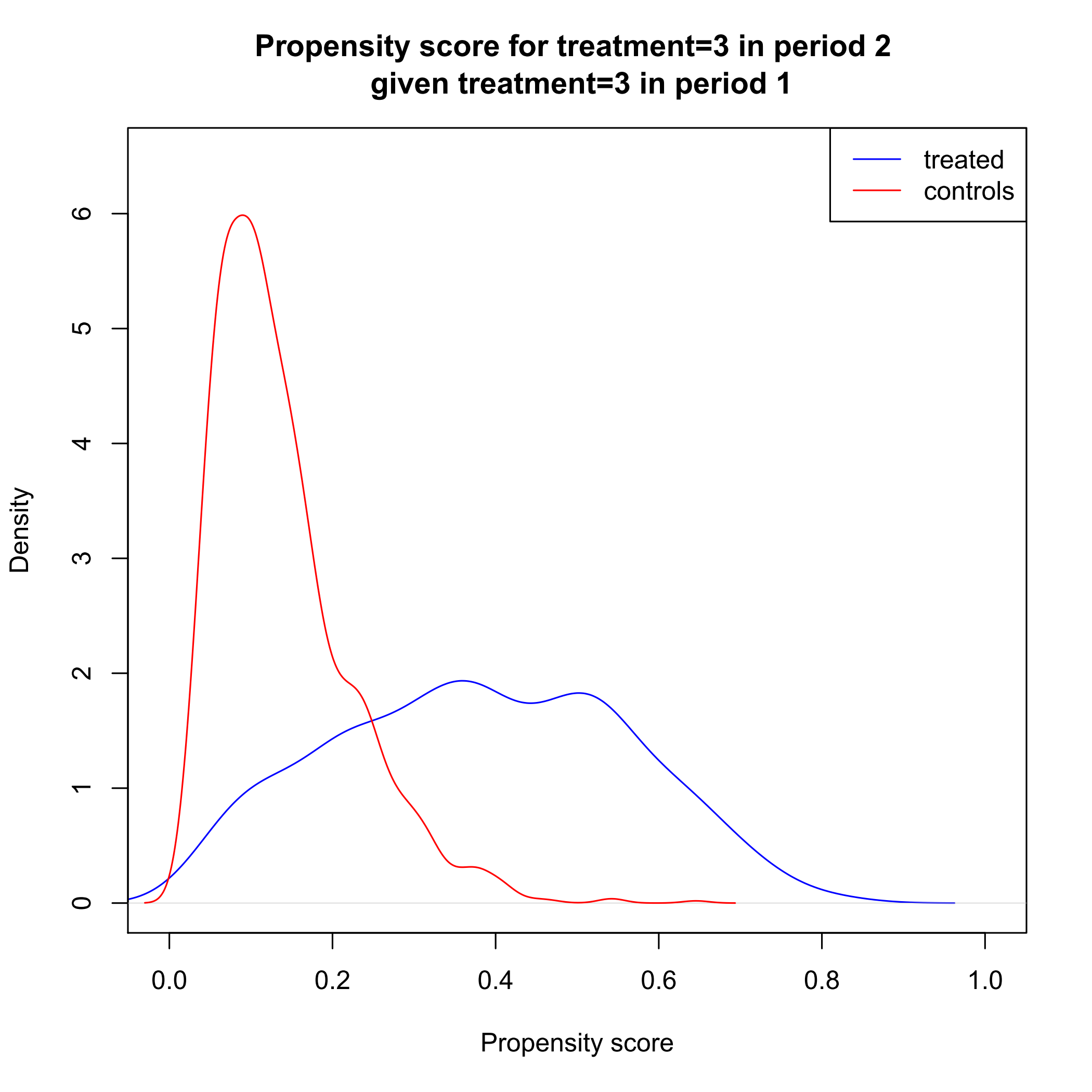}
\includegraphics[width=0.5\textwidth]{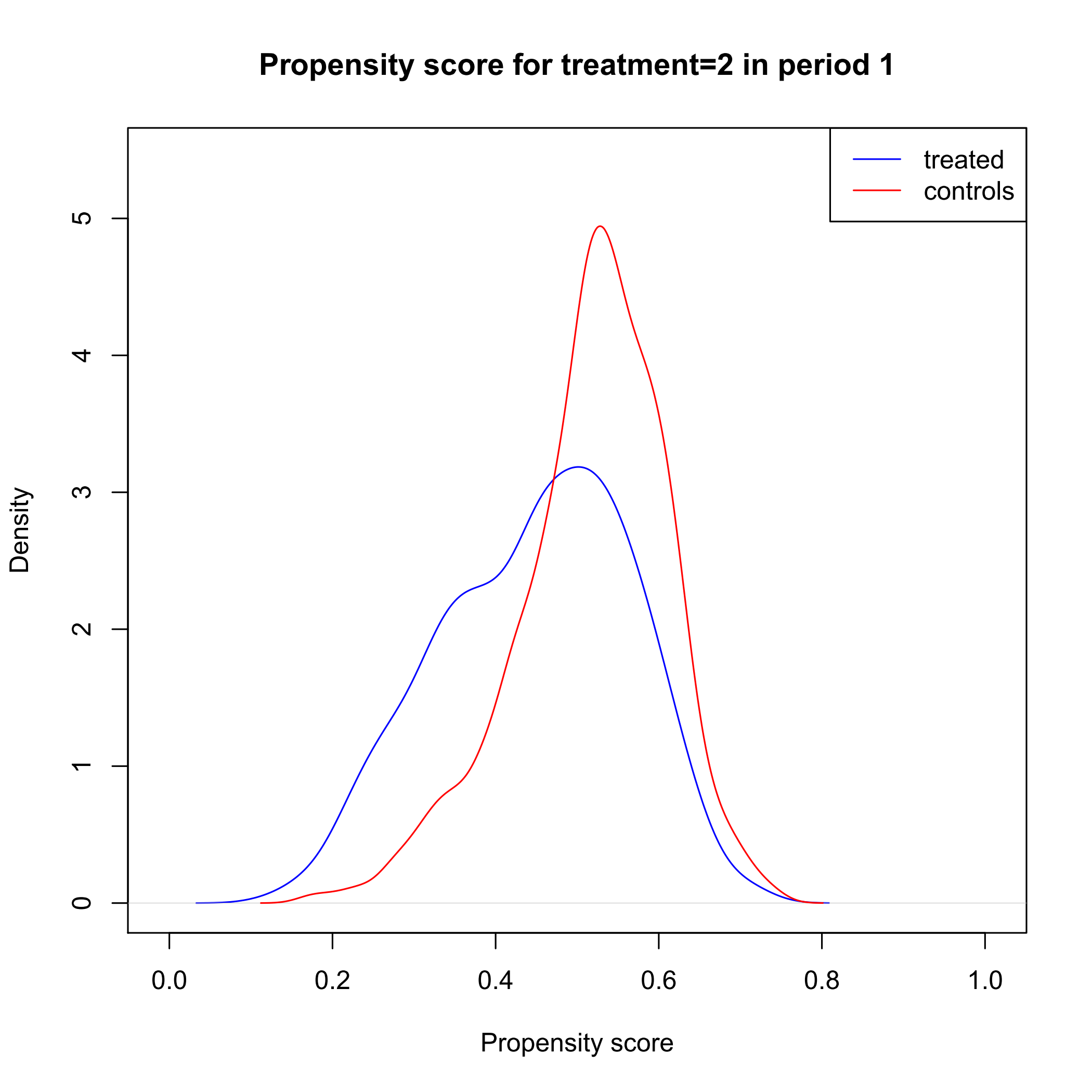}
\includegraphics[width=0.5\textwidth]{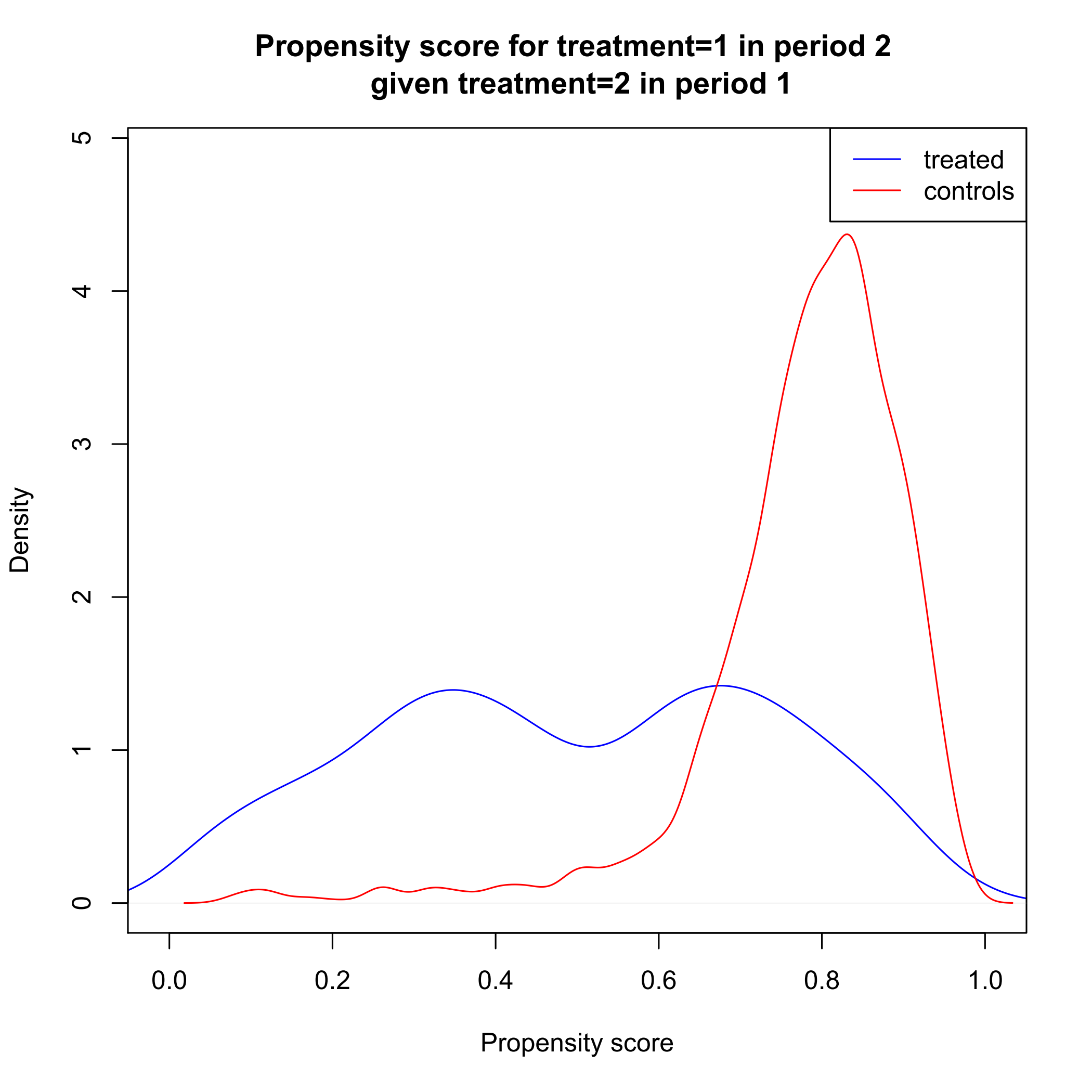}
\end{figure}

\newpage
\begin{figure}[htbp]
\caption{Support for treatment sequences 33 vs.\ 11 with a trimming threshold of 0.01}
\label{fig:support1}
\includegraphics[width=0.5\textwidth]{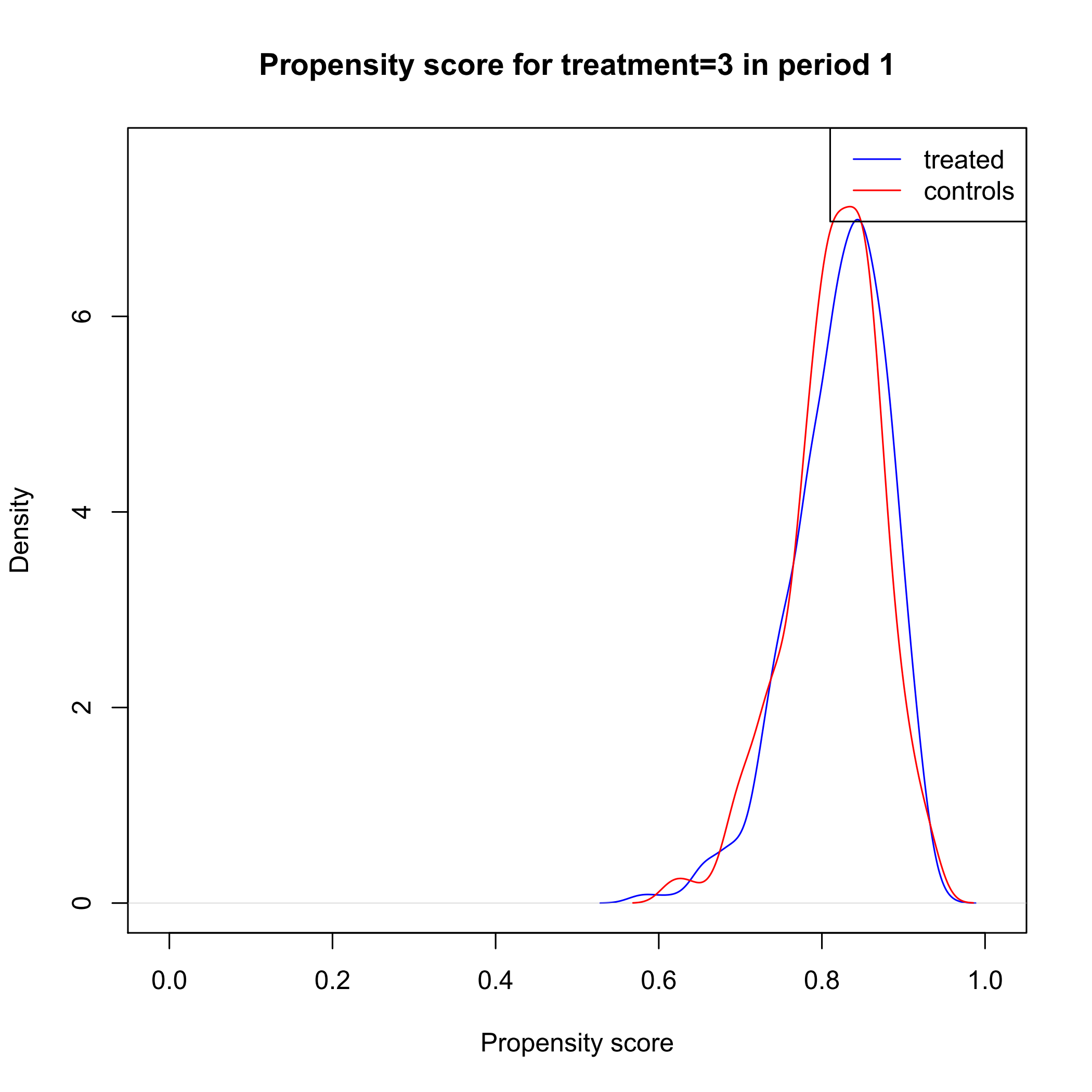}
\includegraphics[width=0.5\textwidth]{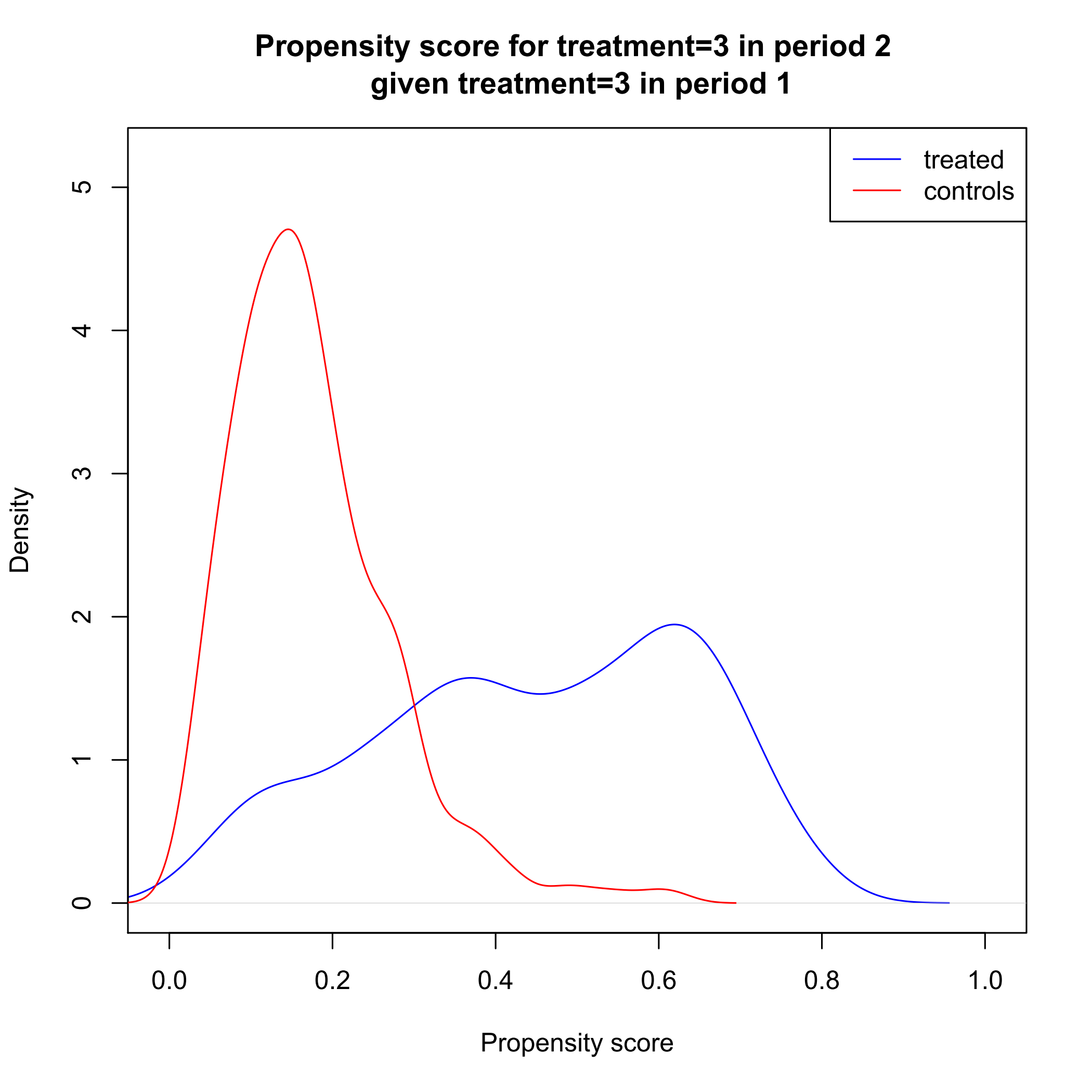}
\includegraphics[width=0.5\textwidth]{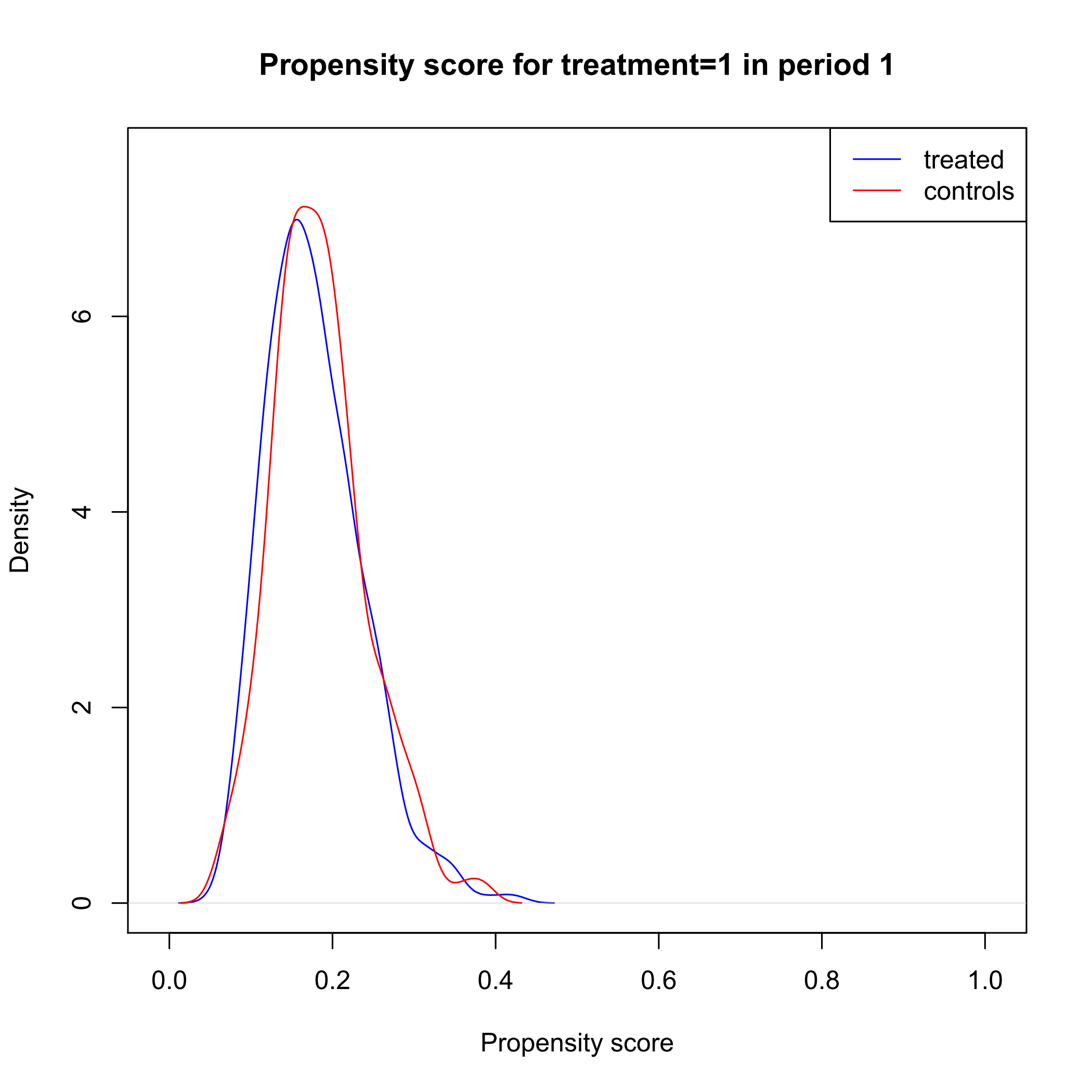}
\includegraphics[width=0.5\textwidth]{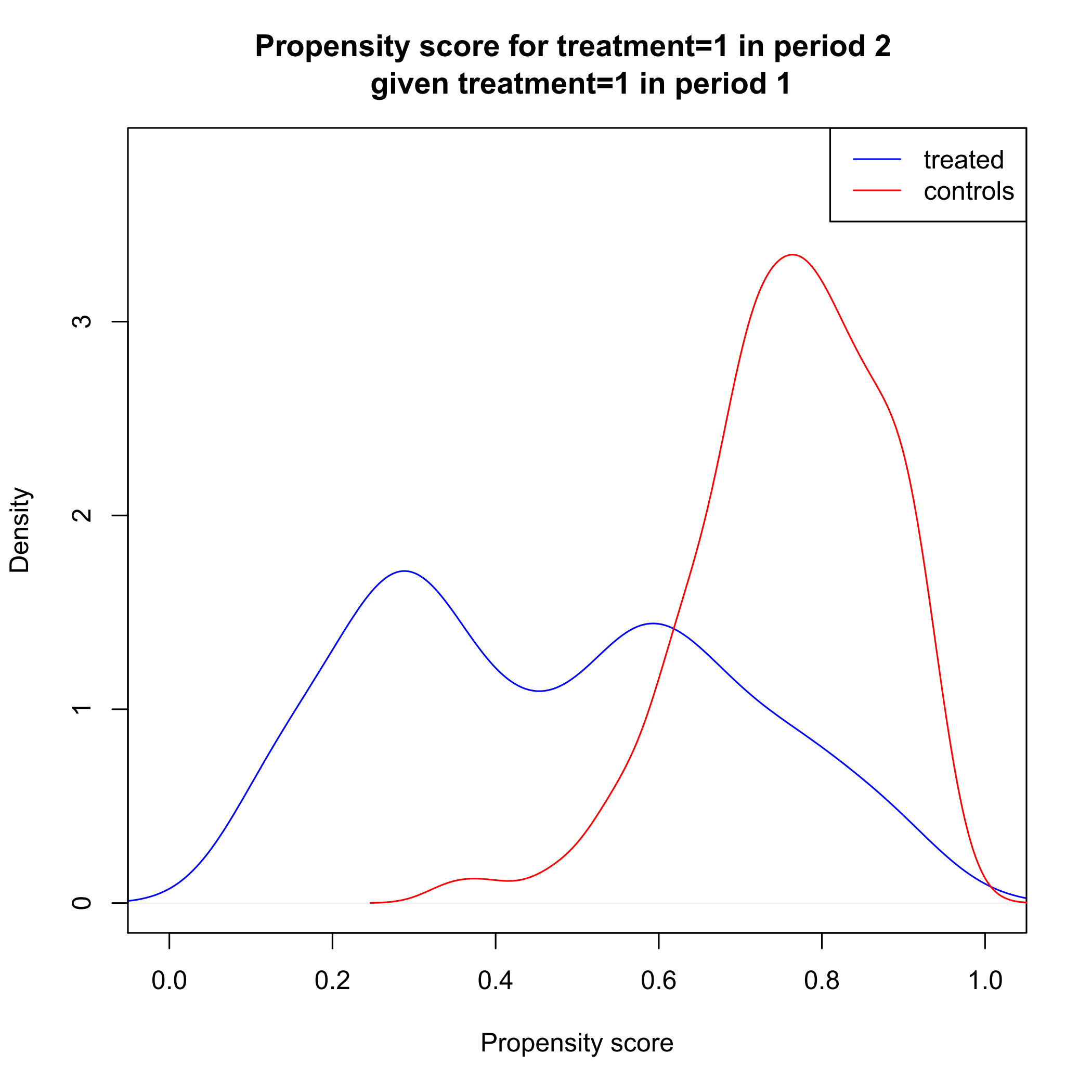}
\end{figure}

\newpage
\begin{figure}[htbp]
\caption{Support for treatment sequences 33 vs.\ 22 with a trimming threshold of 0.03}
\label{fig:support1}
\includegraphics[width=0.5\textwidth]{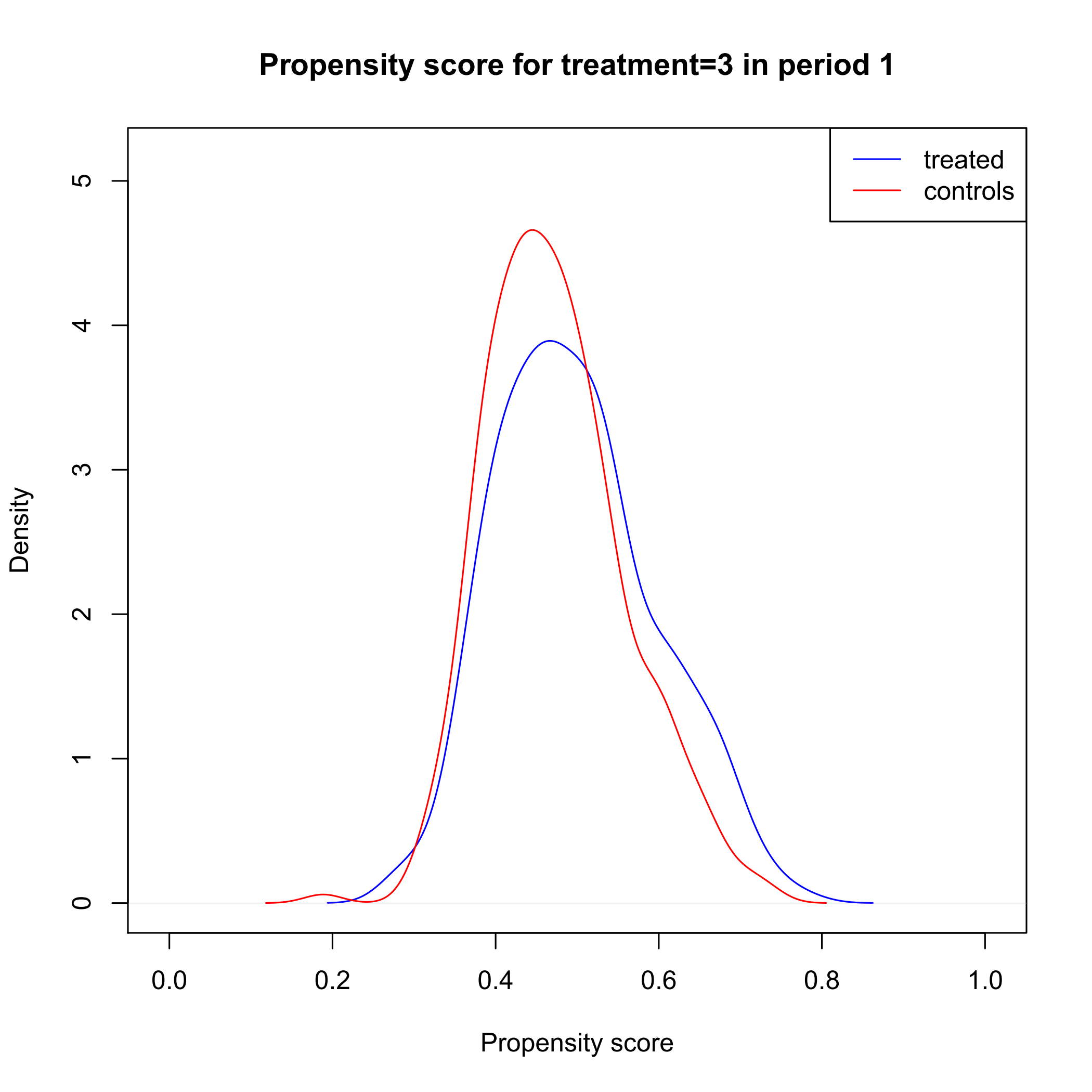}
\includegraphics[width=0.5\textwidth]{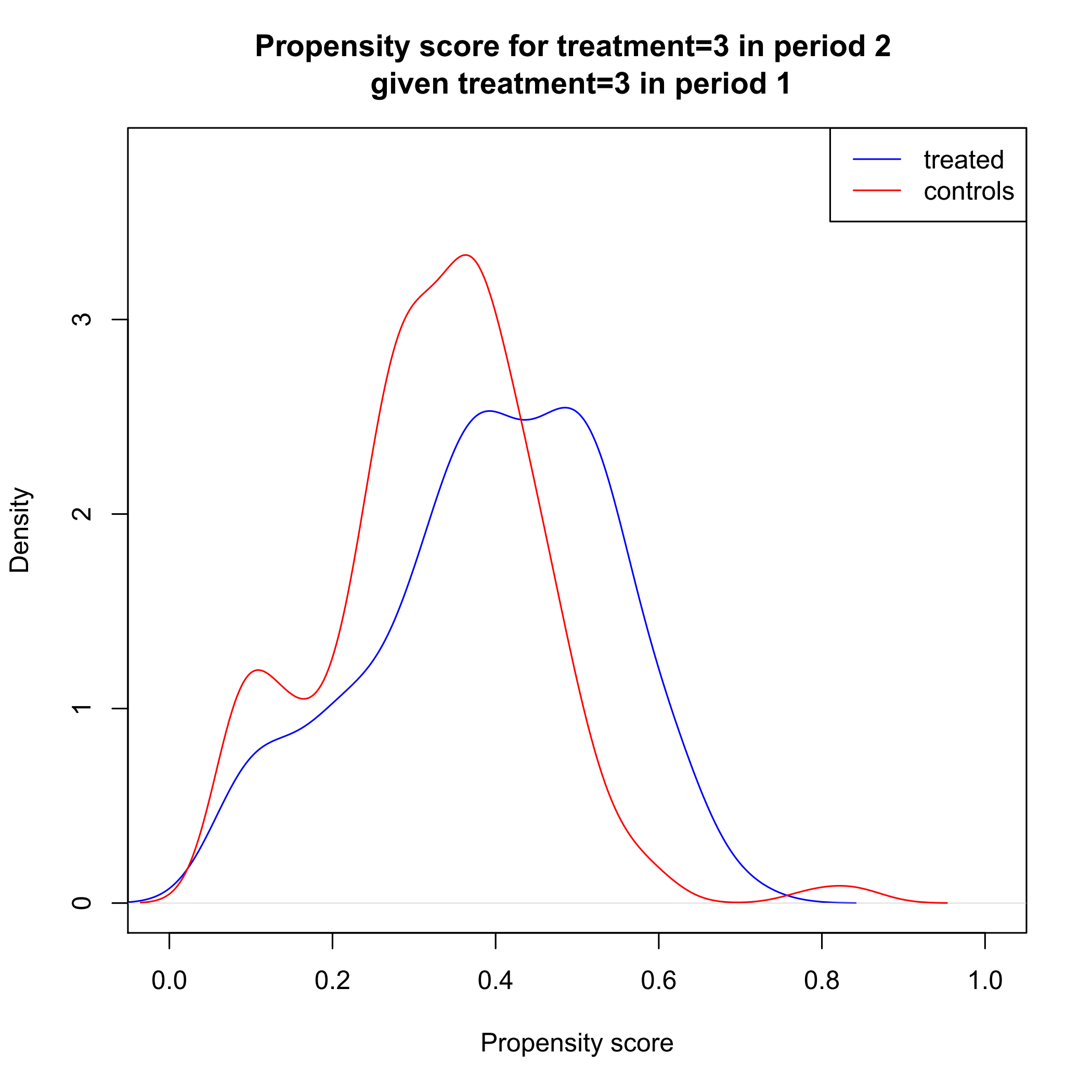}
\includegraphics[width=0.5\textwidth]{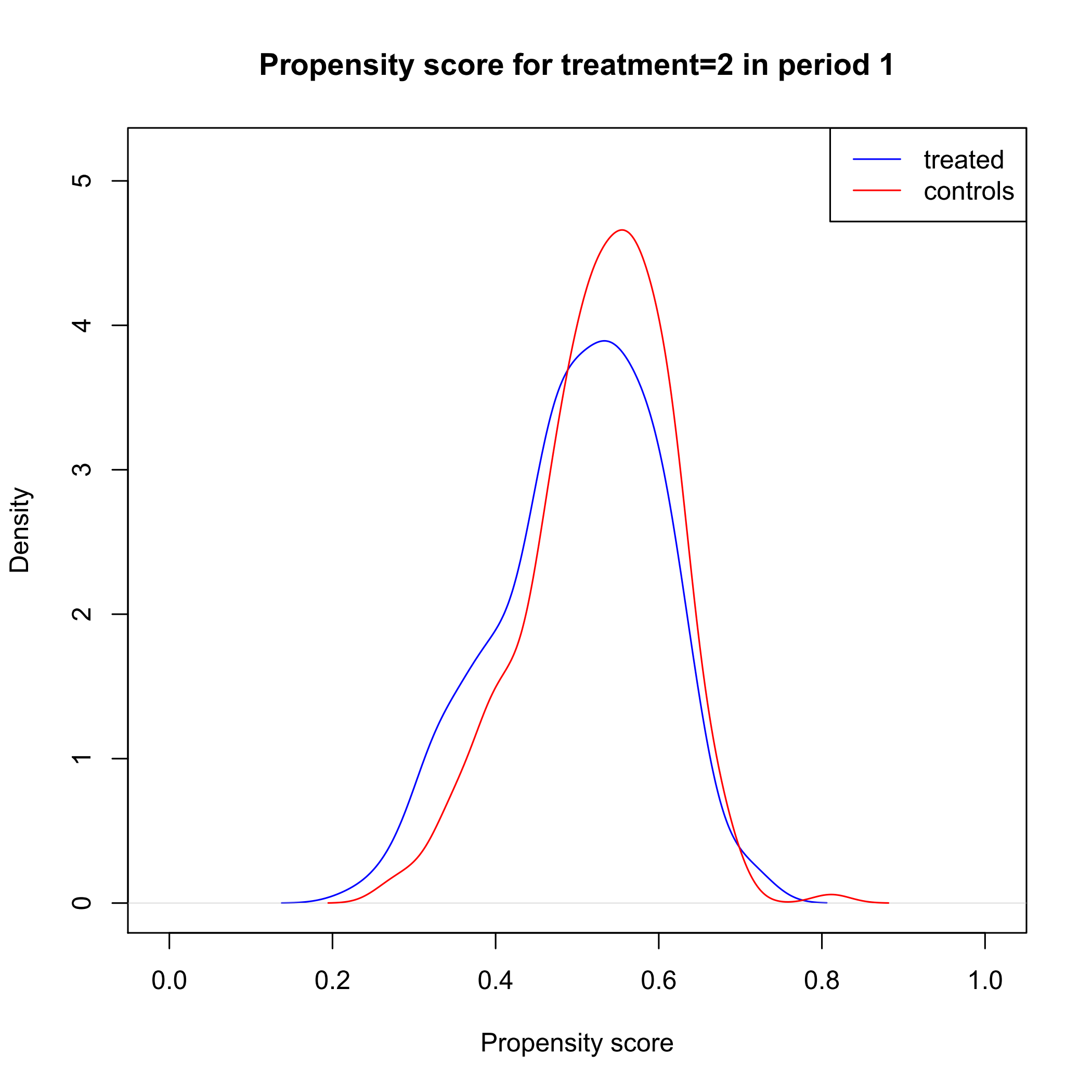}
\includegraphics[width=0.5\textwidth]{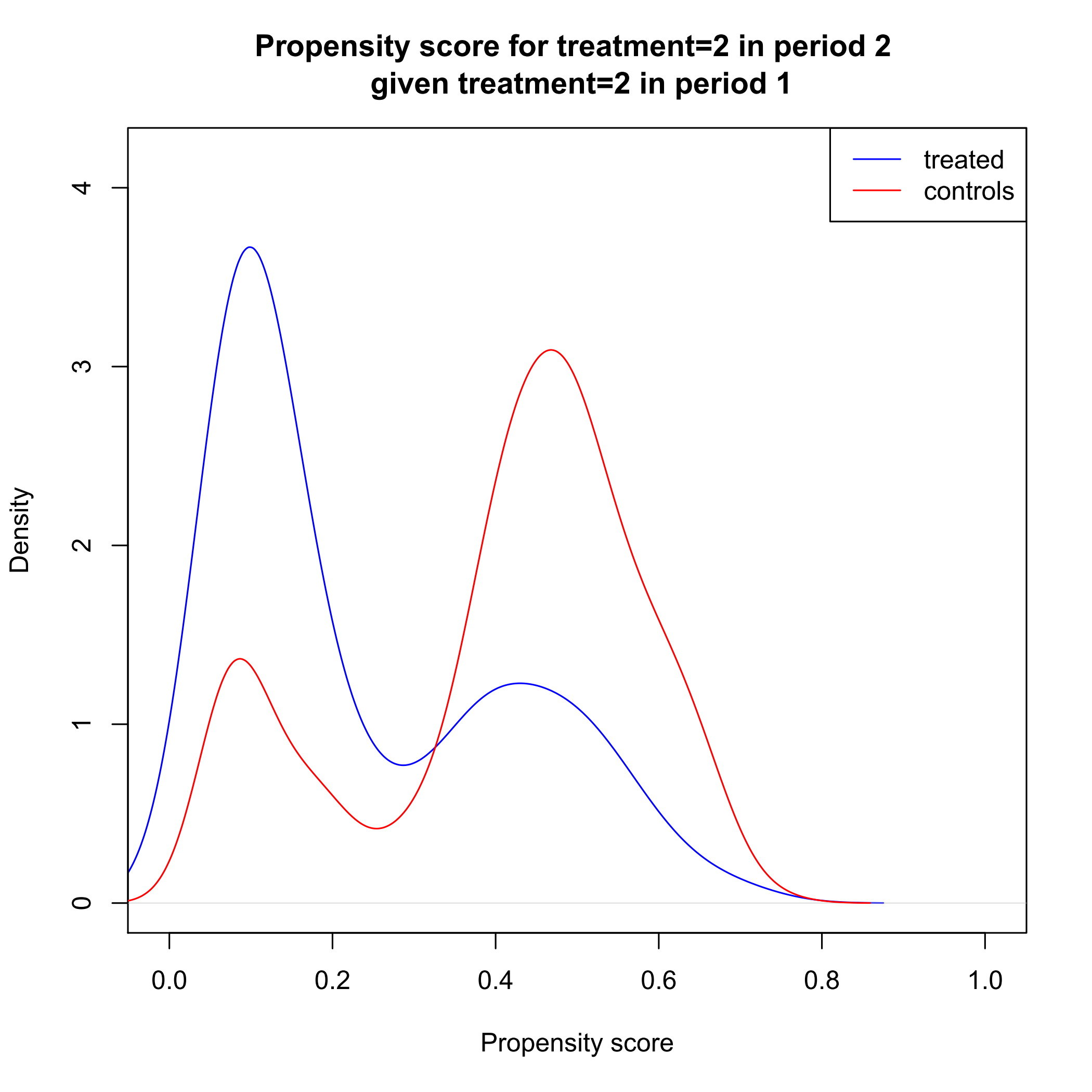}
\end{figure}

\newpage
\begin{figure}[htbp]
\caption{Support for treatment sequences 33 vs.\ 21 with a trimming threshold of 0.03}
\label{fig:support1}
\includegraphics[width=0.5\textwidth]{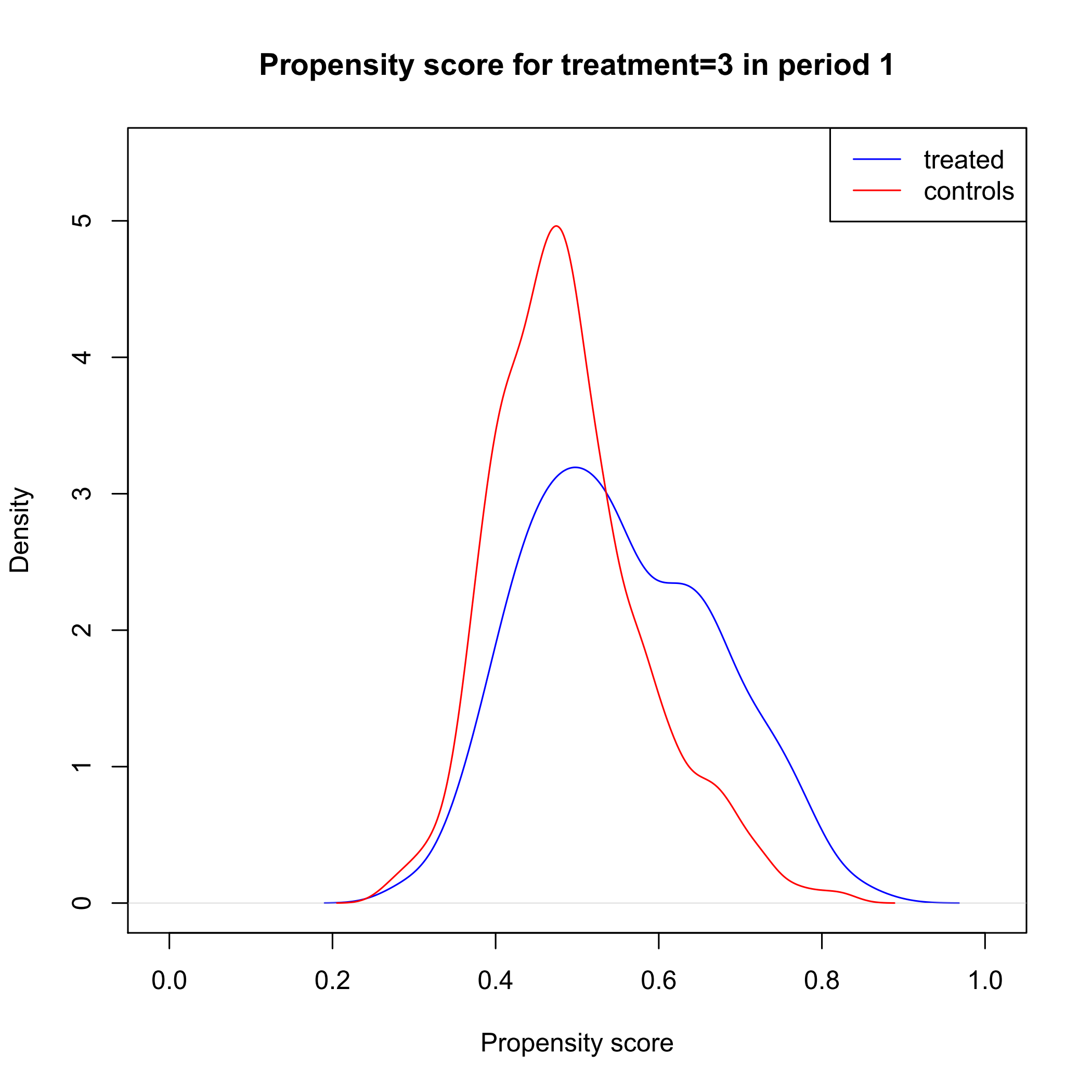}
\includegraphics[width=0.5\textwidth]{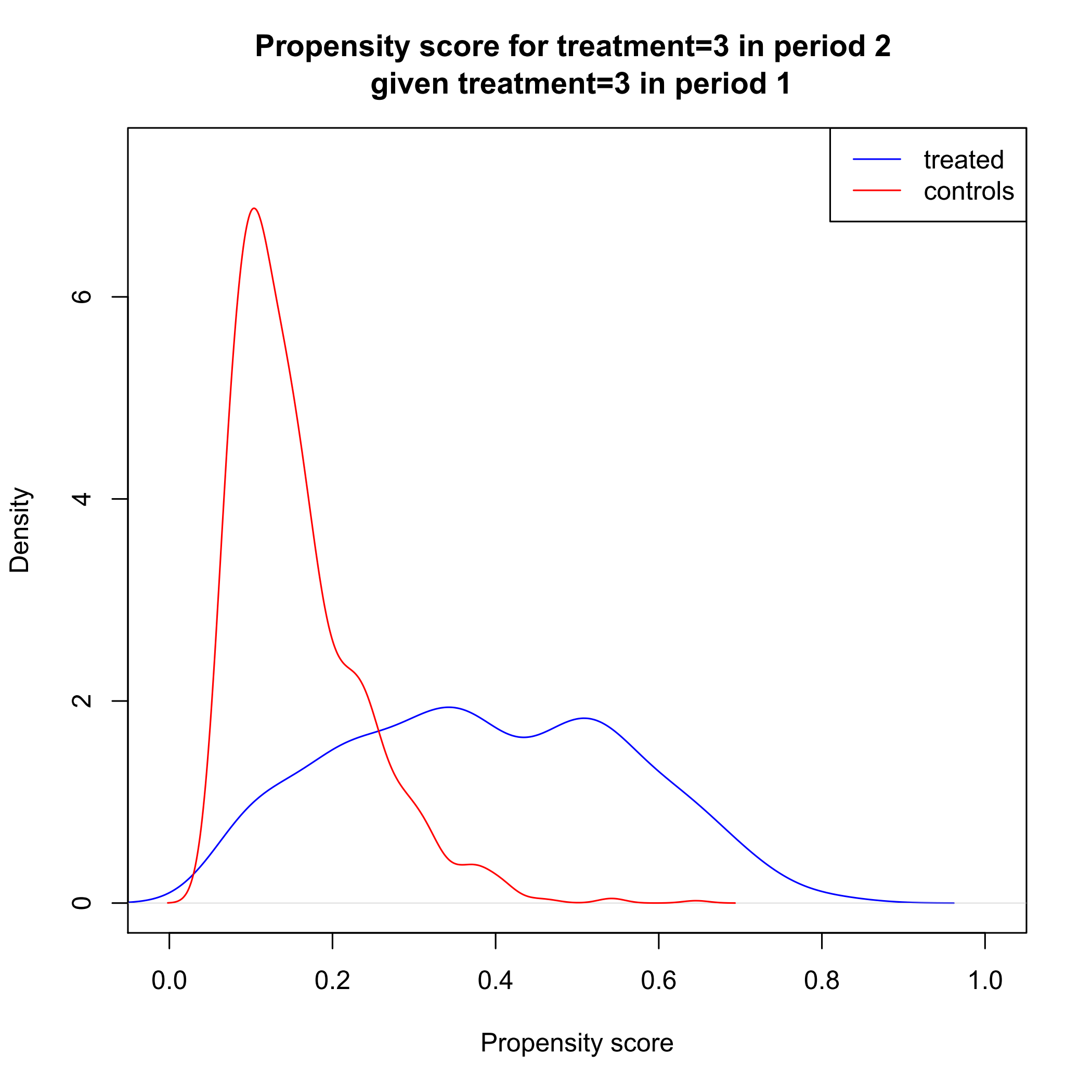}
\includegraphics[width=0.5\textwidth]{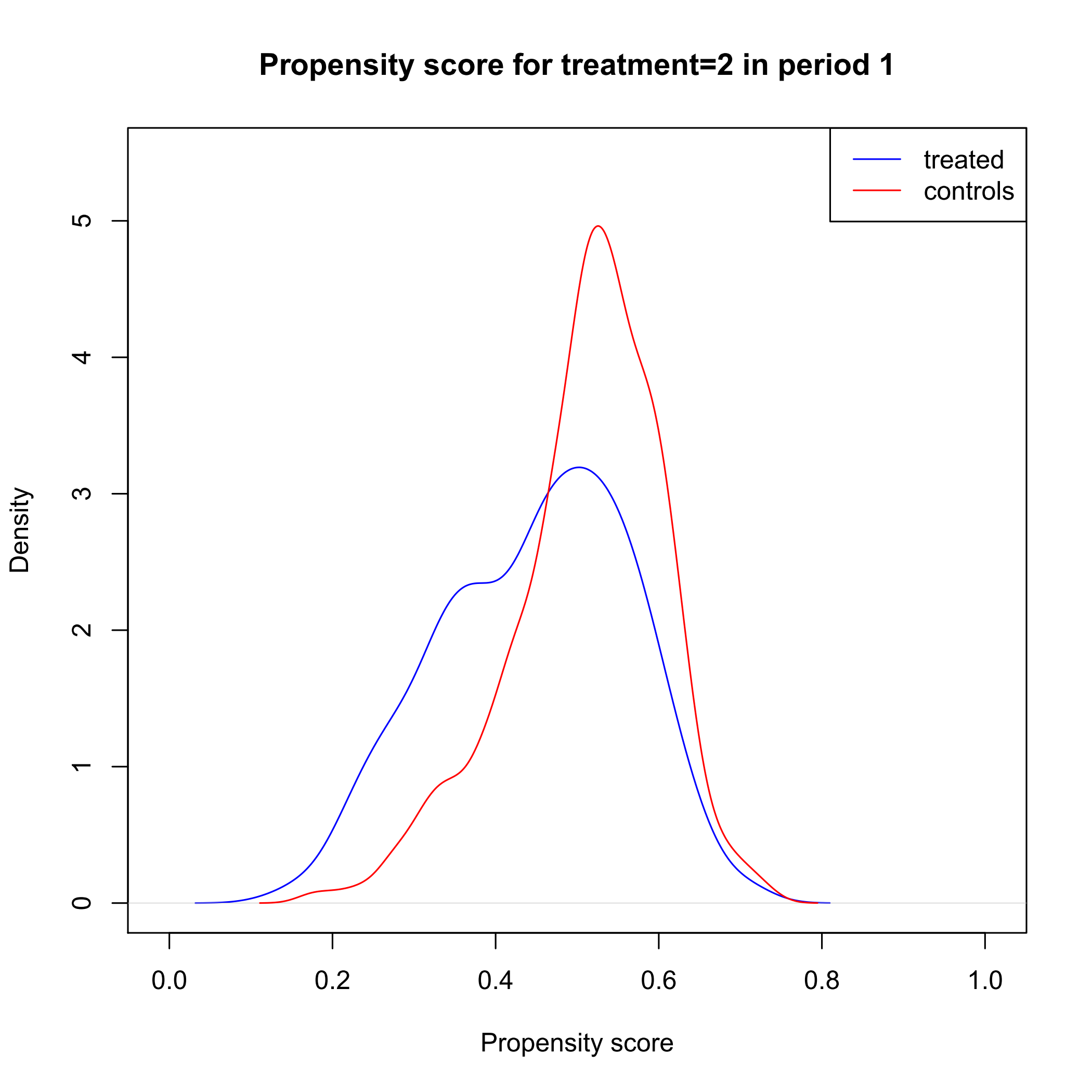}
\includegraphics[width=0.5\textwidth]{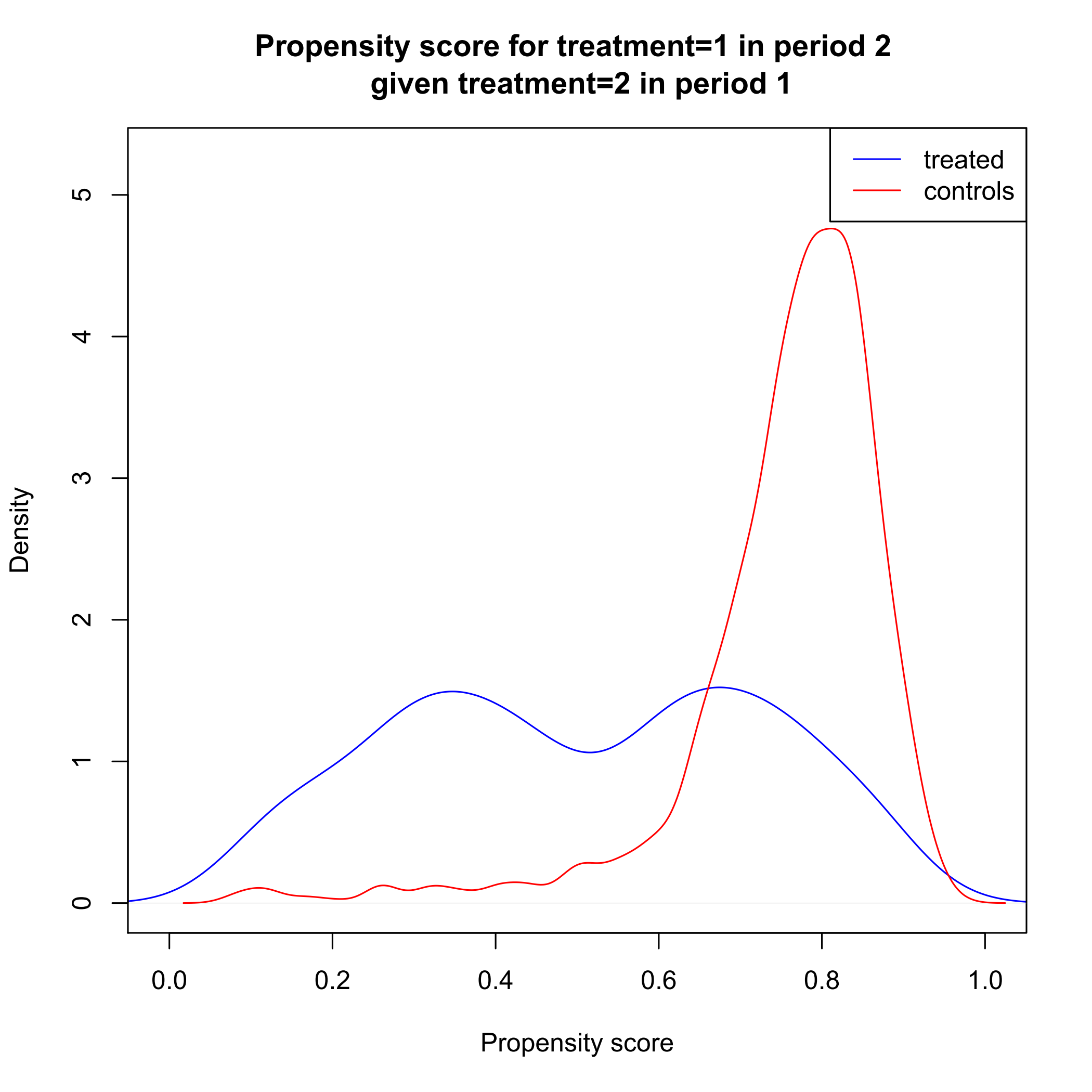}
\end{figure}

\newpage
\begin{figure}[htbp]
\caption{Support for treatment sequences 33 vs.\ 11 with a trimming threshold of 0.03}
\label{fig:support1}
\includegraphics[width=0.5\textwidth]{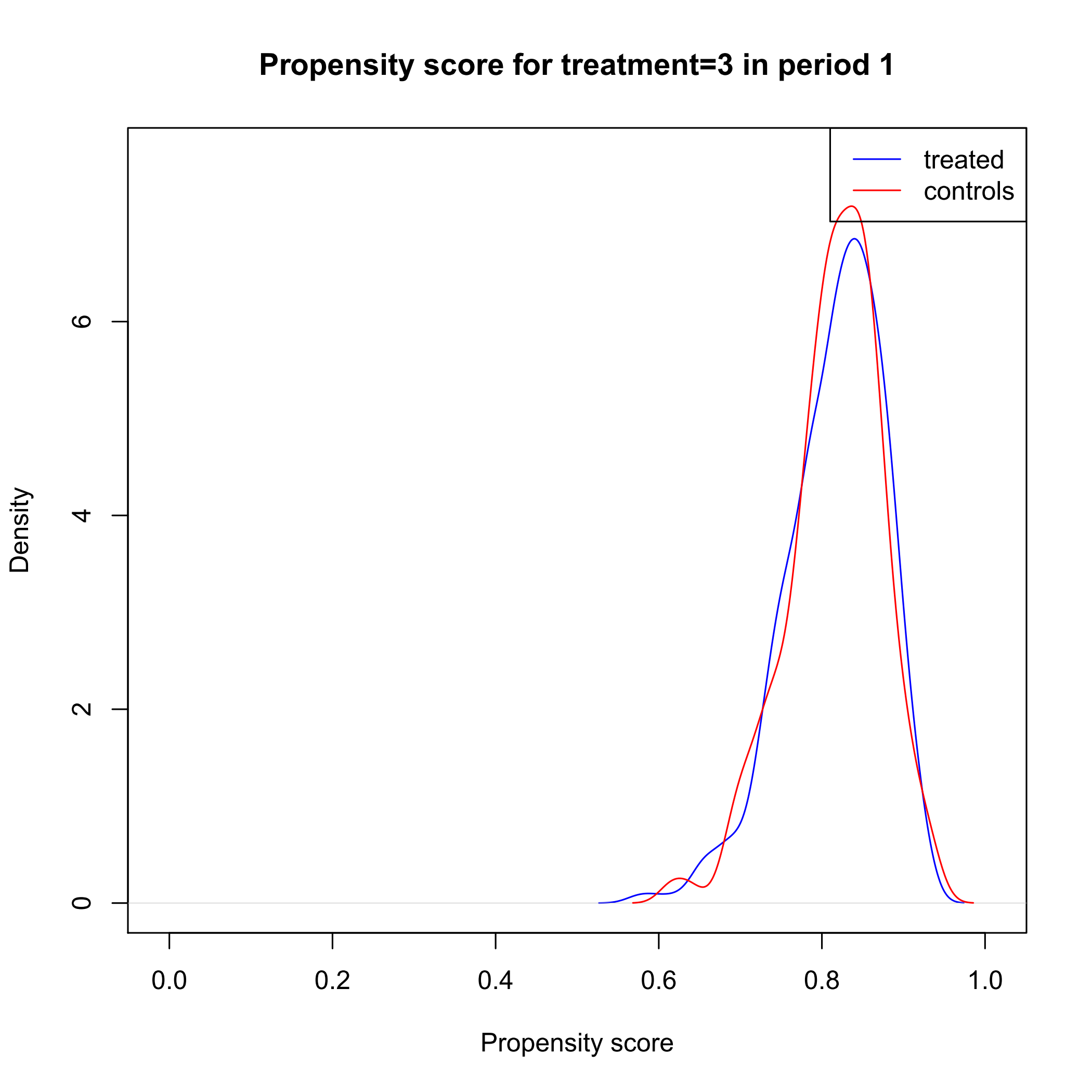}
\includegraphics[width=0.5\textwidth]{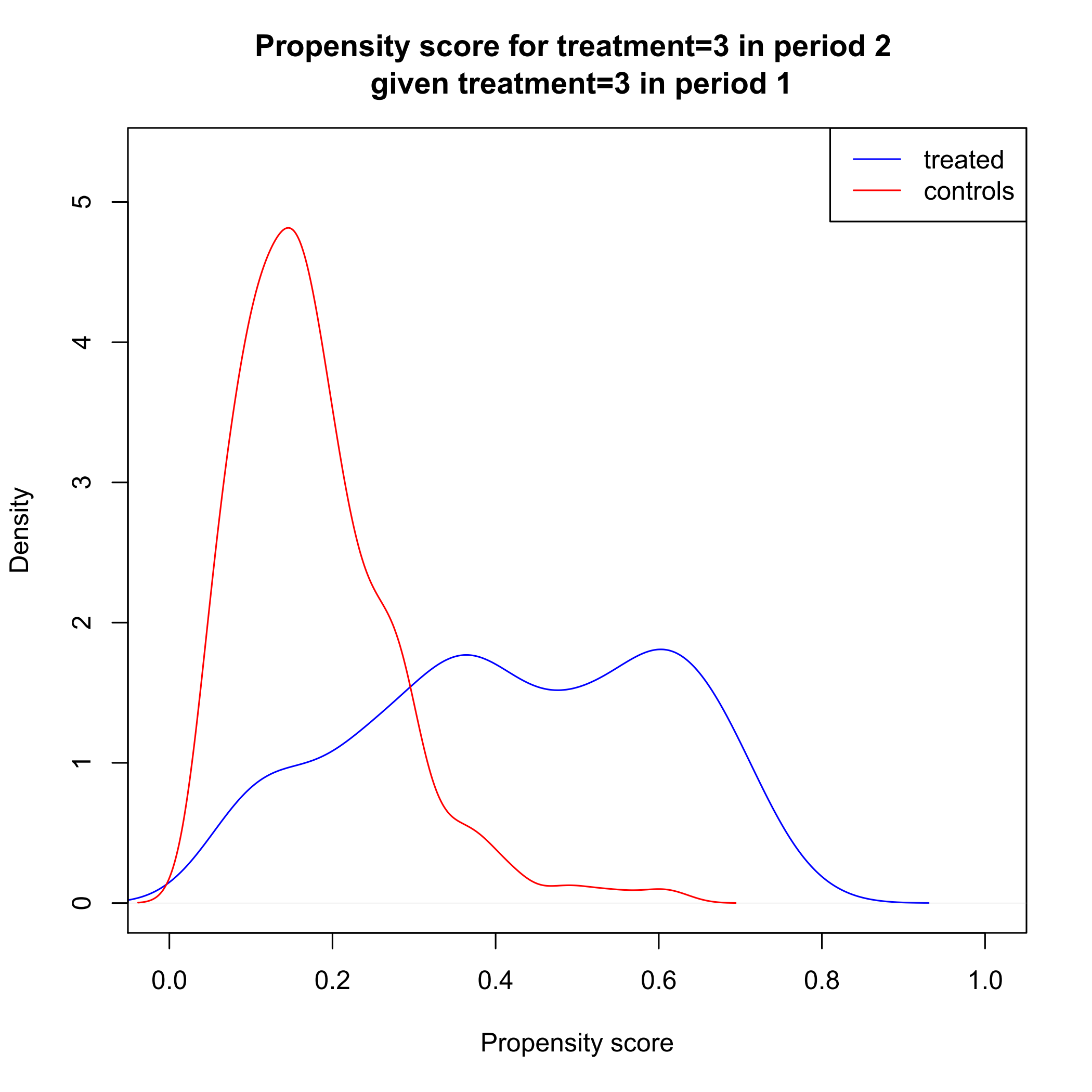}
\includegraphics[width=0.5\textwidth]{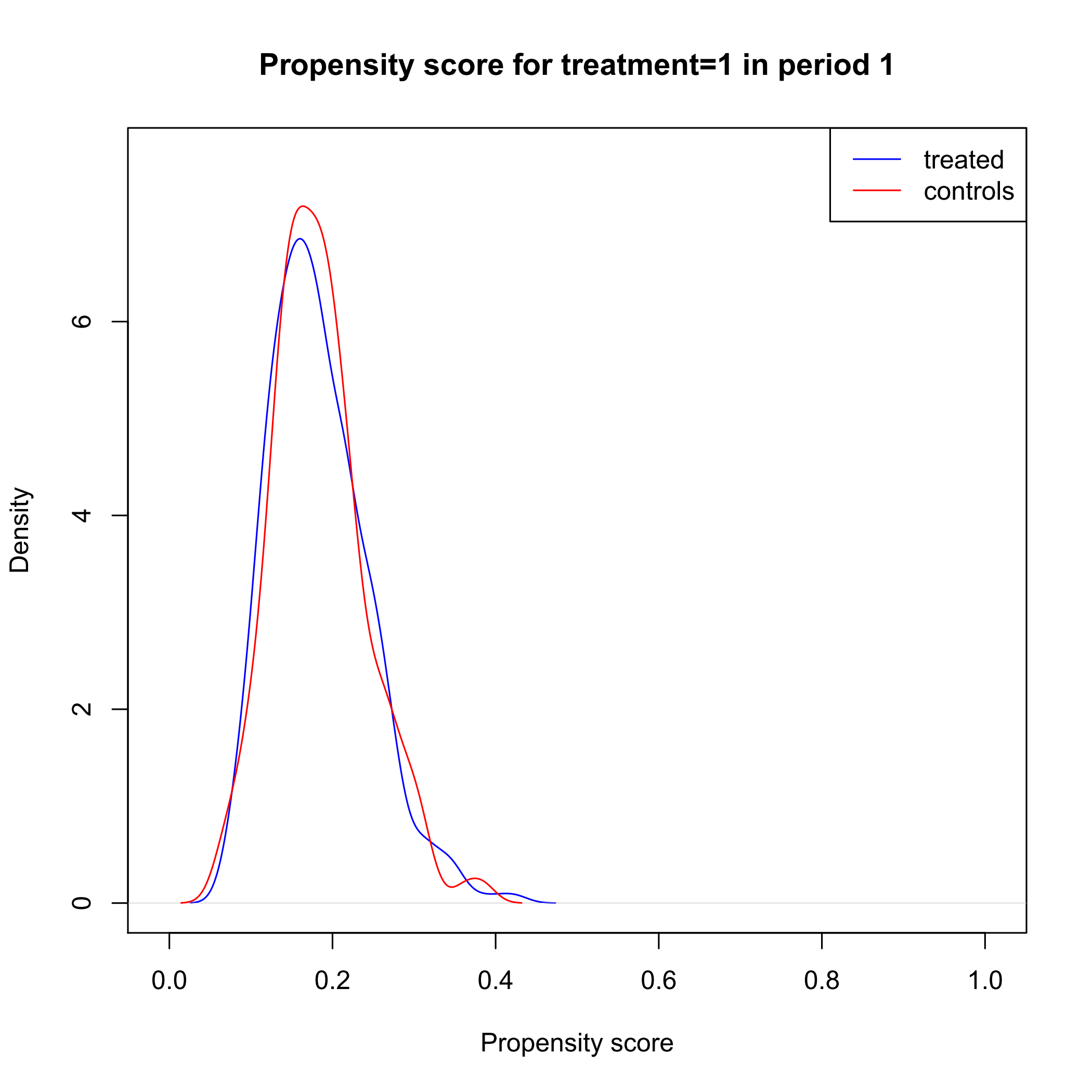}
\includegraphics[width=0.5\textwidth]{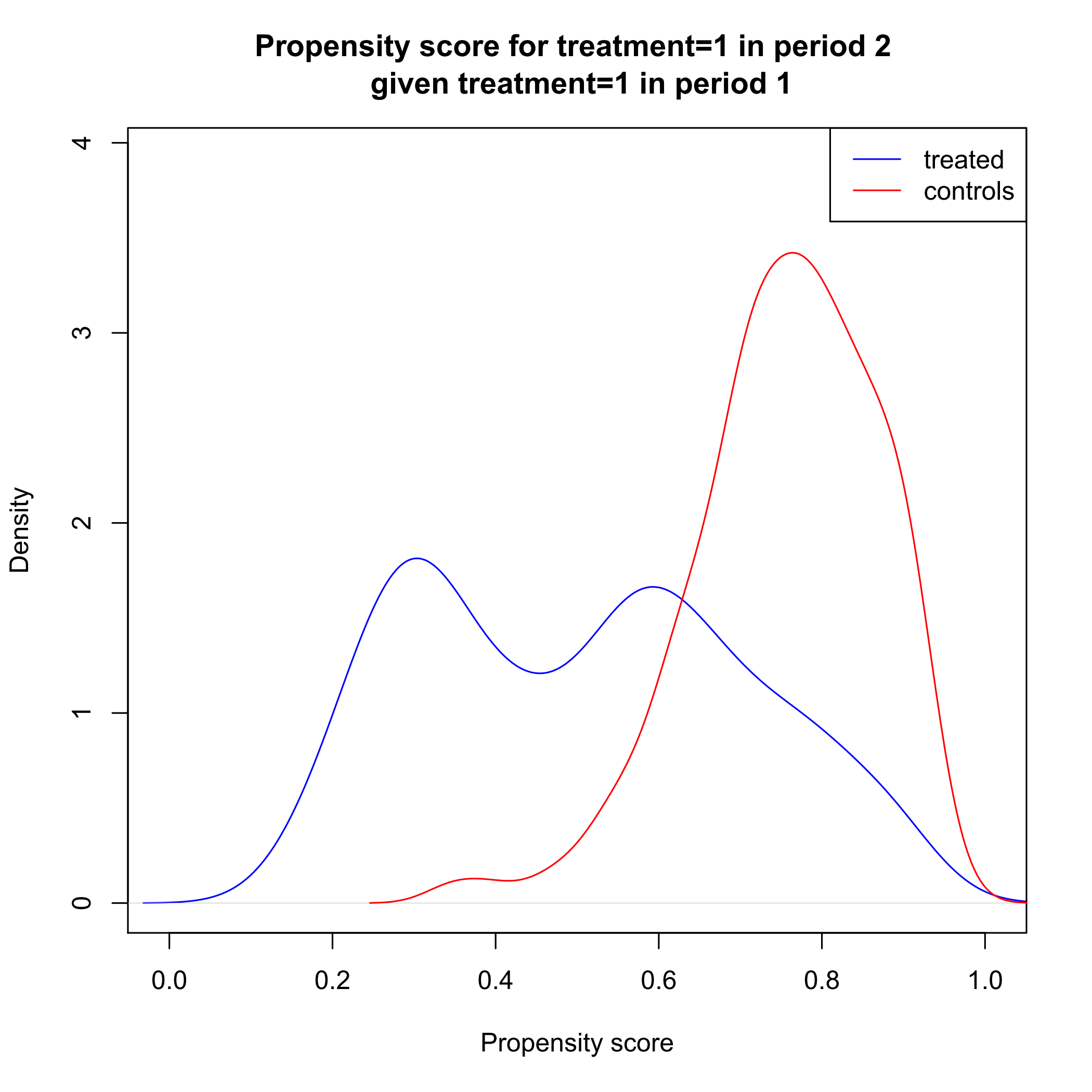}
\end{figure}

\end{appendix}
\end{document}